\documentclass[fleqn,usenatbib]{mnras}

\usepackage{newtxtext,newtxmath}

\usepackage[T1]{fontenc}
\usepackage{ae,aecompl}

\usepackage{graphicx}	
\usepackage{amsmath}	

\usepackage{hyperref}
\usepackage{makecell}

\title[CoSANG I. Dark matter halo structure]{Coupling Semi-Analytic and N-body Galaxies (CoSANG) for cosmological stellar halo simulations I- Methods and the structure of dark matter halos}

\author[S. Talei et al.]{
Shahram Talei$^{1}$,\thanks{E-mail: stalei@crimson.ua.edu (UA)}
Krista Mccord$^{2}$,
Jeremy Bailin$^{1}$,
Darren J. Croton$^{3,4}$ \and
Alexandra Mannings$^{5}$,
Michael Sitarz$^{6}$,
Annelia Anderson $^{1}$ \and
Brooke Bailey $^{1}$,
Alexis Rollins $^{1}$
\\
\\
$^{1}$Department of Physics and Astronomy, University of Alabama Box 870324, Tuscaloosa, AL,
35487-0324, USA\\
$^{2}$Deep Space Systems Inc, 8100 Shaffer Pkwy, Littleton, CO 80127, USA\\
$^{3}$Centre for Astrophysics and Supercomputing, Swinburne University of Technology, P.O. Box 218, Hawthorn, VIC 3122, Australia\\
$^{4}$ARC Centre of Excellence for All Sky Astrophysics in 3 Dimensions (ASTRO 3D)\\
$^{5}$Department of Astronomy \& Astrophysics, University of California Santa Cruz, 1156 High Street, Santa Cruz, CA 95064, USA\\
$^{6}$Department of Physics \& Astronomy, University of Kansas, Lawrence, KS 66045, USA
}

\date{Accepted XXX. Received YYY; in original form ZZZ}

\pubyear{2022}
\begin{document}
\label{firstpage}
\pagerange{\pageref{firstpage}--\pageref{lastpage}}
\maketitle

\begin{abstract}

We present CoSANG (Coupling Semi-Analytic and N-body Galaxies), a new hybrid model for cosmological dark matter and stellar halo simulations. In this approach a collisionless model (Gadget3) is used for gravitational interactions while a coupled semi-analytic model (SAGE) calculates baryonic effects at each time-step. This live self-consistent interaction at each time-step is the key difference between CoSANG and traditional semi-analytic models that are mainly used for post-processing. By accounting for the gravitational effect of the baryons, CoSANG can overcome some of the deficiencies of pure N-body simulations, while being less computationally expensive than hydrodynamic simulations. Moreover CoSANG can produce stellar halo populations via tagging tracer dark matter particles. We demonstrate the performance and dynamical accuracy of this approach using both controlled test simulations and a set of three cosmological zoom-in simulations of Milky Way (MW) mass halos. We simulate each target halo both without the coupling (dark matter only, hereafter DMO) and with the coupling (CoSANG). We compare the internal structure (subhalo distribution, shape and orientation) of the halos. The following changes are observed in the CoSANG model compared to the DMO model: 1) the total number of subhaloes close to the center of the halo is reduced, 2) the $V_{\mathrm{max}}$ distribution peaks at a lower value and lies below the DMO model, 3) the axis ratio is smaller. The difference between DMO and CoSANG simulations is more significant in early forming halos.

\end{abstract}




\section{Introduction}
\label{sec:intro}

Over recent decades a combination of precise cosmological measurements and accurate theoretical models have revolutionized our understanding of galaxy formation.
The $\Lambda$CDM paradigm has become the most popular model and has been developed into a successful framework that explains many different cosmological observations with a remarkable precision, from the cosmic microwave background \citep[e.g.][]{Bennett_2013,Planck_2016}, to galaxy formation \citep[e.g.][]{Eggen_1962,White_1984,Cole_2000,Benson_2010,Silk_2012}. Nevertheless many aspects of galaxy formation are still highly debated.

Hierarchical structure formation play an important role in the $\Lambda$CDM scenario and galaxies build up their dynamical mass via dynamical accretion and mergers. Since dark matter acts as a potential well in the assembly of galaxies \citep{White_1978}, the distribution of the substructures and shape of the dark matter halo reflect their formation history. 
Therefore understanding the inner structure of the dark matter halo can help us to understand galaxy formation within our cosmological framework. 
Detection of these substructures can confirm key predictions of $\Lambda$CDM theory and rule out many alternative theories.

According to this bottom-up scenario, structures form out of primordial Gaussian fluctuations in the density distribution in the early universe and then grow to the current size. Although structures grow linearly in the early universe (\citealp{zeldovich_1970}), growth quickly becomes nonlinear with no known analytic solution, requiring numerical techniques to study. Gravitational interactions between all components govern structure formation \citep{Peebles_1974,Peebles_1982,Davis_1985}.

A simplified two stage model can capture many aspects of this interaction between baryons and the dark matter (\citealp{White_1978}, for a more recent review see \citealp{Oser_2010}). 
According to this model, the majority of stars are formed while falling in the dark matter potential well and therefore the invisible dark matter leaves a fingerprint on the visible stellar population. The gravitational potential of the baryons can alter the dark matter distribution in return \citep[e.g.][]{Garrison_Kimmel_2014,Bryan_2013,Zhu_2016,Graus_2019,Kelley_2019}; the mutual interactions between these components is crucial. 

Since dark matter does not interact via the electromagnetic force, any visible component that traces the dark matter can provide a valuable probe of this distribution and be a bridge between observation and formation history.
Galaxies undergo different mechanisms like violent relaxation, phase mixing, chaotic mixing and Landau damping (\citealp{Lynden-Bell_1967},\citealp{Shu_1978}, \citealp{Tremaine_1986}, \citealp{Binney_2008}) and most of the formation history is lost in the main body of the galaxy. We can study galaxy formation either by looking at early galaxies at higher redshifts or we can extract the formation history through cosmological archaeology in the local universe at lower redshifts. Galactic components with a very long dynamical time can help with the later approach because unlike the other components, they are not relaxed. 

Stellar halos are a clear example of hierarchical accretion. The stellar halo is the most extended stellar component of a galaxy and it is the remnant of disruption. Although the stellar halo is a small fraction of stars in a galaxy, it displays a rich multi-component inner structure. The stellar halo of the Milky Way (MW) contains very old and metal-poor stars. It has a much longer relaxation time than the disk, and plays a key role as a time capsule preserving the formation history of the galaxy. The stellar halo can interact with the dark matter halo directly, and therefore can help us to understand the dark matter and determine the stellar content of the galaxy (\citealp{Eggen_1962,Zaritsky_1994,Helmi_1999,Johnston_2008,Kuhlen_2009,Vera_Ciro_2013,Bonaca_2014,Dai_2018,Bonaca_2018,DSouza_2018,Zaritsky_2020,Nadler_2021}; see \citealp{Helmi_2020} for a recent review).
Therefore a combined study of the dark matter halo and the stellar halo can shine a light on galaxy formation and bridge the gap between theory and observation.

Numerical simulations are extensively used to study galaxy formation and they are powerful tools to examine various models. Today these models can include multiple physical processes to reproduce more realistic galaxies; however despite many improvements in technology and the implementation of new efficient algorithms, simulating galaxies and their stellar halos is still extremely computationally expensive. This regime is very sensitive to the numerical parameters we choose (\citealp{Power_2003}) and limited by numerical resolution
and systematic errors \citep{Bailin_2014}. 

Any computational modeling is an attempt to improve resolution, cosmological sample size and physics (the number of processes that are included). Increasing any of these factors is constrained by computational resources. 
Depending on the questions that one wants to answer, different combinations of the simulation set-up can be optimized. However, most of the time this optimum requires compromising at least one other aspect of the simulation that could be important to explain observations. 

Reproducing the relevant physical processes with fidelity is often the highest priority. Although dark matter is well described with collisionless gravitational interactions (even though the true nature of the dark matter is still unknown) (\citealp{Kuhlen_2012}), baryonic matter interacts via multiple channels. Several recent studies include these processes, with a very high computational cost \citep[e.g.][]{Sanderson_2018,Mackereth_2019,Yu_2020}. 

Many dark matter only (DMO) studies ignore baryons to reduce the computational cost. They may attempt to recover the missing structures through post-processing with an empirical model, but the missing baryon-dark matter interaction has a significant dynamical impact on dark matter structure and eventually on the stellar population. Full physics hydrodynamic models are not immune from these systematic errors either. In full physics simulations a sub-grid model can capture baryon-baryon interactions at sub-resolution scales.

A new fast and more accurate model that can produce a large galaxy sample at very high resolution would be an extremely powerful tool to understand galaxy formation. In this series of papers, we present one such approach: Coupled Semi-Analytic and N-body Galaxies (CoSANG), which takes advantage of a fast N-body code at its core, coupled with a well calibrated semi-analytic model (SAM) of galaxy formation. This model has limits but it is well suited for stellar halo simulations. 
It includes dynamical coupling between dark and baryonic components, which can produce more realistic dynamics, and tags stellar properties onto tracer dark matter particles do produce stellar halo populations.

CoSANG is used in three major studies: dark matter kinematics, stellar tracers kinematics and stellar halo populations.
In this work we focus on the dark matter kinematics with CoSANG by studying dark matter halos, focusing on their shape; the abundance and the distribution of substructures; stellar halos produced by CoSANG will be discussed in upcoming work.

This paper is structured as follows: We start with a general overview of dark matter in Section \ref{sec:DM} then summarize the observational results and theoretical models and simulations
in Section \ref{sec:halosimulations}. 
In Section \ref{sec:cosang} we present CoSANG and different components of the model and how we test them. In Section \ref{sec:results} we present dynamical results with both the DMO model and CoSANG and compare substructures and internal structure of the dark matter halos. Particle tagging with CoSANG and a stellar halo analysis will be discussed in a separate paper.

\section{Dark matter halos}
\label{sec:DM}

The gravitational collapse of dark matter halos can produce a universal profile (\citealp{Einasto_1965}, \citealp{Einasto_1989}, \citealp{Navarro_1996}), but the internal structure can vary between halos as a result of different formation histories and baryonic content (\citealp{White_1978}). These differences might manifest in the distribution of subhaloes, or the shape and orientation of the halos.

Subhaloes around the main halo have been studied extensively. Substructures can form and survive in small clumps even in very inner regions of the galactic halo (\citealp{Diemand_2008}). 
The mass distribution of subhaloes can be described by a power-law distribution \citep[e.g.][]{Helmi_2002,Springel_2008} or Schechter function (\citealp{Schechter_1976}).

Although many early studies showed discrepancies between observations of the substructures and predictions from N-body and hydro models (e.g., \citealp{Klypin_1999}, \citealp{Moore_1999}),
many of these discrepancies are reduced or disappear when including more and better baryonic physics in simulations \citep[e.g.][]{Bullock_2000,Brooks_2013,Wetzel_2016}.
On the other hand the population of the observed satellites is growing rapidly and observational incompleteness could be a major concern 
\citep{Tollerud_2008,Walsh_2009,Hargis_2014,Kim_2018}.

Substructures within MW mass galaxies have been studied with ELVIS (Exploring the Local Volume in Simulations) in models of both zoom-in Local Group analogues and isolated  structures (\citealp{Garrison_Kimmel_2014}). They find Local Group-like pairs average almost twice as many companions and the velocity field is kinematically hotter and more complex on Mpc scales.
They report no difference in the abundance or kinematics of substructure within the virial radii of isolated versus paired hosts. 
Extension of this study into hydrodynamic models show that energetic feedback from supernovae and subsequent tidal stripping in hydro models significantly reduces the DM mass in the central regions of luminous satellite galaxies (\citealp{Brooks_2014}).

Another important parameter of dark matter haloes that can be affected by baryonic physics is their ellipsoidal shape.
The most common method to define the shape of a halo  
is using the second moment of the mass distribution \citep{Katz_1991,Dubinski_1991,Warren_1992,Bailin_2005,Allgood_2006,Kuhlen_2007,Zemp_2011}. 

Dark matter halo shapes have been studied in many recent simulations. 
\citet{Vera_Ciro_2011} used the Aquarius N-body simulation \citep{Springel_2008} to demonstrate how the shape of individual halos changes with time, evolving from a typically prolate configuration at early stages to a more triaxial/oblate geometry at the present day.
Baryonic effects were explored in the Illustris simulation by comparing Illustris with the Illustris dark model, and show that condensation of baryons results in significantly rounder and more oblate halos, with the median minor-to-major axis ratio $\langle c/a \rangle \approx 0.7$ (\citealp{Chua_2019}). A similar study with 30 zoom-in MW mass galaxies in the Auriga simulations confirms the same conclusion, and also finds a strong alignment between the disk and halo shape (\citealp{Prada_2019}; see also \citealp{Bailin_2005}).

\section{Halo simulations}
\label{sec:halosimulations}

Today highly scalable algorithms start from a random initial condition consistent with the CMB distribution and they are capable of reproducing a large variety of observable features. 
For detailed review of different codes and their capabilities, see \citet{Benson_2010,Vogelsberger_2020,Angulo_2022}. 
The linear phase is evolved analytically in order to start the simulation at lower redshift where the nonlinear phase begins \citep[e.g.][]{Hahn_2011,Knebe_2009}.

Simulations can generally be categorized as either collisionless (N-body) or full physics (hydrodynamic) models.  We provide here a summary of the simulation techniques that have been used to date that inform how CoSANG is constructed.

\subsection{N-body and hydro models}
\label{sec:nbodyhalosimulations}

N-body models only include gravitational interactions and ignore dissipational baryonic physics. 
In general N-body models determine forces by solving Poisson's equation. 
Although N-body models might include baryonic components, such as stars, they do not take into account non-gravitational effects.

N-body models are not restricted by sub-grid physics; they are much faster than hydrodynamic models, thus for a specific amount of computational power, they can achieve a higher resolutions.
N-body models have been used extensively to study the local group \citep[e.g.][]{Besla_2010, Erkal_2018, Garavito_Camargo_2020, Vasiliev_2021, Naidu_2021}.

The main disadvantage of N-body models is the lack of baryonic processes, therefore they are suitable for dark matter simulations \citep[e.g.][]{Navarro_1996,Navarro_1997,Moore_1999,Gao_2004,Springel_2008,Klypin_2016} but they fail modeling baryonic contributions (e.g. galactic potentials, different feedback mechanisms, dissipation). 

N-body models lacking baryons do not predict the abundance of subhaloes accurately due to effects like adiabatic contraction, tidal disruption, and reionization, which shape the dark matter distribution in both the main halo and its subhaloes (\citealp{Zhu_2016}). 
In a comparison between DMO and hydrodynamic versions of the Illustris simulations \citep{Vogelsberger_2014,Nelson_2015}, \citealp{Graus_2018} found that not only is the overall number of subhaloes reduced in the hydrodynamic simulation, but their radial distribution is less centrally concentrated, which suggests that dynamical disruption of substructure from the central host galaxy can reduce substructure counts compared to the DMO simulation. 

One may improve N-body models adding baryonic effects. N-body models can include baryons such as stellar components (no interactions other than gravity, e.g. \citealp{Garavito_2019}), and if those components are themselves accurate then the N-body simulation will faithfully reproduce the dynamics.

Hydrodynamic models include full physics. The higher computational cost is the main disadvantage of these models. 
Subsequently small scale sub-physics is not resolved and an alternative sub-grid prescription is used to include important sub-grid processes such as star formation and cooling (e.g., \citealp{Springel_2003}). Sub-grid models are calibrated with different constraints \citep[e.g.][]{Schechter_1976} and they are capable of reproducing more realistic galaxies \citep[e.g.][]{Vogelsberger_2014_2,Schaye_2015,Wang_2015,Grand_2017,Hopkins_2018}. Sub-grid models can impose a big restriction on the predictability of the model \citep{Munshi_2019,Li_2020}.

Examining 30 galaxies in the Auriga simulation shows a diverse inner structure and shapes and a connection between population gradients and mass assembly histories: galaxies with few significant progenitors have more massive halos, possess large negative halo metallicity gradients, and steeper density profiles \citep{Monachesi_2016_2,Monachesi_2019}.

\subsection{Semi-analytic and hybrid models}
\label{sec:SAMhalosimulations}
Beside explicit simulation, implicit empirical models, also known as semi-analytic models (SAMs), have been extensively used as an independent technique or in combination with an explicit model to understand the underlying physical processes. In general SAMs calculate the evolution of the baryonic component analytically while the evolution of the dark matter component is either imported from an N-body model or produced using Monte Carlo technique \citep[e.g.][]{White_1978,Cole_1991, Cole_2000,White_1991,Baugh_2006, Croton_2006, Croton_2016}, or take advantage of machine learning (e.g., \citealp{Behroozi_2019}).

SAMs can produce remarkably consistent results at much lower computational cost compared to hydrodynamic models (\citealp{Benson_2001}). The main advantage of using SAMs is that these models are not restricted by resolution. SAMs are complimentary models and they rely heavily on observational data or explicit models as their input and for calibration.

Any combination of SAMs and N-body models can belong to one of these categories:

\emph{Passive SAMs:} No interaction; SAMSs are used independently or for post-processing without any live dynamical interactions \citep{Cole_2000}\\
\emph{Active SAMs:} One side interaction; In these models an analytic baryonic galaxy is included in the dynamics but this galaxy does not evolve according to the dark matter halo evolution even though there might be a prescription for the time evolution. This technique is more common in non-cosmological simulations, for example stellar streams kinematic, such as Sagittarius, LMC, and the Milky Way (e.g. see \citealp{Vasiliev_2021} and references there).\\

\emph{Interactive SAMs:} Mutual interaction; There is a mutual interaction between an explicit simulation and the SAM. There is not many example in this category and perhaps the closest examples are \citet{Garrison-Kimmel_2017} and \citet{Kelley_2019}, however, in those cases the growth of the galaxy is calculated using halo abundance (\citealp{Behroozi_2010}) rather than a SAM.

Empirical models can open a new window into low cost stellar halo modeling, when they are coupled with an N-body model using a technique known as particle tagging \citep{Bullock_2005,LeBret_2017}. 
Justified by the similarities between accreted dark matter particles and stellar halo populations, a sample of the virialized particles are modified and tagged as stars in a non-cosmological DMO simulation (\citealp{Bullock_2005}). This model has been successfully used, for example in recovering accretion histories with substructure abundance (\citealp{Johnston_2008}) and chemical space abundance (\citealp{Font_2006}) using a prescription to correlate accretion history and enrichment due to different enrichment mechanisms (\citealp{Robertson_2005}).
This prescription is extended into cosmological simulations \citep{Cooper_2010,Rashkov_2012,Cooper_2017}.

Comparison between observations and hydrodynamic simulations confirms that accretion-only models produce promising results at much lower costs.
Application of particle tagging in both hydrodynamic models and N-body models, and side by side comparison with the stellar halo population directly produced by a hydrodynamic model, shows the existence of systematic errors (\citealp{Bailin_2014}) although it is possible to reduce the error by improving the SAM and tagging scheme (\citealp{Cooper_2017}). 

These accretion based approaches have three assumptions in common (\citealp{LeBret_2017}):
\begin{enumerate}
  \item Stars form tightly bound to their parent halos.
  \item Recently formed star particles and their tagged DM analogues subsequently follow similar phase space trajectories.
  \item Baryonic effects are not important in shaping the stellar halo.
\end{enumerate}

The third assumption is more problematic. Direct comparison between hydrodynamic models and N-body models confirms the significance of these baryonic effects \citealp{Brooks_2014,Wetzel_2016, Zhu_2016, Samuel_2020}.
We can test the effect of a baryonic disk by adding a galactic potential into a DMO model and compare the results with the same model with no disk. This potential make itself apparent in substructure distributions and the inner structure and the shape of the halo. Nearby satellite halos within the virial radius of the host halo have lower survival rates in the presence of a disk \citep{Donghia_2010,Bauer_2018, Graus_2019, Kelley_2019} and this baryonic potential combined with different feedback mechanisms could possibly explain the discrepancy between the number of dwarf satellites in DMO and hydrodynamic simulations.

A hybrid scheme can significantly improve the accuracy and the range of problems we can study with fast N-body models. Adding an evolving analytic galactic potential into a collisionless model using abundance matching (\citealp{Behroozi_2013}) results in significantly less substructures going back $\approx 8$ Gyr that is a better fit to pericentres (minimum separation between the host and the subhalo) calculations from Gaia observations and demonstrate an opposite missing satellite problem \citep{Kim_2018,Kelley_2019}, rather than too many subhaloes within 50 kpc of the MW that had $V_\mathrm{peak}\geq 20$ km/s to account for the number of ultra-faint galaxies already known within that volume today (\citealp{Graus_2019}).
There have been recent attempts towards hybrid stellar halo modeling (e.g. \citealp{Ricotti_2022}),
but CoSANG is the first self-consistent hybrid model for stellar halo simulations with a full baryon prescription consistent with dark matter evolution.

\begin{figure*}
	\includegraphics[width=\textwidth]{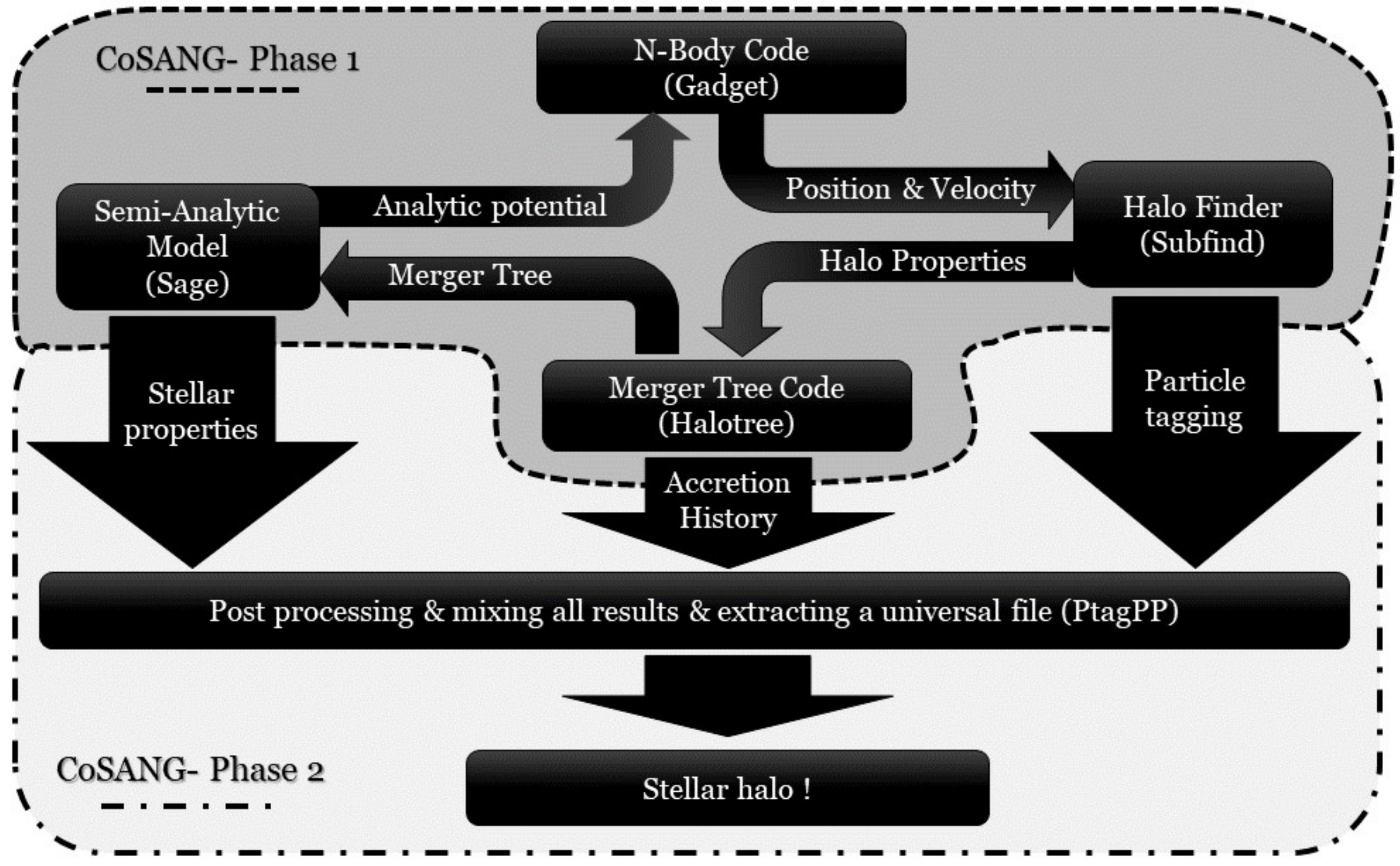}
    \caption{Anatomy of CoSANG. A CoSANG simulation consists of two phases: The live phase (grey shaded) and post-processing (white). In each time-step the N-body code Gadget calculates gravitational interactions. The halo finder Subfind extracts dark matter halo information, while Halotree extracts the merger tree and SAGE calculates the galaxy population using halos and the merger tree. An analytic potential is applied to affected particles and this happens in all time-steps. CoSANG produces the stellar halo in the second phase.}
    \label{fig:cosang_figure}
\end{figure*}

\section{CoSANG, structure and methods}
\label{sec:cosang}

CoSANG is a computationally inexpensive algorithm that takes advantage of a fast collisionless model for the explicit component of the simulation. 

There are two major factors that contribute to systematic errors in accretion-only models (\citealp{Bailin_2014}): 
\begin{enumerate}
    \item \emph{Inaccurate dynamics:} Ignoring baryonic potentials (e.g. bulge, disk, supermassive black hole) will result in a less accurate dynamical evolution (The importance of baryonic potentials is extensively discussed in Section \ref{sec:results}). 
    \item \emph{Particle tagging scheme:} The spatial and kinematic distribution and properties of stellar tracer particles will have an impact on the predicted halo structure \citep{Naidu_2021}. 
\end{enumerate}

CoSANG is an attempt to overcome these deficiencies. CoSANG is an interactive hybrid model.

The two phase development of CoSANG addresses these errors.
\begin{enumerate}
    \item \emph{Phase 1:} Self-consistent dynamics: CoSANG uses a live interaction between the SAM and the N-body model in each time-step in a live simulation (Figure \ref{fig:cosang_figure}).
    \item \emph{Phase 2:} Particle tagging: CoSANG implements a comprehensive SAM (Section \ref{sec:sage}) and a flexible tagging scheme (Talei et al. in preparation).
\end{enumerate}

CoSANG merges several different pre-existing code components so that they work together as one cohesive whole. CoSANG includes: Gadget3 (see \citealp{Springel_2005}; \citealp{Springel_2020} for descriptions of v2 and v4 respectively) as the N-body code, SUBFIND (\citealp{Springel_2001}) for the halo finder, L-Basetree and L-Halotree as the merger tree code, and the semi-analytic model SAGE (\citealp{Croton_2006}, \citealp{Croton_2016}). This allows the dark matter results from the N-body simulation to influence the results
in the SAM. Importantly the SAM results are fed back into the N-body simulation as it runs to perturb the dark matter particles.  This feedback
loop is continuous throughout a CoSANG simulation allowing both the dark matter
particles and baryonic matter to evolve together.

In each time-step Gadget3 (Figure \ref{fig:cosang_figure}) calculates the gravitational force. Subfind is implemented in Gadget3 and provides halo information. At each snapshot after the first, which are output approximately every 50 million years, L-Halotree derives the merger tree.
The SAM, SAGE, determines the galactic properties of each halo and is used to calculate an analytic gravitational force of the bulge, the black hole and the disk for all halos and subhaloes. We add this new term to the total force. 
This cyclic interaction continues at each time-step.

We discuss each code component in the next few sections.
Beside the main algorithm, CoSANG is equipped with many new analysis codes that are developed and used for the first time and are publicly available\footnote{\url{https://github.com/stalei/}}.

\subsection{N-body model Gadget}
\label{sec:gadget3} 
The objective of N-body codes is to solve Poisson\textquotesingle s equation to compute the
forces from the particle positions, advance the particles, and update their positions. Gadget is a well known parallel N-body TreePM code that uses MPI (message passing interface), which combines the gravity tree algorithm for calculating short-range forces with the particle mesh (PM) method for calculating long-range forces.

There are three important characteristics of the TreePM code: the grouping algorithm, the multipole expansion, and the opening criterion (\citealp{Springel_2005}). The grouping method used is an oct tree \citep{Barnes_1986}, which starts with a cube containing the entire mass distribution. It then breaks up the main cube into smaller and smaller cubes until each cube holds only one particle. Gadget uses this tree to incorporate gravitational forces using a hierarchical multipole expansion that groups the particles spatially such that the gravitational force on the particles is computed with $\mathcal{O}(N\log{}N)$ interactions.

The potential calculation is split into two parts, one for particles at short range and the other for more distant particles \citep{Ewald_1921,Hernquist_1991}. A spline kernel 
\citep{Monaghan_1985} 
is implemented so we have a Newtonian potential at long range and a modified Newtonian potential at short range.
At short range, softening becomes important since the potential cannot be evaluated with finite timesteps when the acceleration diverges, so Newtonian potential of a point mass at zero separation becomes $-Gm/\epsilon$ where $\epsilon$ is the force softening.
The softening for a halo with a particular mass is represented by equation
\begin{equation}
    \epsilon=4\frac{R_{\mathrm{vir}}}{\sqrt{\frac{M_{\mathrm{vir}}}{m_{\mathrm{particle}}}}}.
\end{equation}

Gadget3 is not public but a new version (version 4) was published recently with new improvements such as the multi-range force calculation (\citealp{Springel_2020}). CoSANG was developed before Gadget4 was published and there are fundamental changes in the new version (such as using C++ instead of C and object oriented structure instead of functional structure), which are inconsistent with other components we use.

\subsection{Halo finder SUBFIND and Rockstar}
\label{sec:halofinder}
Dark matter halo properties are retrieved by halo finders.
The halo finder SUBFIND was chosen since it is conveniently merged with Gadget3. 
SUBFIND (\citealp{Springel_2001}) combines two methods to find dark matter halos and their subhaloes.
The first method is Friends-of-Friends (FOF) which links nearby particles together
to locate large groups of dark matter particles \citep{Davis_1985} and then local overdensities are used to identify the substructures. Once SUBFIND locates all of the subhaloes within a parent halo, it uses an unbinding method to eliminate particles that are not bound to the halo. Particles with total positive (binding + kinetic) energy are iteratively removed from the particle list \citep{Springel_2001}.
After the unbinding process, halo properties such as velocity, virial mass, virial radius and angular momentum are calculated for the SUBFIND halos.

We use the Rockstar halo finder \citep{Behroozi_2012} in conjunction with SUBFIND for post-processing. Rockstar is a modern adaptive hierarchical refinement of friends-of-friends groups in 6D phase-space that avoids a few known issues with SUBFIND.

\subsection{Merger trees}
\label{sec:merger}

The merger tree code used in CoSANG is comprised of two components, L-Basetree and L-Halotree. These codes were created for constructing merger trees from SUBFIND and therefore readily accept the halo finder\textquotesingle s output. L-Basetree tracks halo particles between two snapshots to determine a halo\textquotesingle s descendant (\citealp{Springel_2005b}). A descendant halo can have multiple progenitors, but a progenitor can only be linked to one descendant. Particles that survive in a descendant halo are weighted based on their binding energy within the halo (\citealp{Springel_2005b}). The group of particles that have the highest binding energy are labeled as the descendant halo (\citealp{Srisawat_2013}). 
This guards against halos that disappear in a snapshot but reappear in the next due to numerical effects by checking for halos across three consecutive snapshots (\citealp{Springel_2005b}; \citealp{Srisawat_2013}). L-Halotree takes output from L-Basetree that contains the descendants for each snapshot and links them together to construct a complete merger history of the halos across the simulation. The merger tree can then be used by a SAM to determine when a galaxy forms and how it should evolve based on the halo merger history.

\subsection{Semi-analytic model (SAM) SAGE}
\label{sec:sage}
SAMs are developed as companions to direct simulations and they are widely used in different scenarios \citep{Baugh_2006}. 
We use SAM to refer to any non-explicit calculation in the simulation. 
SAGE (Semi-Analytic Galaxy Evolution) is the SAM code used by CoSANG to determine when and where galaxies should form during a simulation. SAGE uses a galaxy formation model that incorporates gas infall, AGN heating, gas cooling, ejected gas, satellite galaxies, and galaxy mergers \citep{Croton_2006,Croton_2016}.

Gas infall into each dark matter halo is based on the fraction of baryonic matter that is in the dark matter halo. As a dark matter halo forms, the baryons tend to fall into the halo due to their gravitational interaction. SAGE determines and tracks the growth of baryons for the halo. It does this by using the cosmic baryonic fraction to determine how many baryons should be present in the halo ($f_b \Delta M_{\mathrm{vir}} $) and compares it to the previous timestep to track the growth (\citealp{Croton_2016}). As a way to conserve the baryonic fractions during mergers, which can lead to artificial quenching, no more infalling baryons are allowed if the hot halo mass exceeds the cosmic fraction.
The updated version has also more sophisticated prescription to avoid instant stripping.

SAGE follows the hot halo model where the halo gas has an isothermal density
profile at the virial temperature,
\begin{equation}
    T_{\mathrm{vir}}=39.5\left( \frac{V_{\mathrm{vir}}}{\mathrm{km\hspace{0.1cm} s^{-1}}}\right)^2 \mathrm{K},
\end{equation}
where $V_{\mathrm{vir}}$ is the virial velocity of the halo (\citealp{Croton_2006,Croton_2016}). 
The gas then cools and condenses down to the center of the halo. The cooling rate for this
process is calculated using the cooling time,

\begin{equation}
\label{eq:T_cool}
    t_{\mathrm{cool}}=\frac{3}{2}\frac{\overline{\mu} m_p kT}{\rho_g(r)\Lambda(T,Z)},
\end{equation}
where $\rho$ is the gas density, $\overline{\mu} m_p$ is the mean mass and $\Lambda(T,Z)$ is the cooling function.
The cooling radius and its relation to the halo virial radius, determines if the infalling gas is in the cold accretion regime or the hot halo regime.

Stellar mass has dynamical impacts in CoSANG, and so the details of how it is determined are given below. The use of the stellar populations for particle tagging will be discussed in a future paper.

In CoSANG, SAGE
has been incorporated as an interactive part of the simulation that receives parameters from the N-body code and calculates baryonic features and returns this information into the next step of the simulation. The main assumption is that as each dark matter halo collapses, its own baryonic counterparts collapse with it and the mass fraction in baryons associated with every halo is taken to be $f_b=0.17$ \citep{Bennett_2013}.

The required information per halo is:
\begin{enumerate}
    \item $M_{\mathrm{vir}}$, the halo virial mass and the number of particles in the halo or subhalo;
    \item $V_{\mathrm{vir}}$ and $R_{\mathrm{vir}}$, virial velocity and virial radius;
    \item $V_{\mathrm{max}}$, the maximum circular velocity of the halo;
    \item The Cartesian position, velocity and spin vector of each halo.
\end{enumerate}
The baryon content in each halo grows by $f_bM_{\mathrm{vir}}-m_b$ where $m_b$ is the total mass of baryons present in the previous time-step. 

Star formation is implemented assuming that only cold gas above a critical surface density threshold can form stars, where:

\begin{equation}\label{eq:mcrit}
m_{\mathrm{crit}}=3.8 \times 10^9 \left(\frac{V_{\mathrm{vir}}}{200 \,\mathrm{km s^{-1}}}\right)\left(\frac{r_{\mathrm{disk}}}{10 \,\mathrm{kpc}}\right) \mathrm{M_\odot},
\end{equation}
and
\begin{equation}\label{eq:rdisk}
r_{\mathrm{disk}}=\frac{3}{\sqrt{2}}\lambda R_{\mathrm{vir}}.
\end{equation}
This star formation prescription is based on the work of \citealp{Kennicutt_1998}
(see also \citealp{Springel_2001,DeLucia2004,Croton_2006})
and proceeds at a rate
\begin{equation}\label{eq:mstardot}
\dot{m}_*=\alpha_{SF}\frac{(m_{\mathrm{cold}}-m_{\mathrm{crit}})}{t_{\mathrm{dyn,disk}}},
\end{equation}
where $\alpha_{SF}$ is the star formation efficiency, $m_{cold}$ is the total mass of cold gas, and $t_{dyn,disk}=r_{disk}/V_{vir}$ is the disk dynamical time.

AGN feedback in SAGE (\citealp{Croton_2006}) consists of two modes, a quasar mode and a radio mode. 
The purpose of the quasar mode is to grow the black hole at the center of the galaxy. Supermassive black holes grow during galaxy mergers both by merging with each other and by accretion of cold disc gas.
The updated SAM in \citealp{Croton_2016} adds in feedback from quasar winds by the energy equation:
\begin{equation}
    E_{BH,Q}=\kappa_Q \frac{1}{2}\eta \Delta M_{BH,Q}c^2,
\end{equation}
where $\Delta M_{BH}$ is the change in the black hole mass, $\kappa$ is the wind efficiency parameter, $\eta =0.1$ is the standard efficiency with which inertial mass is liberated upon approaching the event horizon, and $c$ the speed of light.

The low-energy radio mode is the result of hot gas accretion on to a central supermassive black hole once a static hot halo has formed around the host galaxy of the black hole. This accretion is described by
\begin{equation}\label{eq:accretion}
\dot{m}_{BH,R}=\kappa_{\mathrm{AGN}}\left(\frac{m_{BH}}{10^8\,\mathrm{M_{\odot}}}\right)\left(\frac{f_{\mathrm{hot}}}{0.1} \right)\left(\frac{V_{\mathrm{vir}}}{200\mathrm{\,Km\, s^{-1}}} \right)^3,
\end{equation}
where $m_{BH}$ is the black hole mass, $f_{\mathrm{hot}}$ is the fraction of the total halo mass in the form of hot gas, $V_{\mathrm{vir}}\propto T_{\mathrm{vir}}^{1/2}$ is the virial velocity of the halo, and $\kappa_{AGN}$ is the efficiency of accretion (with units of $M_{\odot}yr^{-1}$).

This feedback affects the evolution of the baryonic components, and therefore has dynamical consequences, but there is no \textbf{direct} feedback on individual particles in CoSANG.

Galaxy mergers play a major role in the evolution of galaxies in terms of their
overall growth and morphology. SAGE treats mergers as either a
minor merger or major merger, depending on the mass ratio ($f_{\mathrm{major}}=0.3$). If the merging galaxy\textquotesingle s mass is below a certain percentage of the main galaxy\textquotesingle s mass, then it is considered to be a minor merger. In this case, the satellite galaxy\textquotesingle s stellar mass is incorporated into the bulge mass and its cold gas mass is added to the disk of the main galaxy (\citealp{Croton_2006}). In the case of a major merger, where both galaxy masses are similar to each other, the disk can be destroyed to form an elliptical galaxy and a major starburst begins.

\citet{Croton_2016} have tested this model to be in good agreement with observations and hydrodynamic simulations. SAGE has been used for studies such as galactic morphology, galaxy luminosity function, stellar mass\text{-}gas metallicity relationship, black hole\text{-}bulge mass relation and was recently upgraded with dust modeling \citep{Triani_2020}.

\subsection{Galactic potentials}
\label{sec:potentials}
The gravitational effects of baryonic galaxies are taken into account by including additional forces on particles within the N-body simulation. We include forces due to the bulge, the black hole and the galactic disk.

We adapt the same force softening length as Gadget. For separation we include the softening length $\epsilon$ and we use $r_\epsilon$ for $\mid \vec{r}\mid$ where:
\begin{equation}
    r_\epsilon=\sqrt{\mid\vec{r}\mid^2+\epsilon^2},
\end{equation}

\emph{Bulge:} The bulge is represented by a Hernquist potential (\citealp{Hernquist_1990}) which is spherically symmetric. The potential follows the equation

\begin{equation}
    \Phi_{bulge}=-\frac{GM_{\mathrm{bulge}}}{\mid\vec{r}\mid+ a_{\mathrm{bulge}}},
\end{equation}
where $a_{bulge}$ is the scale radius, $M_{\mathrm{bulge}}$ is the bulge
mass and $\vec{r}$ is the vector from the galaxy center to the dark matter particle that the
force is being calculated for. The bulge scale radius is assumed to follow a constant
fraction of the disk scale length that SAGE derives (\citealp{Carollo_2004}).
The bulge-to-disk scale length ratio has been observed to fall within the range of
0.2 to 0.02 (\citealp{Carollo_1999}). For CoSANG, an intermediate value of 0.1 is used (\citealp{Carollo_2004}).

Including the softening, this force becomes
\begin{equation}
    \vec{F}_{\mathrm{bulge}}=-\frac{GM_{\mathrm{bulge}}\vec{r}}{r_\epsilon(r_\epsilon+a_{\mathrm{bulge}})}.
\end{equation} \\
\emph{Supermassive black hole:} The Plummer model is used to model the central black hole as a softened point source.
\begin{equation}
    \vec{F}_{BH}=-\frac{GM_{BH}\vec{r}}{r_\epsilon^3},
\end{equation}
where $M_{BH}$ is the black hole mass and $r_\epsilon$ is the softening included separation.\\
\emph{Galactic disk:} Although galactic disks generally have exponential radial profiles \citep[e.g.][]{Courteau_1996,deJong_1996,MacArthur_2003,Pohlen_2006}, there is no analytic solution for the gravitational force from an exponential disk. A good approximation is the sum of three Miyamoto-Nagai \citep{Miyamoto_1975,Smith_2015} potentials. Each Miyamoto-Nagai potential has the following form:
\begin{equation}
    \Phi(R,z)=\frac{-GM_{MN}}{\sqrt{R^2+(a+\sqrt{b^2+z^2})^2}},
\end{equation}
where $M_{MN}$ is the total disc mass, $a$ is the radial scalelength, and $b$ is the vertical scaleheight, and the parameters of the three MN potentials as a function of disk scalelength and scaleheight are as described in \citet{Smith_2015}.

For every galaxy disk mass $M_d$, disk scale length $r_{\mathrm{disk}}$, and disk scale height $h_{\mathrm{disk}}=r_{\mathrm{disk}}/9$ are taken from SAGE and used as input to calculate other parameters in Table 1 of \citet{Smith_2015}.
Multiple stability test have been performed to ensure that orbits are stable under the influence of this force both in isolation and inside CoSANG (see section \ref{sec:stabilitytests}).

Galaxies in CoSANG are assumed to be oriented perpendicular to the angular momentum or spin of the dark matter halo. Rotation matrices are applied to the particles to align the angular momentum vector to lie along the z axis. The galaxy forces are then calculated on the particles within the halo and the forces are rotated back to the original frame.

\subsection{Mass adjustment}
\label{sec:massadjustment}
CoSANG, as an N-body simulation, contains particles that initially represent both
the dark matter and baryonic matter. By adding in a galactic potential based on the SAM, a fraction of mass needs to be subtracted from the dark matter particles to account for these baryons. Only particles that are within the halo virial radius have a percentage of mass subtracted for baryons. This
percentage is based on the fraction of mass that an individual particle accounts for in the halo (Equation \ref{eq:fm}). This fraction is multiplied by the galaxy mass and subtracted from the current mass of the particle (equations \ref{eq:m_adjust}, \ref{eq:m_adjust2})
\begin{equation}
\label{eq:fm}
    f_{m}=\frac{m_{\mathrm{particle}}}{M_{\mathrm{vir}}}.
\end{equation}

The total mass incorporates both the dark matter and baryonic
matter. CoSANG distinguishes between the two by subtracting a fraction of mass from the dark matter particles. The mass subtracted from the dark matter particles represents the baryonic matter due to the galactic potential,
\begin{equation}
\label{eq:m_adjust}
    \Delta m=f_m M_{\mathrm{galaxy}},
\end{equation}
where $M_{\mathrm{galaxy}}=M_{BH}+M_{\mathrm{cold gas}}+M_{\mathrm{stellar}}$. and where $M_{\mathrm{coldgas}}$ is the cold gas mass and $M_{stellar}$ is the galaxy stellar mass from
SAGE. Finally,
\begin{equation}
\label{eq:m_adjust2}
    m_{\mathrm{particle}} 	\rightarrow m_{\mathrm{particle}}-\Delta m.
\end{equation}
A particle can reside in both the main halo with a central galaxy and a
halo with a satellite galaxy. Our method accounts for this case since the adjusted
particle mass is preserved over the time-step. If a particle\textquotesingle s mass was adjusted once
for the main halo, when it is adjusted again for a satellite, the adjusted mass is
used in equation \ref{eq:m_adjust} and \ref{eq:m_adjust2}. After each time-step, the particle masses are
also reset with the initial masses from the ICs. The masses are reset before the halo finder SUBFIND runs. The halo finder needs to run on the original particle
masses since it needs to account for the total mass in the halo, otherwise the halo
parameters will be inaccurate.

\subsection{Time resolution and extrapolation}
\label{sec:profiler}
SAGE is run at each snapshot, which is when the galaxies and their parameters are updated. Snapshots are produced on average 50 million years apart. Between each snapshot, there are multiple time-steps when the forces are calculated for each dark matter particle. 
The galaxy properties should continue to evolve between snapshots, but until the next snapshot, we do not know what values of those properties should be. It is, however, important to take this evolution into account because large non-adiabatic jumps in the galaxy parameters can impact the halo unrealistically, therefore a smooth transition between snapshots is required. One way to do this is to extrapolate the next point for the galaxy parameter. Based on how the galaxies behave during the simulation up to a certain point we can predict how they should continue to evolve. A linear extrapolation is carried out on the SAGE galaxies from the previous two snapshots to determine how they should evolve to the next snapshot. The parameters extrapolated by CoSANG are Cold Gas Mass, Stellar Mass, Bulge Mass, Black Hole Mass, Disk Scale Radius and Angular Momentum Vector.

SAGE is run independently at each snapshot and has no connection to a previous SAGE run in terms of linking galaxies across snapshots. CoSANG therefore must determine which previous SAGE galaxy to use to define its extrapolation. CoSANG stores in memory the old galaxy data from the previous snapshot and the galaxy data from the current snapshot, and searches through the galaxy data from the two most recent snapshots to match galaxies within 60 kpc of each other.

The galaxy positions are based on the iterative centers of mass(Section \ref{sec:com}), which are updated at each time-step and so they do not need to be extrapolated. Once galaxy matches have been found, all of the other properties are linearly extrapolated. The slope is calculated for each galaxy between the current snapshot and the previous one and is recalculated at each following snapshot. Galaxy properties are not extrapolated if the galaxy is new, no galaxy match is found or if the galaxy mass increases by more than 20\% indicating a major merger occurred. Extrapolation for major mergers can cause an over extrapolation of the galaxy parameters. 
Major mergers are defined as having the mass ratio of 0.3 and above. SUBFIND, like many halo finder, fundamentally has problems finding host halos, especially during major mergers. This problem can result in mass fluctuations and ambiguity in distinguishing the host halo from merging subhaloes \citep{Behroozi_2013}.

\subsection{Center of mass}
\label{sec:com}
In order for the baryonic potential to be stable, it must reside at the deepest point of the potential regardless of how the dark matter particles sample the halo structure. Tests with isolated spherically-symmetric halos reveal that placing the baryonic potential at the naive center of mass, or at the most bound particle, is not stable \citep{McCord_2017}.

In order to calculate an accurate center of mass for the halo, we use an iterative process. The algorithm searches for any particles that are within the virial radius of the halo and calculates the initial center of mass position. 
At each iteration, the sphere is shrunk by 20\%, and a new center of mass for the particles within that sphere is calculated. The iteration continues until the difference between two consecutive iterations are less than a 1\% threshold. Calculating the center of mass for all haloes at each time step adds significant computational expense, but is necessary for numerical stability. The placement of galaxies at the center of mass of their host halos is a good approximation for most of the evolution but this assumption may not be very accurate during merger events, especially during major mergers in which dark matter and baryons may have different centers of mass.

Merger events pose a real challenge to N-body models with a FOF halo finder. These algorithms have difficulties distinguishing halos at close proximity. Moreover SAMs have difficulties extrapolating parameters during mergers in a live simulation and they may consider a close encounter as a merger and vice versa. SAMs can correct themselves in the next time-step and this ambiguity has insignificant dynamical effect but any merger events should be treated with more caution.

\subsection{Tests and performance}
\label{sec:tests}

To ensure that CoSANG performs as expected, we have performed a number of tests, consisting of:
\begin{itemize}
    \item Technical tests (performance, scalability, compatibility, efficiency)
    \item Scientific tests (dynamics, numerical errors)
\end{itemize}
Technical tests consist of using different hardware configurations, memory load balance, and scaling. Scientific tests include integrating the orbit of a test particle, stability under small perturbations, evolution of an isolated NFW halo, and performing a small box cosmological simulation at different mass resolution. We present one example of the most important tests in this section: the scaling test and an isolated NFW halo. See also the test particle orbital integration in Appendix \ref{sec:OrbitTest}.

For further details on all tests that have been performed on CoSANG, see \citealp{McCord_2017}. \footnote{Test examples can be found at the following github repository: \url{https://github.com/stalei}}

\subsubsection{Scalibility test}
\label{sec:scabilitytests}

CoSANG is built on Gadget, which scales well as both the particle number and number of compute cores increase \citep{Springel_2005}. In order to determine the efficiency of the other parts of the code, such as the halo finder, SAM, center of mass calculation, and analytic gravitational force, we have performed a number of timing tests.

We test the scalibility of the model by running the same model with Gadget (DMO) and CoSANG, using different number of processors and record the time they accomplish specific tasks. 
For ideal scaling, there would be a linear relationship between the number of processors and performance of the code, doubling the number of processors halving the runtime; however, in real world performance is always worse than ideal.  

 The most time consuming new piece of the algorithm is the iterative calculation of the center of mass which technically scales as $\mathcal{O}(N)$. This test is done with a zoom-in box around one MW mass halo, starting at redshift 10 and continuing for less than 2 Gyr. This was enough to compare different models and running a full simulation for a longer time would have exceeded our benchmark allocation.

The required time for lower redshifts is much longer and it slows down more in CoSANG because there are more resolved structures and full grown halos contain very high numbers of particles.

\begin{figure}
	\includegraphics[width=\columnwidth]{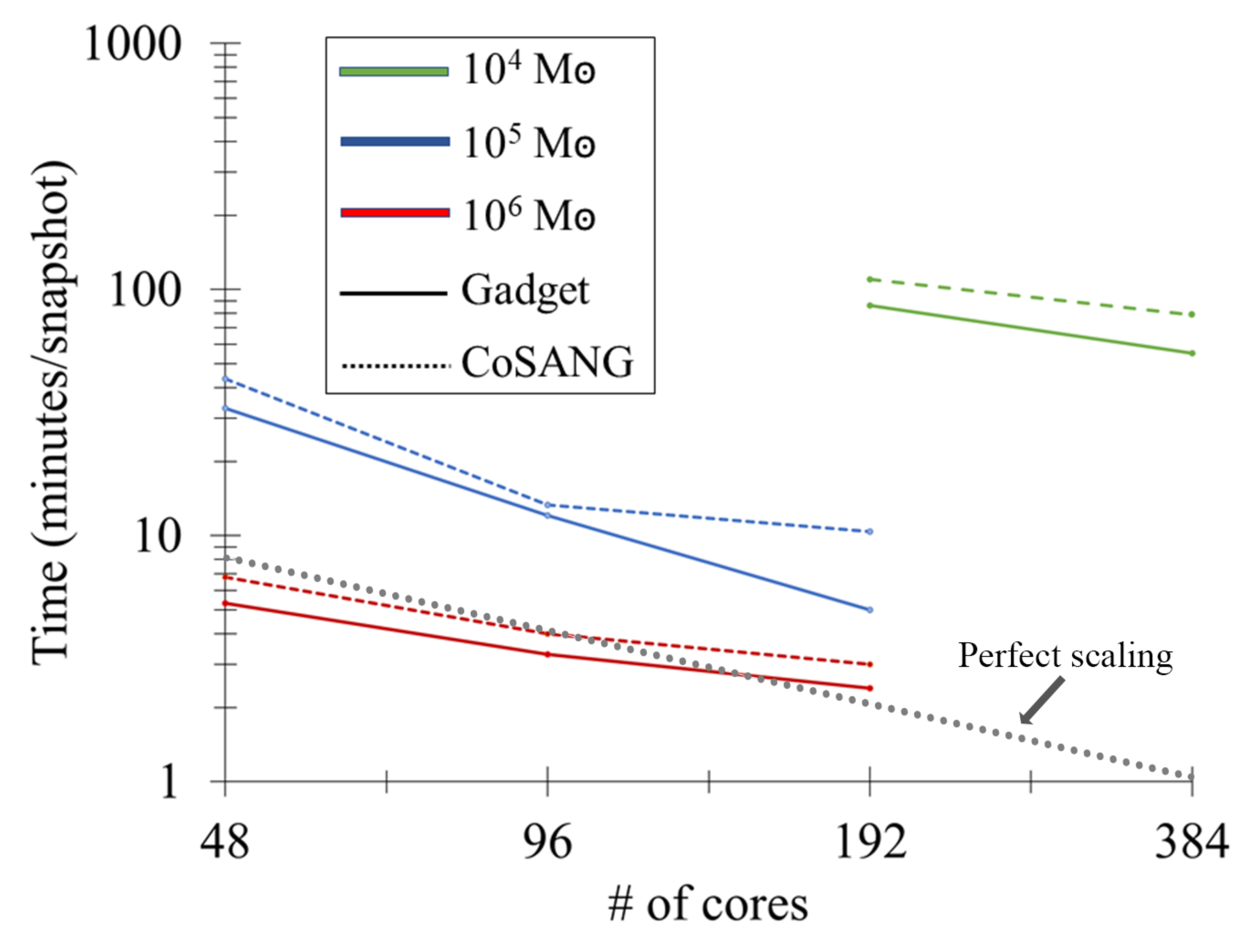}
    \caption{Performance of CoSANG as a function of mass resolution and number of compute cores. We run a zoom-in simulation of a target halo at different mass resolutions (red $10^6 M_{\odot}$, blue $10^5 M_{\odot}$, green $10^4 M_{\odot}$) and record the required time to accomplish specific tasks between two redshifts. Linear (perfect) scaling is shown in gray for comparison. Gadget simulations are shown as solid lines and CoSANG is shown with dashed lines. CoSANG is slower than Gadget by 50\% or less. The highest resolution simulations cannot successfully run in the compute queue for lower numbers of cores.}
    \label{fig:scaling}
\end{figure}

We perform scaling tests using different number of processors and different resolutions. Results are shown in Figure \ref{fig:scaling}. This plot shows the compute time versus the number of processors. We tested three mass resolutions, $10^{4-6} M_\odot$ (different colors). Using small number of processors for the highest mass resolution it was not possible to complete the task within the supercomputer queue time limit; therefore, there is no point for finest mass resolution for fewer number of processors. No data is shown for $10^{5-6} M_\odot$ at high number of cores because simulations run very fast and the recorded time is mostly the hardware input/output rate not numerical calculations. CoSANG requires 50\%-100\% more memory than Gadget due to using new variable and iterative calculations. CoSANG scales well up to $\sim 500$ cores for a mass resolution of $10^5M_\odot$; further improvement in scaling is the subject of ongoing work.

\subsubsection{Stability test}
\label{sec:stabilitytests}
A critical test of CoSANG is that a disk potential embedded in an isolated dark matter halo will remain stable, with particles orbiting correctly in the combined dark matter + disk potential. (\citealp{Binney_2008}) and does not create numerical artifacts or non-physical orbits. 
The orbit of a single particle is discussed in Appendix \ref{sec:OrbitTest}.

We generate an isolated NFW (\citealp{Navarro_1996}) halo IC with Galstep \footnote{\url{https://github.com/ruggiero/galstep}} (e.g. \citealp{Ruggiero_2017}). This simulation is a non-cosmological model using CoSANG.
We run the simulation using CoSANG and let particles evolve under the influence of the analytic potential and look for evolution in the density profile.
We also run a cosmological MW mass halo, both with DMO and CoSANG and compare the density profile. The structure of the cosmological halo remains stable during entire simulation from redshift 20 to redshift 0.

\section{Results And Discussion}
\label{sec:results}
CoSANG is designed for stellar halo studies, and is therefore aimed at MW mass galaxies. We select a sample of initial conditions that will produce galaxies in the same mass range, but possibly with different histories. Our original complete sample covers an extended mass resolution $10^{5-7} M_\odot$. Here we present the highest mass resolution while lower resolutions were used for convergence tests and calibrations. We focus on the dynamical effects of the baryonic potential on the dark matter substructures.

In this work we present three MW mass halos. We use zoom ICs from the FIRE project \citep{Hopkins_2018} that are known to produce MW mass galaxies in hydro simulations.

We reproduce the initial condition using MUSIC \citep{Hahn_2011}. The parameters are identical to those used by FIRE\footnote{\url{http://www.tapir.caltech.edu/~phopkins/publicICs/}} \citep{Wetzel_2022}, except that we use a starting redshift of 20, and do not create baryonic particles.

We run each simulation twice, one full CoSANG run and one where we turn off all CoSANG features (this is effectively Gadget3) as the DMO model.
This allows us to understand the effect of the galactic potential on the dark matter distributions.

There are slight differences in the calculated parameters between SUBFIND and Rockstar (e.g. virial mass) but no significant differences between the two methods.

\subsection{Target halos and initial conditions}
\label{sec:simulations}

The cosmological zoom-in technique (\citealp{Katz_1994},\citealp{Navarro_1994}, \citealp{Hahn_2011}) is used for efficiency. In summary, the zoom-in technique starts with a uniform resolution box at low resolution. Target halos are selected and their particles are tracked back in the initial condition file. We then create a new set of initial condition with high resolution for the region surrounding these particles, leaving all particles outside this region at lower resolution. 

Our cosmological parameters are as follows: $\Omega_M=0.272, \Omega_{\Lambda}=0.728, \Omega_{\mathrm{baryon}}=0.0455, h_0=0.702$
and a periodic box of size 60 $h^{-1}\mathrm{Mpc}$. The highest mass resolution has particles of mass $2.37266 \times 10^5 M_\odot$ and force softening length 2 kpc, starting at redshift 20 and running down to redshift 0. We extract halos at redshift 0. Our target MW mass halos have $M_{\mathrm{halo}} \sim10^{12} M_\odot$. 

Particles outside the Lagrange volume have coarser resolution. 
Contamination and different biases can affect the zoom-in sample \citep{Onorbe_2020} therefore we examine and make sure there is no contamination by low resolution particles inside the virial radius. 
Table \ref{tab:halos_table} lists our sample halos, labled m12b, m12f and m12i.
Since there are two versions of each model, dark matter halo parameters are slightly different in each pair and all parameters are reported from the DMO model, unless otherwise mentioned (e.g., stellar mass, which is reported from SAGE). These numbers may also be different than what is reported in the original FIRE paper (\citealp{Hopkins_2015}) both because the simulation code is different and also different halo finders can characterize halos in different ways (\citealp{Knebe_2011}) and cosmological simulations are chaotic in nature \citep{Kandrup_1991,Keller_2018,Genel_2019}.
We produce 264 snapshots between redshift 20 and 0. Time increments are not linearly spaced but there is in average $\sim 50 \mathrm{Myr}$ between successive snapshots. This is an optimum time resolution so we can follow the changes in the baryonic galaxy properties (see Section \ref{sec:profiler}) without an overwhelming data overhead. We run the same initial condition and same configuration with both Gadget and CoSANG.

\begin{table*}
	\centering
	\caption{Zoom in halos in this work. All numbers are from DMO simulation except stellar mass and disk scale radii, which come from the CoSANG version of the simulation.}
	\label{tab:halos_table}
	\begin{tabular}{lccccccc} 
		\hline
		ID & Mass resolution $[M_\odot h^{-1}]$ & $M_v [M_\odot h^{-1}]$ & $R_v[h^{-1} kpc]$ & DMO subhaloes & CoSANG subhaloes & $M_\star [M_\odot]$ & $R_d [kpc]$\\
		\hline
		m12b & $2.4 \times 10^{5}$ & $6.6 \times 10^{11}$ & 142 & 1635 & 1106 & $6.1 \times 10^{10}$ & 3.2\\
		m12f & $2.4 \times 10^{5}$ & $7.8 \times 10^{11}$ & 150 & 2834 & 2052 & $7.8 \times 10^{10}$ & 3.8\\
		m12i & $2.4 \times 10^{5}$ & $5.5 \times 10^{11}$ & 134 & 1584 & 1733    & $5.0 \times 10^{10}$  & 2.8\\
		\hline
	\end{tabular}
\end{table*}

After we finished the simulations we perform the same analysis with the same configuration on the corresponding pairs.

\subsection{Halo and galaxy growth}
\label{sec:galaxyevolution}
\begin{figure}
	\includegraphics[scale=0.21]{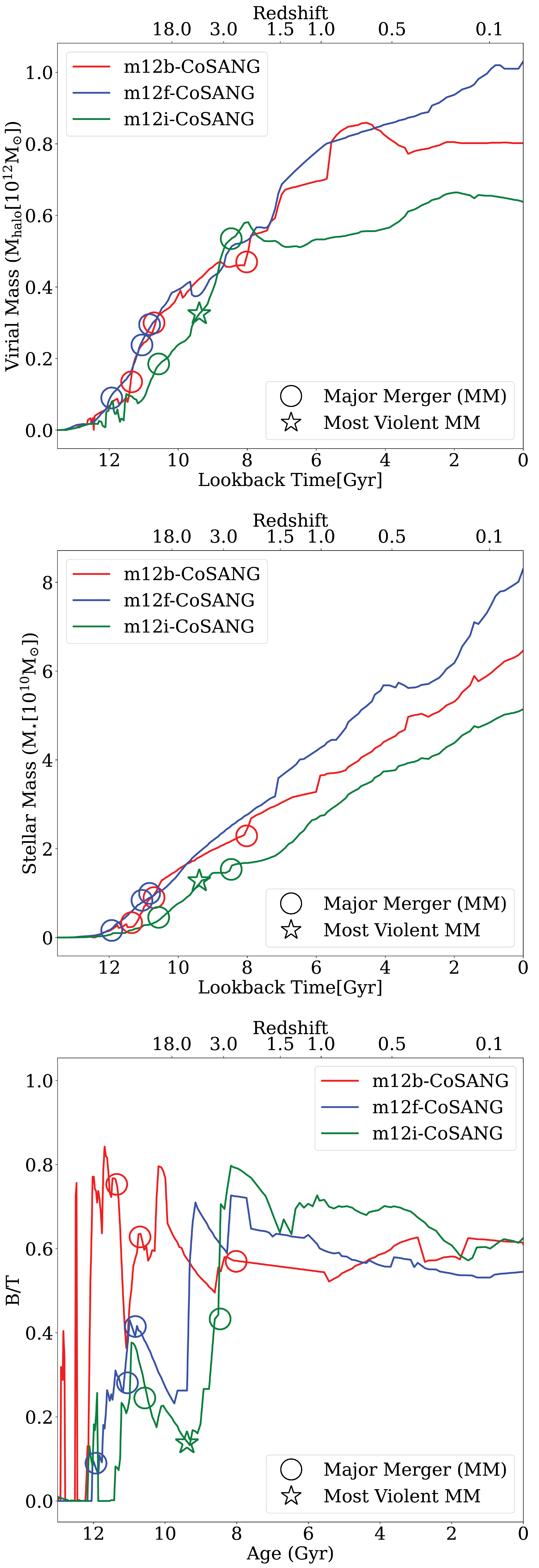}
    \caption{Growth of the virial mass (top) and stellar mass (middle) and bulge to total stellar mass ratio (bottom) for individual halos with CoSANG.
    The circles and star represent major merger events. m12b and m12f have similar halo growth, while the growth and major mergers in m12i occur slightly later.
    The star is the most violent event of all major merger events. Stellar mass follows a similar pattern but in this case the similarity between m12b and m12f is not as significant as for the virial mass.}
    \label{fig:growth}
\end{figure}

In section \ref{sec:sage} we explained how SAGE calculates galactic parameters as halos grow and evolve in the simulation, using both the halo properties and merger history.
In this section we discuss how m12b, m12f, m12i, and their associated galaxies grow their mass over time. Figure \ref{fig:growth} shows the mass growth versus time. Colors represent different halos and circles and the star represent the major merger events. Merger events are defined as mass ratio of $M_{\mathrm{ratio}}>0.3$. The star is the merging of three massive halos with total mass of 1.5 times the mass of the main halo. The merger events are inferred by SAGE (\citealp{Croton_2016}) and confirmed by tracking the merger tree.

m12b and m12f follow similar growth in their virial mass for most of the simulation and they both grow their mass earlier than m12i. m12b and m12f are early forming halos in our model while m12i is a late forming halo gaining much of its mass via a more violent merger that continues for a longer time therefore m12i catches up in mass to be a MW mass halo at z=0. 
m12b has the last major merger, at $z \sim 2$, but it is a smaller event than m12i's and does not dominate its mass growth in the same way.
Note that halos do not truly decrease significantly in virial mass and the oscillation seen is due to the numerical problems with the halo finder \citep[e.g.][]{Knebe_2011,More_2011,Ludlow_2019,Leroy_2020}.

The growth in stellar mass mostly mirrors the virial mass but in this case the similarity between m12b and m12f is not as significant. Stellar mass is less susceptible to fluctuations in the virial mass due to the semi-analytic prescription. Bulge mass to total mass ratio is shown in the bottom panel. In average all three galaxies reach $B/T\sim 0.6$. 

The dynamical impact of the analytic baryonic potential versus dark matter potential is affected by the time and the scale of merger events. As we will see in section \ref{sec:Abundances} a violent relaxation that occurs over a longer time scale on a late forming halo can wash out the importance of a baryonic potential on the final structure. Hydrodynamic models demonstrate similar diversity (\citealp{Engler_2021}) and correlations between formation history and substructures (\citealp{Rey_2022}).

\subsection{Subhalo mass function and radial distribution}
\label{sec:Abundances}

As we mentioned in section  \ref{sec:halosimulations} 
the underlying dark matter kinematics is the key indicator to differentiate between different models. 
We study whether the baryonic galaxy affects the distribution of subhaloes (position, mass, number) in the model.

Every halo inside the virial radius is treated as a subhalo independent of its binding energy. 
We exclude all halos with fewer than 20 particles, corresponding to a subhalo mass limit of $M_\mathrm{vir} >~ 7 \times 10^6 M_\odot$.
Halos with less than 100 particles should be treated carefully.

We show Local Group observations in comparison with both models. The MW and M31 data are taken from \citet{McConnachie_2012,Simon_2019,Drlica-Wagner_2020,Putman_2021}.

\subsubsection{Mass function}
\label{sec:MassAbundance}

\begin{figure*}
	\includegraphics[width=\textwidth]{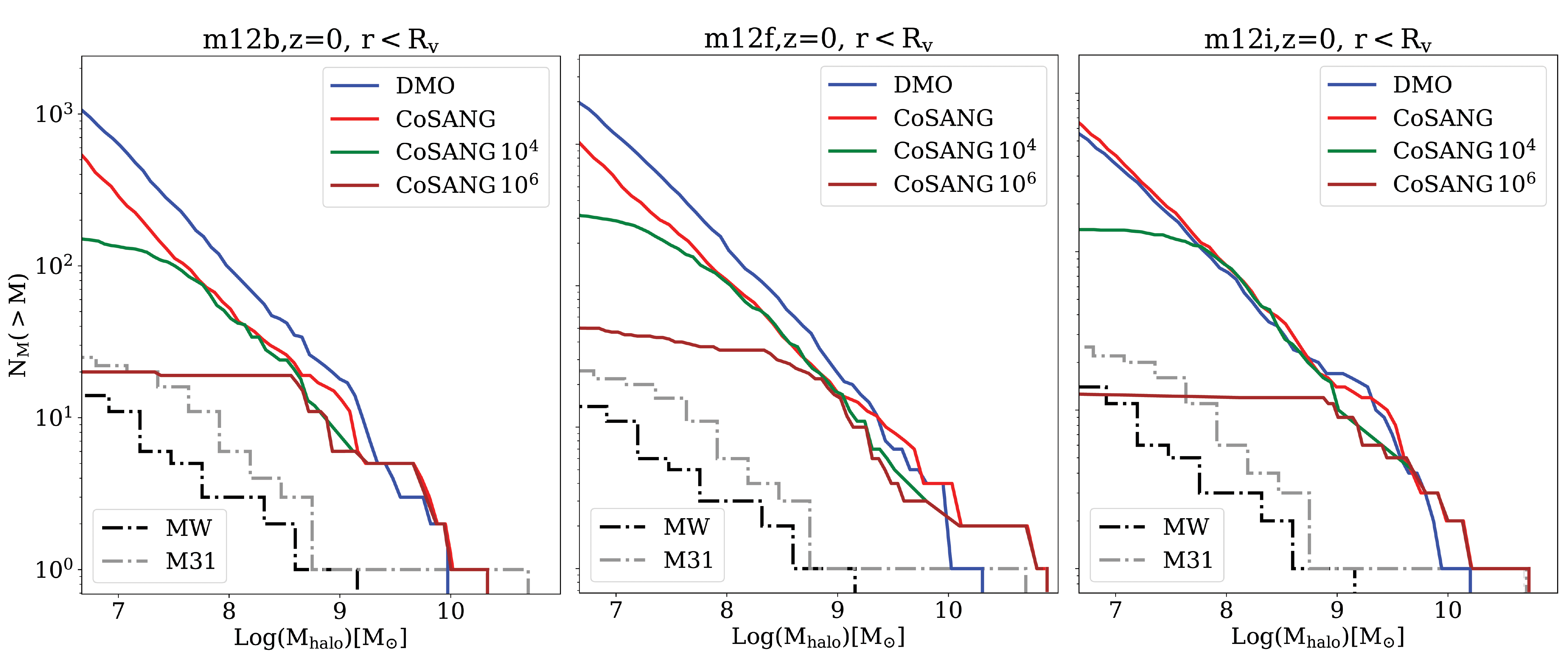}
    \caption{Cumulative mass function of subhaloes in halo m12b (left), m12f (middle) and m12i (right). The blue line represents DMO, the red line represents CoSANG, the black dashed line is the MW, and the gray dashed line is M31. CoSANG $10^4$ (green) and CoSANG $10^6$ (brown) represent CoSANG with stellar mass cut $>10^4M_\odot$ and $>10^6M_\odot$ respectively. This figure contains all subhaloes within the virial radius. In general the CoSANG simulations contain fewer subhaloes than the DMO simulations. m12b and m12f display a more significant difference; m12i, which has a different merger history, shows no difference between CoSANG and DMO simulations. CoSANG shows reduced substructure but not enough on its own to explain the missing satellite problem.}
    \label{fig:m12b-Mdist}
\end{figure*}

Figure \ref{fig:m12b-Mdist} shows the cumulative mass function for m12b (left), m12f (middle) and m12i (right). We examine the subhaloes mass abundance in both the DMO (solid blue) and CoSANG (solid red) simulations.

In m12b and m12f, the CoSANG simulation shows a reduced number of subhaloes for $M_{\mathrm{subhalo}}<10^9 M_\odot$. In other words, the baryonic potential of the galaxy can reduce the number of satellite galaxies and this effect is more significant on smaller satellites. At higher mass there is less significant differences and the most massive halos are slightly more massive in CoSANG compared to DMO.

We also plot the observed data for the MW (dashed black) and M31 (dashed gray). We used \citet{Putman_2021} for most of the data. For galaxies without a reported mass, we estimate the dynamical mass using its effective radius and velocity dispersion. For a given velocity dispersion ($\sigma$), the mass interior to radius $r$ in an isothermal sphere is given by \citep{Binney_2008}
\begin{equation}
    M(r)=\frac{2\sigma^2r}{G}.
\end{equation}

It is well known that DMO models produce a higher number of satellites compared to that around the MW \citep[e.g.][]{Klypin_1999, Moore_1999}. Although observations can have an incomplete sample and the real number of satellites is higher than what is observed \citep{Tollerud_2008,Walsh_2009,Hargis_2014,Kim_2018}, baryonic processes can reduce the number of satellites and hydrodynamic models do not produce as many satellites as DMO models  \citep[e.g][]{Libeskind_2006,Wetzel_2016,Font_2021}.   

CoSANG simulations produce fewer subhaloes than DMO models, but still significantly more than the number of observed satellites around Local Group galaxies. This confirms the reduction of the number of subhaloes due to baryonic potentials.

\citealp{Jahn_2019} show number counts of subhaloes in simulations with
 baryons, within $0.2 r_{200m}$ $\sim 70\%$ of substructure depleted by the MW mass hosts, compared to simulations without baryons, and a factor of $\sim3.5$ reduction (from DMO to baryonic runs) in the number of subhaloes at $V_\mathrm{max}$ = 10 km/s
in MW mass hosts. Dynamical interactions with the disk can increase the likelihood of the presence of the subhaloes at closer distance however this can strip the subhaloes and decrease the number of subhaloes at this range. 
 
\begin{figure*}
	\includegraphics[width=\textwidth]{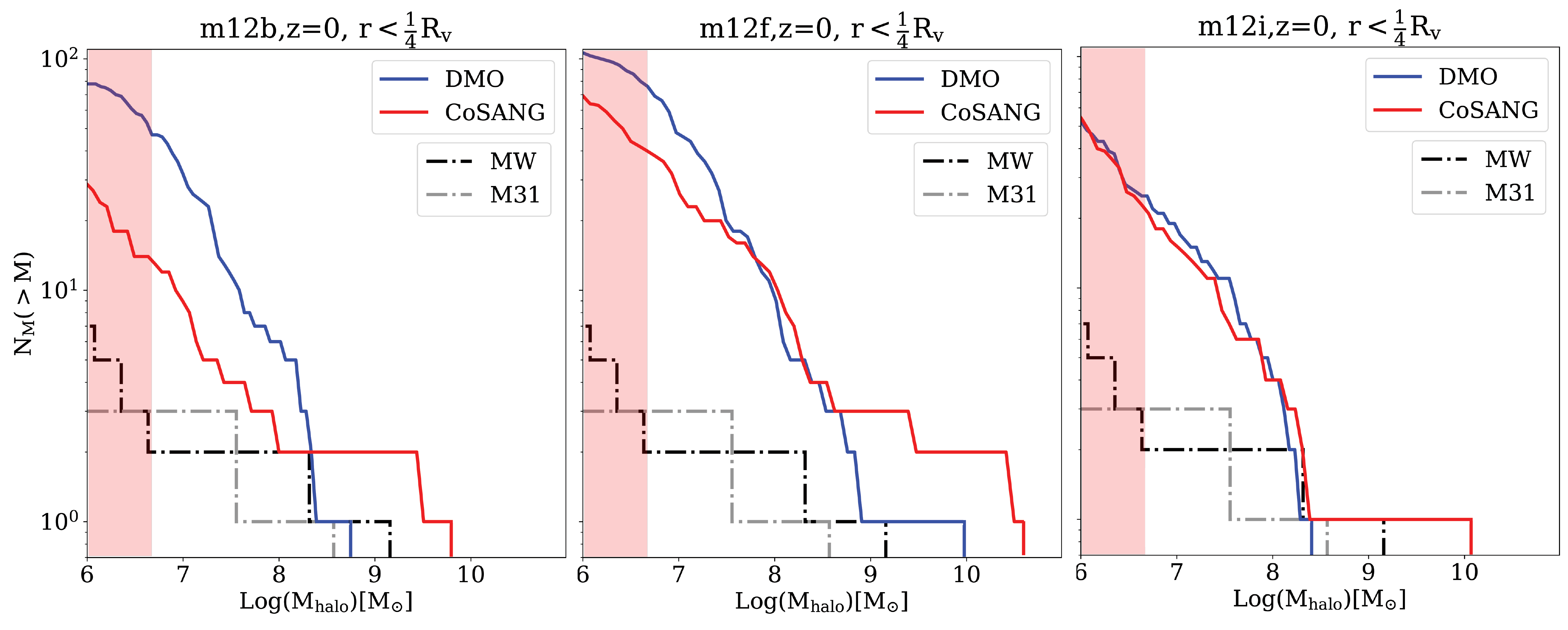}
    \caption{Cumulative mass function of subhaloes at $r<\frac{1}{4}R_v$ in halo m12b (left), m12f (middle) and m12i (right). The blue line represents DMO, the red line represents CoSANG, the black dashed line is the MW and the gray dashed line is M31. The red shaded region is the mass resolution limit. The difference between DMO and CoSANG is more significant at smaller radii, especially for m12b. m12i does not show any difference.}
    \label{fig:M_hist_close}
\end{figure*}

The baryonic potential dominates at small radius and so is expected to have the largest effect there. Figure \ref{fig:M_hist_close} shows that the depletion of subhaloes is indeed more significant within $\frac{1}{4}R_v$.
 CoSANG produces one more massive satellite than DMO and we see a few subhaloes are closer to the host halo in CoSANG compared to DMO.

m12i does not show as significant difference as two other models. The difference in the merger history and the environment can explain this difference between m12i and two other models. We will discuss these differences in more detail in section \ref{sec:m12i_difference}. 

Observed satellites are selected by their stellar mass. Therefore, a better comparison between observed and simulated subhaloes should also incorporate stellar mass limits for the simulated sample. CoSANG $10^4$ (green) and CoSANG $10^6$ (brown) represent CoSANG with stellar mass cut $>10^4M_\odot$ and $>10^6M_\odot$ respectively. This cut-off reduces the lower end of the mass distribution significantly, in all three cases. More massive subhaloes are "too big to fail" \citep{Boylan-Kolchin_2011} and smaller subhaloes are not forming stars efficiently. This implies the stellar mass threshold has a significant impact. This reduced number is still much higher than hydrodynamic models.

\begin{figure*}
	\includegraphics[width=\textwidth]{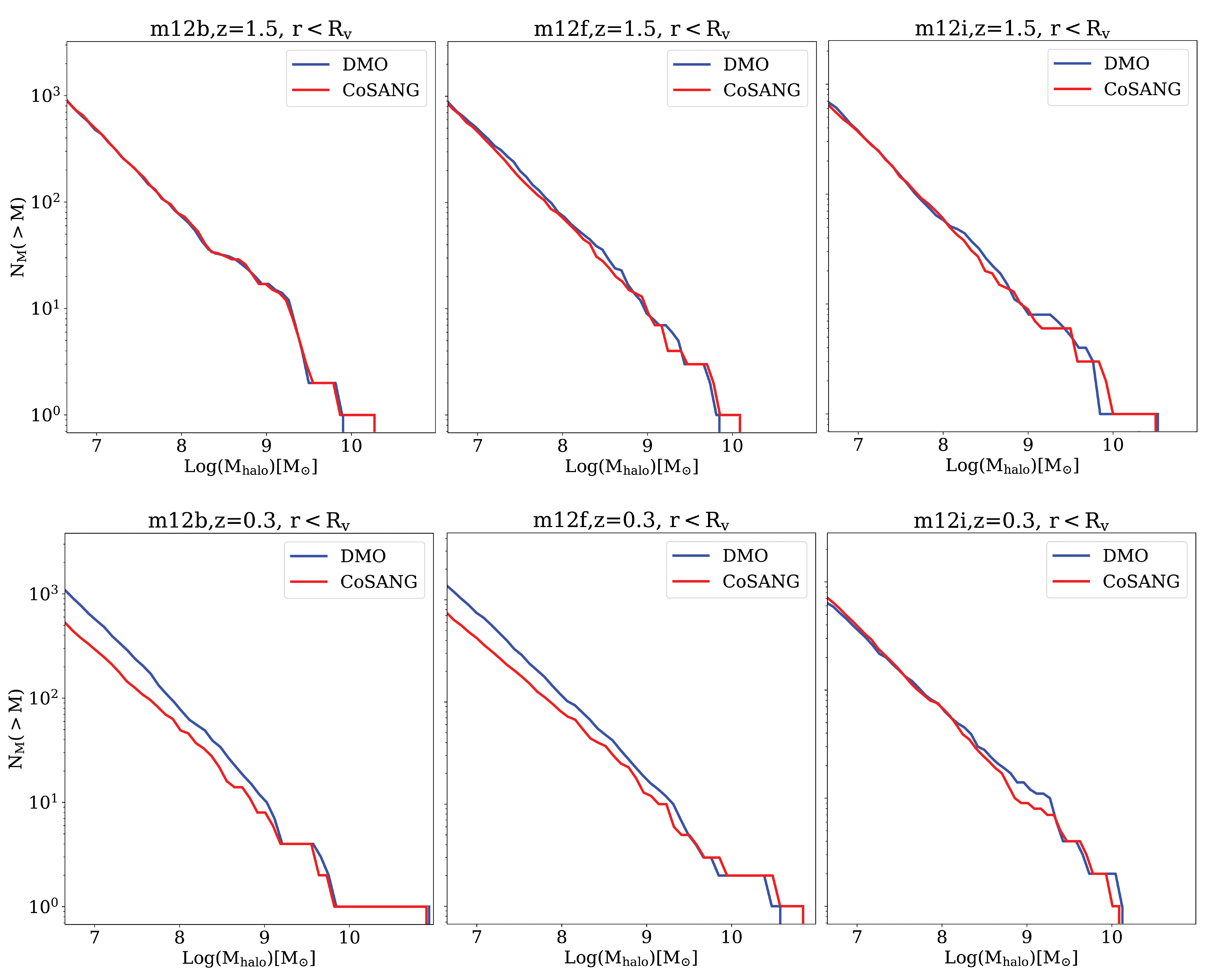}
    \caption{Redshift evolution of cumulative mass function of subhaloes in m12b (left), m12f (middle) and m12i (right). Blue lines represent DMO and red lines represent CoSANG. This figure includes all subhaloes within the virial radius. The shaded red region is our mass limit. The difference is not significant at z>1.5 and the two models are statistically similar at higher redshifts. However, the differences increase with time as the galaxy grows and the dynamical effects have time to act on substructures.}
    \label{fig:MassAbundanceRedshift}
\end{figure*}

We investigate the effect of a growing disk in time by following the evolution of the mass function at different redshifts. Figure \ref{fig:MassAbundanceRedshift} shows the distribution at two different redshifts ($z=1.5$ top, $z=0.3$ bottom). Although the simulation starts at redshift 20 and the galactic potential becomes part of the simulation as soon as galaxies form (which is $z\sim 13$, even though the stellar mass is not significant at high redshifts), still the three models are statistically similar up to redshift $z\approx 2$ when the stellar mass is large enough to produce a significant statistical difference. Since the main increase in the stellar mass is around this redshift, we see the difference becomes more pronounced after this at around $z \approx 1.5$ and the deviation between DMO and CoSANG increases as we progress in time. This indicates a correlation between the growth of a galactic potential and an increasing difference between DMO and CoSANG and not a random difference at different time-steps.

\begin{figure*}
	\includegraphics[scale=0.34]{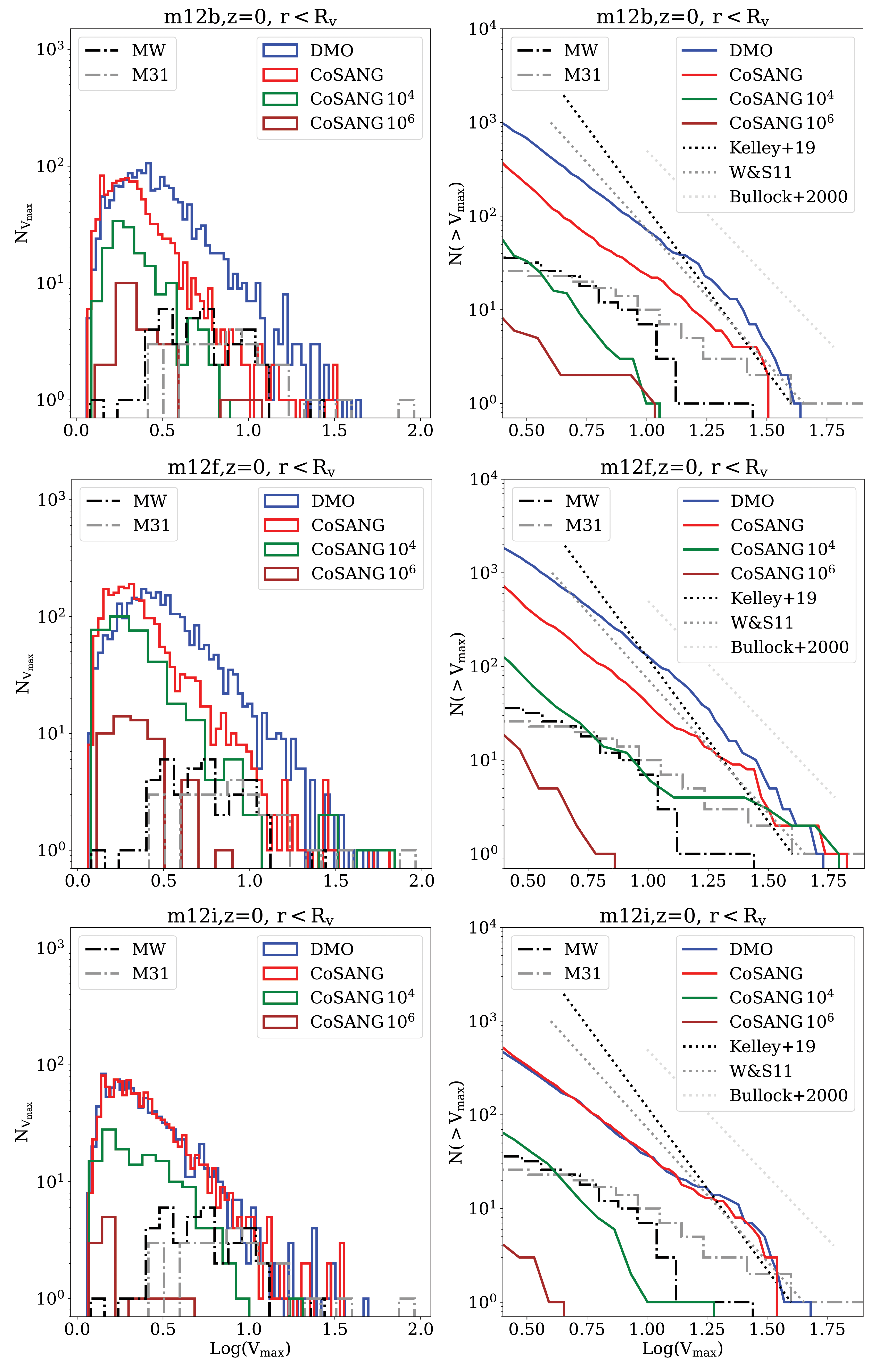}
    \caption{$V_{\mathrm{max}}$ distribution (left) and cumulative $V_{\mathrm{max}}$ distribution (right) for subhaloes within the virial radius of m12b (top), m12f (middle) and m12i (bottom). DMO is in blue and CoSANG is in red. CoSANG $10^4$ (green) and CoSANG $10^6$ (brown) represent CoSANG with stellar mass cut $>10^4M_\odot$ and $>10^6M_\odot$ respectively. Observed satellites of the MW (black dashed) and M31 (gray dashed) are shown for comparison. For a given $V_{\mathrm{max}}$ the CoSANG simulations contain fewer subhaloes than DMO and $V_{\mathrm{max}}$ peaks at a smaller value. Kelley+19 (dotted black line) \citep{Kelley_2019}, Bullock+2000 (dotted gray \citep[][]{Bullock_2000} and W\&S11 (dotted light gray) \citep[][]{Wadepuhl_2011} show the slope from other studies with 12,1,6 halos in their sample respectively.}
    \label{fig:vmax_m12b}
\end{figure*}

Observational comparisons are more robust when using velocities, which can be directly observed, rather than unobservable virial masses \citep{McConnachie_2012,Fritz_2020}.
Figure \ref{fig:vmax_m12b} shows the distribution of $V_{\mathrm{max}}$ from DMO (blue) and CoSANG (red). Observed satellites of the MW (black dashed) and M31 (gray dashed) are shown for comparison. 
The observations are likely incomplete at lower $V_{\mathrm{max}}$ \citep{Tollerud_2008,Hargis_2014,Kim_2018}. The left panel shows the differential distribution and the right panel shows the cumulative distribution for m12b (top), m12f (middle) and m12i (bottom). $V_{\mathrm{max}}$ peaks at smaller velocity in CoSANG compared to DMO and 
in general lies below the DMO distribution, CoSANG produces more lower $V_{\mathrm{max}}$ satellites. 
\citet{Kelley_2019} show that adding a disk potential into a DMO model results in a $V_{\mathrm{max}}$ distribution below the DMO model, consistent with our results, but with a steeper slope (Kelley+19 shown as dotted black line). They sample 12 halos. Bullock+2000 (one halo) in dotted gray \citep[][]{Bullock_2000} and W\&S11 (six halos) in dotted light gray \citep[][]{Wadepuhl_2011} have distributions with shallower slopes. Different processes can affect the slope and hydrodynamic models produce shallower profiles (\citealp{Wadepuhl_2011}), therefore this difference in the slope could be due to underlying physical processes in the model.

While m12b and m12f demonstrate a similar distribution, nevertheless this difference is not very significant in m12i. We will discuss this difference in more detail in Section \ref{sec:m12i_difference}. 

\begin{figure*}
	\includegraphics[scale=0.34]{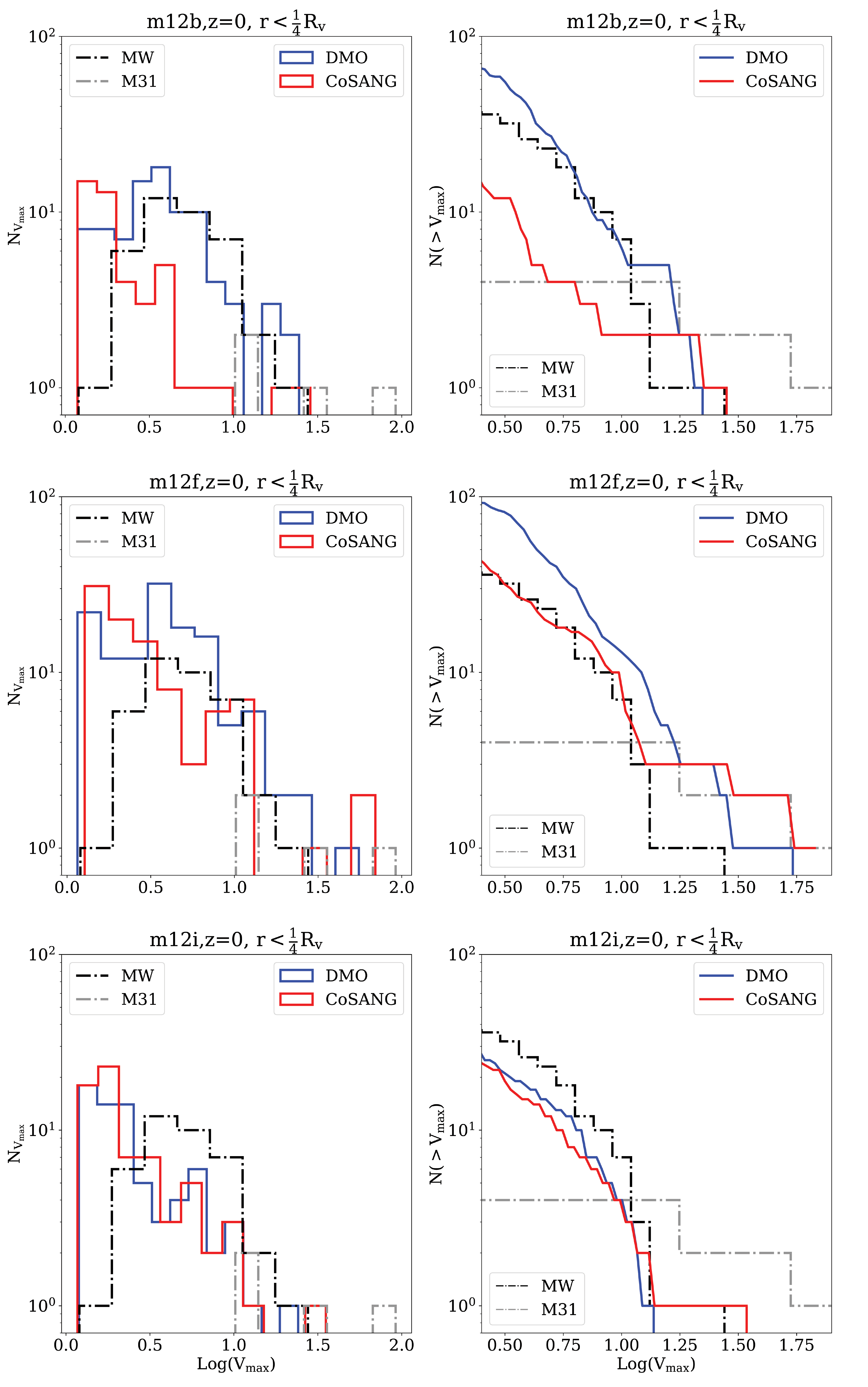}
    \caption{$V_{\mathrm{max}}$ distribution (left) and cumulative $V_{\mathrm{max}}$ distribution (right) for subhaloes at $r<\frac{1}{4}R_v$ of m12b (top), m12f (middle) and m12i (bottom). DMO is in blue and CoSANG is in red. Observed satellites of the MW (black dashed) and M31 (gray dashed) are shown for comparison. For a given $V_{\mathrm{max}}$ CoSANG simulations contain fewer subhaloes compared to DMO.}
    \label{fig:vmax_close}
\end{figure*}

Observations are certainly incomplete at large radius, so the comparison at smaller radius is likely more useful.
Figure \ref{fig:vmax_close} shows $V_{\mathrm{max}}$ distribution at closer distance ($\frac{1}{4}R_v$).
In this regime the CoSANG predictions bracket the observations well while the DMO predictions are higher.

Figure \ref{fig:vmax_z} shows the redshift evolution of $V_{\mathrm{max}}$. As with the mass function, the difference between the Vmax distribution in CoSANG versus DMO becomes significant at z<1.5.

\subsubsection{Radial profile}
\label{sec:RadialDistribution}

\begin{figure*}
	\includegraphics[width=\textwidth]{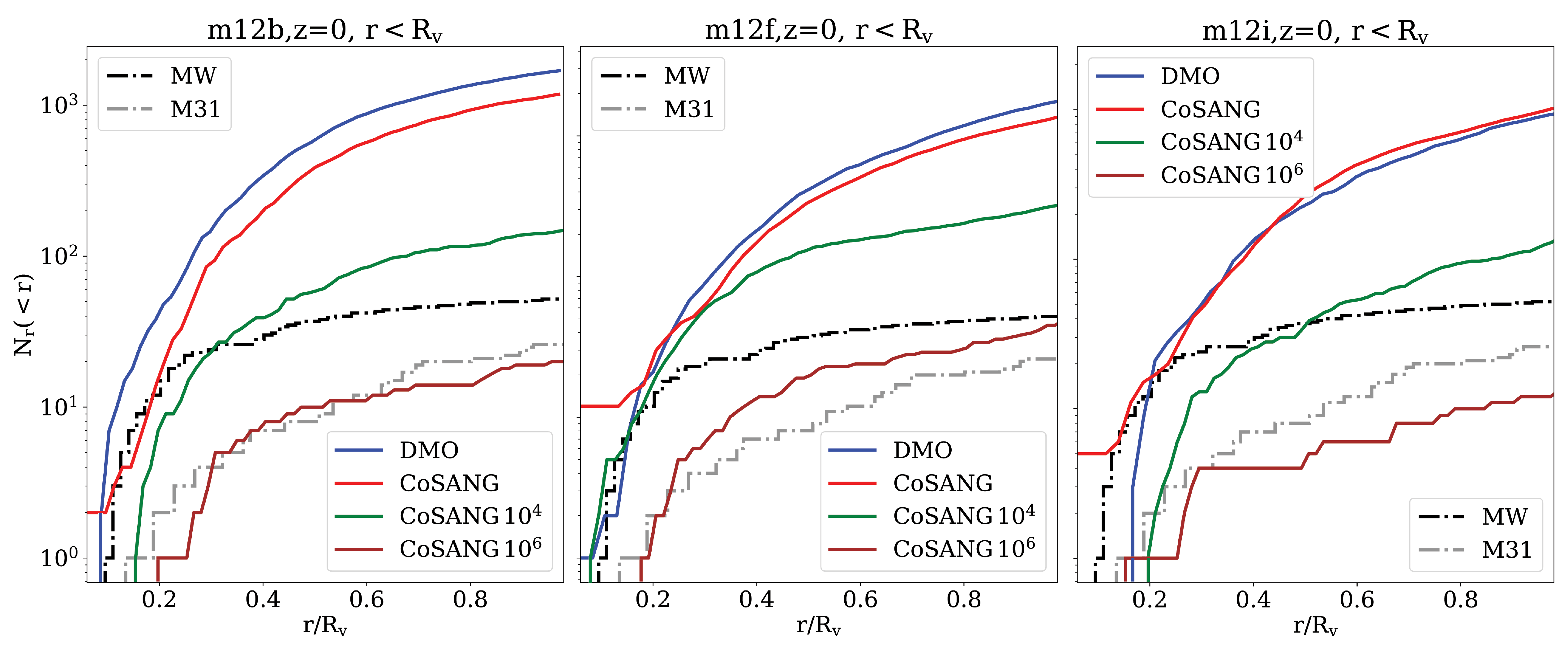}
    \caption{Cumulative radial distribution of subhaloes of m12b (left), m12f (middle) and m12i (right). This shows all subhaloes within the virial radius and DMO is in blue and CoSANG is in red. Observed satellites of the MW (black dashed) and M31 (gray dashed) are shown for comparison. CoSANG $10^4$ (green) and CoSANG $10^6$ (brown) represent CoSANG with stellar mass cut $>10^4M_\odot$ and $>10^6M_\odot$ respectively. The number of subhaloes in CoSANG is reduced compared to DMO at most distances, though it shows an increase in the very innermost region. CoSANG appears quite consistent with the MW data within 50 kpc, beyond which the observational data is likely quite incomplete.}
    \label{fig:RadialAbundance}
\end{figure*}

An analytic potential not only can change the distributions of subhalo masses but also their positions. In this section we study the radial distribution of subhaloes. Fig \ref{fig:RadialAbundance} shows the number of subhaloes enclosed in a given distance as a function of distance from the center, DMO (blue) and CoSANG (red) are the substructure distribution within the virial radius of m12b (left), m12f (middle) and m12i (right). Observed satellite galaxies of the MW (black dashed) and M31 (gray dashed) are shown for comparison. 
We see a uniform decrease in CoSANG relative to DMO at all radii, plus an increase at the closest distances. 

The observed satellites of the Milky Way are quite consistent with CoSANG and lower than the DMO predictions at radii <50 kpc, beyond which incompleteness in the observations becomes a major issue \citep{Tollerud_2008,Hargis_2014,Kim_2018,Samuel_2020}. 

Like the mass distribution, the redshift evolution of the radial distribution starts at z~1.5 and grows with time (Figure \ref{fig:RadialAbundanceRedshift})

A comparison between a DMO and DMO+disk; in PhatELVIS, shows a similar trend \citep{Kelley_2019}. 
\citet{Carlsten_2020} compare the radial distribution of subhaloes in variety of models (DMO, DMO+disk, hydro, box and zoom-in) with observations and show subhaloes of the MW mass galaxies are more centrally concentrated compared to these simulations. \citet{Grand_2021} show increasing resolution in their hydrodynamic model produces more subhaloes and the radial distribution of subhaloes is more centrally concentrated.

The implemented potential in CoSANG reduces the total number of subhaloes but does not deplete them at small radii. \citet{Newton_2018,Samuel_2020} predict the discovery of more satellites at large radii, however \citet{Carlsten_2020} believe this will not be enough to account for the lower number of satellites at small radii in observation. 

Incorporating a stellar mass
threshold reduces the total number of satellites significantly and the result is more consistent with the current observational data. The lower stellar mass limit in the M31 sample is $10^4M_\odot$. The distribution of observed galaxies lies below CoSANG $10^4$ in all three models. CoSANG $10^4$ suggests we might expect more satellites at larger radii waiting to be discovered.

\subsection{Halo shape}
\label{sec:InnerStructure}
We adapt the same approach as \citet{Zemp_2011} to measure the triaxial dark matter halo shape.
We use the second moment of the mass distribution as shape:
 \begin{equation}\label{eq:Sij}
 S_{ij}=\frac{\sum_k m_k(r_k)_i(r_k)_j}{\sum_k m_k}
 \end{equation}
 Where $(r_k)_j$ denotes the j-component of the position vector of the kth particle.

For each radial bin we calculate the second moment of the mass distribution and we find its eigenvalues and eigenvectors. Eigenvalues are proportional to the square of axis lengths; we denote them from the longest to the shortest as $a,b,c$ with associated eigenvectors $\vec{A}$ ,$\vec{B}$ ,$\vec{C}$ representing the direction of each axis. In the next iteration for this bin we change the volume from which particles are selected to be an ellipsoidal shell whose shape and direction are consistent with the previous iteration, and iterate until we converge or until we reach the iteration limit. One important problem with determining the shape is the treatment of the subhaloes, which we choose to exclude.

We use the average radius of the bins as corresponding radii and we study the evolution of the shape as a function of the distance from the host.

\begin{figure*}
	\includegraphics[width=\textwidth]{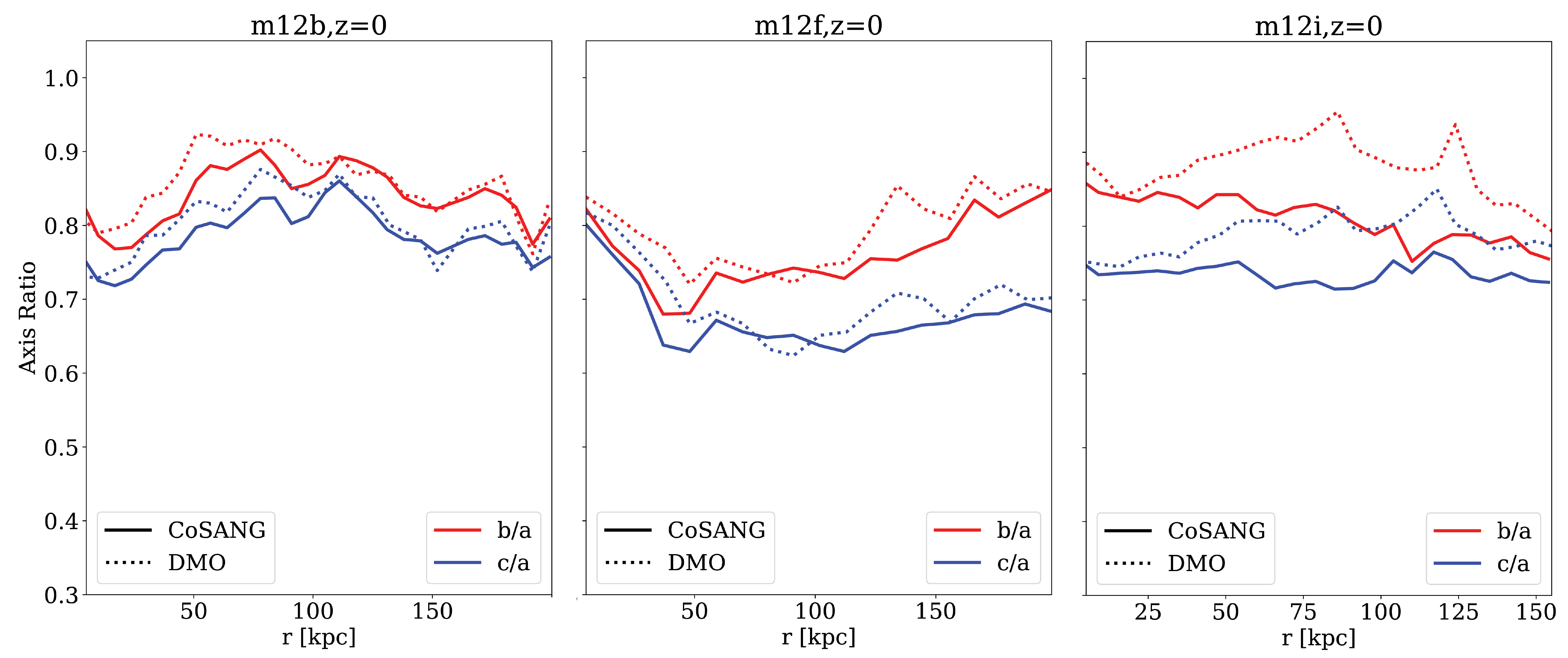}
    \caption{Shapes as function of radial distance from the center of mass of the halo in m12b (left), m12f (middle) and m12i (right). $\frac{b}{a}$ is in red, $\frac{c}{a}$ is in blue, and DMO and CoSANG are shown as the dashed and solid lines respectively. There is a difference between DMO and CoSANG for all halos, always in the sense that the axis ratios are smaller in CoSANG, but the radial dependence varies between halos. }
    \label{fig:shapes_m12bAxis}
\end{figure*}

Figure \ref{fig:shapes_m12bAxis} shows the axis ratios $\frac{b}{a}$ (red) and $\frac{c}{a}$ (blue) in both DMO (dashed) and CoSANG (solid) for m12b (left), m12f (middle) and m12i (right). In m12b the difference between two models are more significant at distances closer than 100kpc and beyond this distance the axes are almost the same. 
In m12f we see axis ratios are lower in CoSANG. The difference in axis ratios in mid-radii is not as significant as at small and large radii.
In m12i the difference is significant everywhere within the virial radius. In all three halos, CoSANG produces smaller axis ratios so it is less spherical than DMO. This could be mainly due to the existence of a disk potential at the center. This is not consistent with similar studies with hydrodynamic model that produce rounder halos compared to DMO models. One possible explanation could be the dynamical effect of the feedback that is stronger than the dynamical effect of the disk. Feedback is accounted for indirectly in CoSANG, via SAGE, but not directly. This could be an interesting update to study the dynamic effect of the feedback in future work.

One interesting result is the significant difference between DMO and CoSANG in m12i. Although this late forming halo with violent relaxation does not show a significant difference in its subhalo distribution, its shape shows a significant difference between DMO and CoSANG.

\begin{figure*}
	\includegraphics[width=\textwidth]{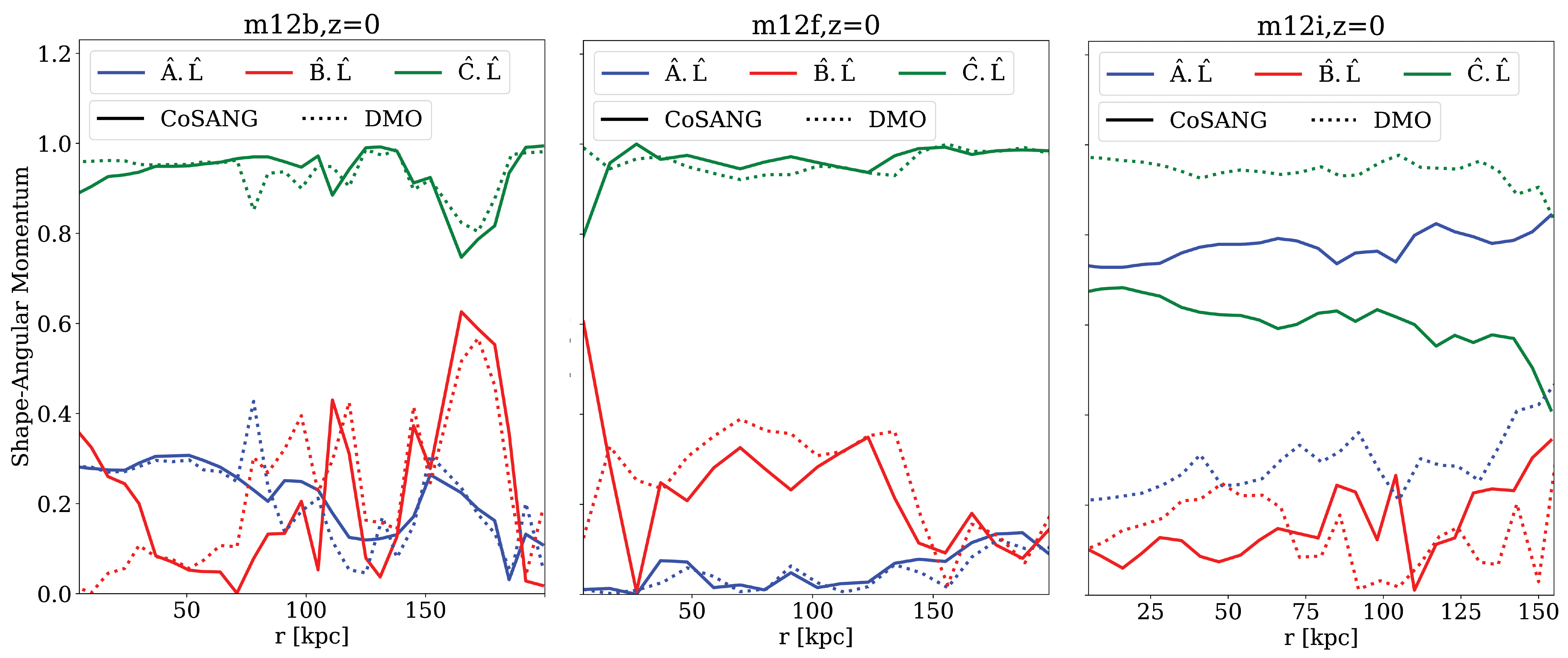}
    \caption{Shape axis vs angular momentum orientation in m12b (left), m12f (middle) and m12i (right). Individual shape axis (A in blue, B in red, C in green) alignment with the angular momentum both in DMO (dashed) and CoSANG (solid). In m12b and m12f CoSANG and DMO halos have similar alignment at distances larger than 40 kpc, while within inner 40kpc, we see more alignment between the disk and B and slightly less alignment between the disk and C. Interestingly m12i displays a very different behavior: its disk is more aligned with A in CoSANG while it is aligned with C in DMO. The B and C mixing at R>125 kpc is due to degeneracy between the axes in this almost-prolate region.}
    \label{fig:shapes_m12bAM}
\end{figure*}

The shortest axis being reduced in CoSANG may reflect the orientation of the angular momentum of the halo, which sets the direction of the disk potential (see Section \ref{sec:potentials}). We can examine this in Figure \ref{fig:shapes_m12bAM} where we compare the direction of each axis to that of the angular momentum vector.  
We use the total angular momentum of the halo to compare the orientation of the shape with the angular momentum and consequently the galactic disk. We plot each component separately ($\vec{A}. \vec{L}$ in blue, $\vec{B}. \vec{L}$ in red, $\vec{C}. \vec{L}$ in green) in DMO (dashed) and in CoSANG (solid) for both m12b (left), m12f (middle) and m12i (right).

For m12b and m12f in CoSANG, the minor axis is aligned with the angular momentum in DMO, as seen in other studies \citep[e.g.][]{Bett_2010,Teklu_2015,Obreja_2022} CoSANG modifies the inner regions and makes them align more with the intermediate axis. Similar weak alignments and a large scatter in the distribution of these alignments have been studied in different models \citep{Bett_2010,Vera_Ciro_2011,Schneider_2012,Bryan_2013} which suggest the angular momentum and subsequently the disk orientation can play an important role in determining the shape. On the other hand, in m12i the alignment switches in CoSANG from being predominanetly along the minor axis to being predominently along the major axis.

CoSANG produces more prolate and less spherical shapes than DMO. This demonstrates that the non-linear effects of baryonic physics can result in more complicated differences, and many earlier studies have found correlations between the angular momentum and environment, assembly history and characteristics of the halo \citep[e.g.][]{Gao_2005,Faltenbacher_2010,Libeskind_2012,Schneider_2012,Sharma_2012,Trowland_2013,Genel_2015,Zjupa_2017}.

\subsection{Halo twists}
\label{sec:halo_twists}
\begin{figure*}
	\includegraphics[scale=0.33]{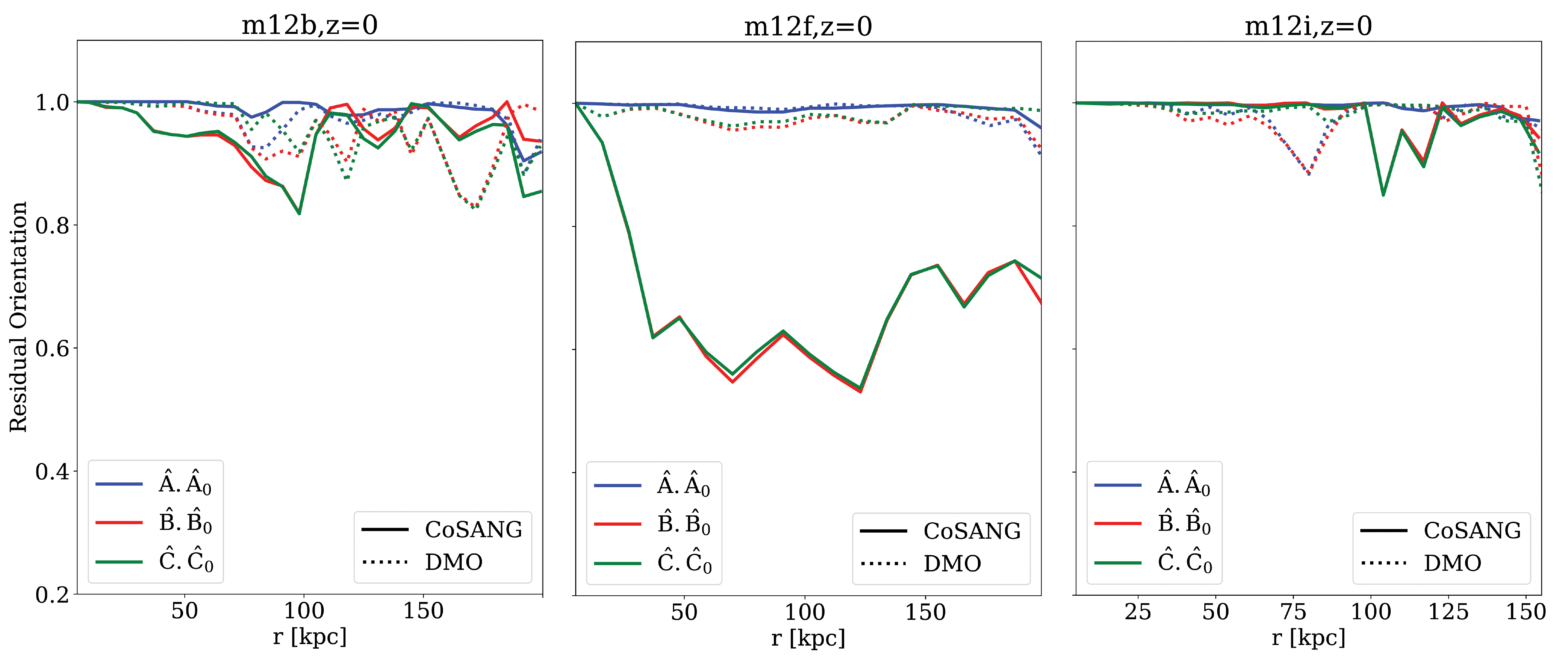}
    \caption{Shape axis residual in m12b (left), m12f (middle) and m12i (right). The residual is the orientation of the axis at each radial bin (A in blue, B in red, and C in green) relative to the orientation of that axis in the innermost bin in both DMO (dashed) and CoSANG (solid), The haloes have a relatively consistent orientation at all distances. m12b and m12i have insignificantly more twists in CoSANG. m12f demonstrates a significant twist }
    \label{fig:shapes_m12bResidual}
\end{figure*}

We can study internal twists in the halo shape to understand if different layers respond to the central potential or if the inter-layer interaction can dominate the kinematics. We compare the orientation in each bin with the first bin (residual orientation). Figure \ref{fig:shapes_m12bResidual} shows the results in DMO (dashed) and CoSANG (solid) for m12b (left), m12f (middle) and m12i (right) for the three axis vectors (in colors) considered previously.
In both m12b and m12i, we do not see a significant difference between DMO and CoSANG and in both halos all layers have relatively well-aligned shape orientations with no sudden changes. The major axis A has the highest self-alignment but B and C experience small changes. m12f, on the other hand, demonstrates a significant twist in the shape in CoSANG.

\subsection{Diversity of substructures }
\label{sec:m12i_difference}
CoSANG allows us to understand the impact of the baryonic potential on substructures. m12b and m12f demonstrate differences between DMO and CoSANG as one might expect; however m12i raises the question why it shows a different behavior to the two other halos? We believe this is an interesting outcome supporting the importance of a self-consistent interactive hybrid models like CoSANG. Although all halos are in the same mass range, they experience different formation histories. Merger history and environment can change the substructures and violent relaxation can dominate other forces. A comparison between different models including full physics hydrodynamic simulations shows a wide range of substructures even for the same mass host (\citealp{Engler_2021}).
\citet{Rey_2022} demonstrated using genetically modified DMO models that early forming and late forming halos can exhibit different structure in the dark matter particles that they tagged with stellar populations.

As discussed in section \ref{sec:galaxyevolution}, m12i is a late forming halo and has a different merger history compared to m12b and m12f. This difference is reported in other similar studies of simulations that use the same initial conditions, where capturing the unusual accretion history of m12i becomes a challenge (\citealp{Cunningham_2021}). SAMs are sensitive to the merger history of the halo and this difference can leave their fingerprint on the final substructure. Therefore we can not ignore this assembly history, and hence any hybrid model should include baryonic potentials that reflect not only the halo abundance but also the associated environment and history. CoSANG has the advantage of including all of these considerations. 

Such a diversity is reported in hydrodynamic models (e.g., \citealp{Samuel_2020}, \citealp{Engler_2021}) but DMO models do not show as significant a diversity (\citealp{Kelley_2019}).
Although there are differences in the sample size in different studies and different methods we may use to identify distributions, we can safely hypothesize the environment and formation history may explain most of the differences between different models and similar studies. We choose these galaxies based on their mass at $z=0$ but they don't necessarily follow the same evolution as the MW and their ultimate properties may fail to capture these differences. This is one of the strongest features of CoSANG, that we implement a full SAM. This is a significant difference compared to similar studies. CoSANG includes different components of the galaxy and the SAM is very sensitive not only to the halo abundance but also the merger tree, so even similar halo abundances can produce different galaxy parameters. This makes CoSANG a very powerful toolkit to create large samples of galaxies having a diversity in properties capturing individual galaxy formation histories.

\section{Conclusions}
We present CoSANG (Coupled Semi-Analytic N-body Galaxies), a new cosmological model developed for stellar halo simulations, but which has many additional applications. In this paper, we use it to study the structure of dark matter halos of Milky Way mass galaxies. CoSANG is a hybrid model with an N-body core and a semi-analytic model (SAM) coupled at every time-step.  
CoSANG allows for more accurate dynamics than a pure dark matter simulation due to the inclusion of a potential corresponding to the baryonic component of the galaxy and these dynamics have significant effect on substructures. CoSANG has been tested using a variety of scenarios from isolated orbit models to large scale cosmological simulations. 

We performed zoom-in simulations of three Milky Way (MW) mass halos and analyzed their substructures in both dark matter only (DMO) and CoSANG cases. We find the following:
\begin{itemize}
    \item The dynamical effect of a baryonic potential is significant on the inner structure of the halo.
    \item CoSANG produces a smaller number of subhaloes compared to the DMO model. This is consistent with similar studies indicating the galactic disk can reduce the number of subhaloes. The baryonic potential can both enhance the tidal field to disrupt substructure, and also bring them in to smaller radius. Thus we see the overall number of subhaloes has decreased in CoSANG and also slightly enhanced at small radii. We still see a higher number of subhaloes compared to hydrodynamic models. This suggest we will need more than baryonic potentials (such as feedback force) to reproduce similar results as hydrodynamic models. Adding a stellar mass threshold reduces the number of satellites significantly, especially at lower mass end.
    \item N-body models produce many subhaloes. At radii where observational samples of Milky Way satellites are expected to be mostly complete, the reduced number of CoSANG subhaloes provides a reasonably good match, but that it still predicts more than the number of M31 satellites.
    \item The SAM SAGE predicts galaxies start forming at very high redshift ($z\sim 13$), and there is no significant difference between CoSANG and DMO until $z\sim 1.5$ but this difference then grows with time.
    \item Although adding a galactic potential can result in differences between DMO and CoSANG, our MW analogue m12i, that formed later and had a more significant major merger does not exhibit much difference between DMO and CoSANG, unlike the others (m12b, m12f). All three halos have similar masses but different merger histories and different environments can change the significance of this variation from one halo to another.
    \item CoSANG produces more stretched (more prolate) shapes compared to DMO. Hydrodynamic models produce more oblate shapes. This may suggest the galactic potential is not the main factor determining the final shape of a halo and feedback may play a more important role in shape of the halo. This could be investigated in future update of CoSANG to include the dynamical effect of the feedback. The minor axis is aligned with the angular momentum of the halo in both models but this alignment is slightly different at closer distances in CoSANG where we see more alignment between the intermediate axis and angular momentum.
    \item The total number of subhaloes in CoSANG is lower than DMO and their radial separation is more centrally concentrated. 
\end{itemize}

A hybrid model like CoSANG with an empirical baryonic potential is computationally efficient. Although this model is not a replacement to a full physics hydrodynamic simulation, it is a complementary model and a powerful tool to study galactic dynamics and the stellar halo. 

CoSANG has been used here to study dark matter kinematics. But CoSANG is also capable of producing stellar halo populations through particle tagging. Its use for stellar halo kinematics and stellar halo populations will be presented in an upcoming paper (Talei et al., in preparation).

\section*{Acknowledgements}
The authors are sincerely grateful for every help we received throughout the development of CoSANG. Astronomy faculty members at the University of Alabama were supportive during entire development process. 
We are also immensely grateful to Monica Valluri (University of Michigan) for her contribution to CoSANG and comments on an earlier version of the manuscript.
We also acknowledge useful conversations with galaxy theory group at Steward observatory. Especially Gurtina Besla, Peter Behroozi and Nicolas Garavito Camargo have offered many insightful comments. We gratefully acknowledge the support of XSEDE Campus Champions program.

We ran all cosmological simulations using UAHPC (University of Alabama High Performance Computing) and TACC (Texas Advanced Computing Center) through XSEDE allocations (TG-AST200026 and TG-PHY200076) and OZ-Star (Swinburne super-computing center). Most post-processing and analysis was done with UAHPC. 

\section*{Software Acknowledgement}
\label{sec:codes_appendix}
All codes we use are publicly available at this URL: \href{http://www.github.com/stalei}{http://www.github.com/stalei}\\
Many of these codes are developed and used for the first time. The authors respect and encourage reproducible results as an important tool for improvement. All first time codes are under public domain licence and using and improving them with reference is encouraged. 
We also acknowledge the use of the following packages:\\
Galstep: \url{https://github.com/ruggiero/galstep}, \\
numpy: \citealp{Harris_2020},\\
Rockstar: \citealp{Behroozi_2012},\\
Scipy: \citealp{Jones_2001},\\
Splash: \citealp{Price_2007},\\
yt: \citealp{Turk_2011},

\section*{Data Availability}
All data used in this work is produced with CoSANG. The simulation output will be public as soon as the full sample is complete. Data is available upon request.




\bibliographystyle{mnras}
\bibliography{refs} 

\begin{thebibliography}{}
\makeatletter
\relax
\def\mn@urlcharsother{\let\do\@makeother \do\$\do\&\do\#\do\^\do\_\do\%\do\~}
\def\mn@doi{\begingroup\mn@urlcharsother \@ifnextchar [ {\mn@doi@}
  {\mn@doi@[]}}
\def\mn@doi@[#1]#2{\def\@tempa{#1}\ifx\@tempa\@empty \href
  {http://dx.doi.org/#2} {doi:#2}\else \href {http://dx.doi.org/#2} {#1}\fi
  \endgroup}
\def\mn@eprint#1#2{\mn@eprint@#1:#2::\@nil}
\def\mn@eprint@arXiv#1{\href {http://arxiv.org/abs/#1} {{\tt arXiv:#1}}}
\def\mn@eprint@dblp#1{\href {http://dblp.uni-trier.de/rec/bibtex/#1.xml}
  {dblp:#1}}
\def\mn@eprint@#1:#2:#3:#4\@nil{\def\@tempa {#1}\def\@tempb {#2}\def\@tempc
  {#3}\ifx \@tempc \@empty \let \@tempc \@tempb \let \@tempb \@tempa \fi \ifx
  \@tempb \@empty \def\@tempb {arXiv}\fi \@ifundefined
  {mn@eprint@\@tempb}{\@tempb:\@tempc}{\expandafter \expandafter \csname
  mn@eprint@\@tempb\endcsname \expandafter{\@tempc}}}

\bibitem[\protect\citeauthoryear{Allgood, Flores, Primack, Kravtsov, Wechsler,
  Faltenbacher  \& Bullock}{Allgood et~al.}{2006}]{Allgood_2006}
Allgood B.,  Flores R.~A.,  Primack J.~R.,  Kravtsov A.~V.,  Wechsler R.~H.,
  Faltenbacher A.,   Bullock J.~S.,  2006, \mn@doi [Monthly Notices of the
  Royal Astronomical Society] {10.1111/j.1365-2966.2006.10094.x}, 367, 1781

\bibitem[\protect\citeauthoryear{{Angulo} \& {Hahn}}{{Angulo} \&
  {Hahn}}{2022}]{Angulo_2022}
{Angulo} R.~E.,  {Hahn} O.,  2022, \mn@doi [Living Reviews in Computational
  Astrophysics] {10.1007/s41115-021-00013-z}, \href
  {https://ui.adsabs.harvard.edu/abs/2022LRCA....8....1A} {8, 1}

\bibitem[\protect\citeauthoryear{Bailin et~al.,}{Bailin
  et~al.}{2005}]{Bailin_2005}
Bailin J.,  et~al., 2005, \mn@doi [The Astrophysical Journal] {10.1086/432157},
  627, L17

\bibitem[\protect\citeauthoryear{Bailin, Bell, Valluri, Stinson, Debattista,
  Couchman  \& Wadsley}{Bailin et~al.}{2014}]{Bailin_2014}
Bailin J.,  Bell E.~F.,  Valluri M.,  Stinson G.~S.,  Debattista V.~P.,
  Couchman H. M.~P.,   Wadsley J.,  2014, \mn@doi [The Astrophysical Journal]
  {10.1088/0004-637x/783/2/95}, 783, 95

\bibitem[\protect\citeauthoryear{{Barnes} \& {Hut}}{{Barnes} \&
  {Hut}}{1986}]{Barnes_1986}
{Barnes} J.,  {Hut} P.,  1986, \mn@doi [\nat] {10.1038/324446a0}, \href
  {https://ui.adsabs.harvard.edu/abs/1986Natur.324..446B} {324, 446}

\bibitem[\protect\citeauthoryear{{Bauer}, {Widrow}  \& {Erkal}}{{Bauer}
  et~al.}{2018}]{Bauer_2018}
{Bauer} J.~S.,  {Widrow} L.~M.,   {Erkal} D.,  2018, \mn@doi [\mnras]
  {10.1093/mnras/sty120}, \href
  {https://ui.adsabs.harvard.edu/abs/2018MNRAS.476..198B} {476, 198}

\bibitem[\protect\citeauthoryear{Baugh}{Baugh}{2006}]{Baugh_2006}
Baugh C.~M.,  2006, \mn@doi [Reports on Progress in Physics]
  {10.1088/0034-4885/69/12/r02}, 69, 3101

\bibitem[\protect\citeauthoryear{{Behroozi}, {Conroy}  \&
  {Wechsler}}{{Behroozi} et~al.}{2010}]{Behroozi_2010}
{Behroozi} P.~S.,  {Conroy} C.,   {Wechsler} R.~H.,  2010, \mn@doi [\apj]
  {10.1088/0004-637X/717/1/379}, \href
  {https://ui.adsabs.harvard.edu/abs/2010ApJ...717..379B} {717, 379}

\bibitem[\protect\citeauthoryear{Behroozi, Wechsler  \& Wu}{Behroozi
  et~al.}{2012}]{Behroozi_2012}
Behroozi P.~S.,  Wechsler R.~H.,   Wu H.-Y.,  2012, \mn@doi [The Astrophysical
  Journal] {10.1088/0004-637x/762/2/109}, 762, 109

\bibitem[\protect\citeauthoryear{{Behroozi}, {Wechsler}  \&
  {Conroy}}{{Behroozi} et~al.}{2013}]{Behroozi_2013}
{Behroozi} P.~S.,  {Wechsler} R.~H.,   {Conroy} C.,  2013, \mn@doi [\apj]
  {10.1088/0004-637X/770/1/57}, \href
  {https://ui.adsabs.harvard.edu/abs/2013ApJ...770...57B} {770, 57}

\bibitem[\protect\citeauthoryear{Behroozi, Wechsler, Hearin  \&
  Conroy}{Behroozi et~al.}{2019}]{Behroozi_2019}
Behroozi P.,  Wechsler R.~H.,  Hearin A.~P.,   Conroy C.,  2019, \mn@doi
  [Monthly Notices of the Royal Astronomical Society] {10.1093/mnras/stz1182},
  488, 3143

\bibitem[\protect\citeauthoryear{Bennett et~al.,}{Bennett
  et~al.}{2013}]{Bennett_2013}
Bennett C.~L.,  et~al., 2013, \mn@doi [The Astrophysical Journal Supplement
  Series] {10.1088/0067-0049/208/2/20}, 208, 20

\bibitem[\protect\citeauthoryear{Benson}{Benson}{2010}]{Benson_2010}
Benson A.~J.,  2010, \mn@doi [Physics Reports]
  {https://doi.org/10.1016/j.physrep.2010.06.001}, 495, 33

\bibitem[\protect\citeauthoryear{Benson, Pearce, Frenk, Baugh  \&
  Jenkins}{Benson et~al.}{2001}]{Benson_2001}
Benson A.~J.,  Pearce F.~R.,  Frenk C.~S.,  Baugh C.~M.,   Jenkins A.,  2001,
  \mn@doi [Monthly Notices of the Royal Astronomical Society]
  {10.1046/j.1365-8711.2001.03966.x}, 320, 261

\bibitem[\protect\citeauthoryear{{Besla}, {Kallivayalil}, {Hernquist}, {van der
  Marel}, {Cox}  \& {Kere{\v{s}}}}{{Besla} et~al.}{2010}]{Besla_2010}
{Besla} G.,  {Kallivayalil} N.,  {Hernquist} L.,  {van der Marel} R.~P.,  {Cox}
  T.~J.,   {Kere{\v{s}}} D.,  2010, \mn@doi [\apjl]
  {10.1088/2041-8205/721/2/L97}, \href
  {https://ui.adsabs.harvard.edu/abs/2010ApJ...721L..97B} {721, L97}

\bibitem[\protect\citeauthoryear{{Bett}, {Eke}, {Frenk}, {Jenkins}  \&
  {Okamoto}}{{Bett} et~al.}{2010}]{Bett_2010}
{Bett} P.,  {Eke} V.,  {Frenk} C.~S.,  {Jenkins} A.,   {Okamoto} T.,  2010,
  \mn@doi [\mnras] {10.1111/j.1365-2966.2010.16368.x}, \href
  {https://ui.adsabs.harvard.edu/abs/2010MNRAS.404.1137B} {404, 1137}

\bibitem[\protect\citeauthoryear{Binney \& Tremaine}{Binney \&
  Tremaine}{2008}]{Binney_2008}
Binney J.,  Tremaine S.,  2008, Galactic Dynamics, 2 edn.
Princeton

\bibitem[\protect\citeauthoryear{{Bonaca} \& {Hogg}}{{Bonaca} \&
  {Hogg}}{2018}]{Bonaca_2018}
{Bonaca} A.,  {Hogg} D.~W.,  2018, \mn@doi [\apj] {10.3847/1538-4357/aae4da},
  \href {https://ui.adsabs.harvard.edu/abs/2018ApJ...867..101B} {867, 101}

\bibitem[\protect\citeauthoryear{{Bonaca}, {Geha}, {K{\"u}pper}, {Diemand},
  {Johnston}  \& {Hogg}}{{Bonaca} et~al.}{2014}]{Bonaca_2014}
{Bonaca} A.,  {Geha} M.,  {K{\"u}pper} A. H.~W.,  {Diemand} J.,  {Johnston}
  K.~V.,   {Hogg} D.~W.,  2014, \mn@doi [\apj] {10.1088/0004-637X/795/1/94},
  \href {https://ui.adsabs.harvard.edu/abs/2014ApJ...795...94B} {795, 94}

\bibitem[\protect\citeauthoryear{{Boylan-Kolchin}, {Bullock}  \&
  {Kaplinghat}}{{Boylan-Kolchin} et~al.}{2011}]{Boylan-Kolchin_2011}
{Boylan-Kolchin} M.,  {Bullock} J.~S.,   {Kaplinghat} M.,  2011, \mn@doi
  [\mnras] {10.1111/j.1745-3933.2011.01074.x}, \href
  {https://ui.adsabs.harvard.edu/abs/2011MNRAS.415L..40B} {415, L40}

\bibitem[\protect\citeauthoryear{Brooks \& Zolotov}{Brooks \&
  Zolotov}{2014}]{Brooks_2014}
Brooks A.~M.,  Zolotov A.,  2014, \mn@doi [The Astrophysical Journal]
  {10.1088/0004-637x/786/2/87}, 786, 87

\bibitem[\protect\citeauthoryear{{Brooks}, {Kuhlen}, {Zolotov}  \&
  {Hooper}}{{Brooks} et~al.}{2013}]{Brooks_2013}
{Brooks} A.~M.,  {Kuhlen} M.,  {Zolotov} A.,   {Hooper} D.,  2013, \mn@doi
  [\apj] {10.1088/0004-637X/765/1/22}, \href
  {https://ui.adsabs.harvard.edu/abs/2013ApJ...765...22B} {765, 22}

\bibitem[\protect\citeauthoryear{{Bryan}, {Kay}, {Duffy}, {Schaye}, {Dalla
  Vecchia}  \& {Booth}}{{Bryan} et~al.}{2013}]{Bryan_2013}
{Bryan} S.~E.,  {Kay} S.~T.,  {Duffy} A.~R.,  {Schaye} J.,  {Dalla Vecchia} C.,
    {Booth} C.~M.,  2013, \mn@doi [\mnras] {10.1093/mnras/sts587}, \href
  {https://ui.adsabs.harvard.edu/abs/2013MNRAS.429.3316B} {429, 3316}

\bibitem[\protect\citeauthoryear{Bullock \& Johnston}{Bullock \&
  Johnston}{2005}]{Bullock_2005}
Bullock J.~S.,  Johnston K.~V.,  2005, \mn@doi [The Astrophysical Journal]
  {10.1086/497422}, 635, 931

\bibitem[\protect\citeauthoryear{Bullock, Kravtsov  \& Weinberg}{Bullock
  et~al.}{2000}]{Bullock_2000}
Bullock J.~S.,  Kravtsov A.~V.,   Weinberg D.~H.,  2000, \mn@doi [The
  Astrophysical Journal] {10.1086/309279}, 539, 517

\bibitem[\protect\citeauthoryear{{Carlsten}, {Greene}, {Peter}, {Greco}  \&
  {Beaton}}{{Carlsten} et~al.}{2020}]{Carlsten_2020}
{Carlsten} S.~G.,  {Greene} J.~E.,  {Peter} A. H.~G.,  {Greco} J.~P.,
  {Beaton} R.~L.,  2020, \mn@doi [\apj] {10.3847/1538-4357/abb60b}, \href
  {https://ui.adsabs.harvard.edu/abs/2020ApJ...902..124C} {902, 124}

\bibitem[\protect\citeauthoryear{Carollo}{Carollo}{1999}]{Carollo_1999}
Carollo C.~M.,  ed. 1999, {The formation of galactic bulges}

\bibitem[\protect\citeauthoryear{{Carollo}}{{Carollo}}{2004}]{Carollo_2004}
{Carollo} C.~M.,  2004, in {Ho} L.~C.,  ed., Coevolution of Black Holes and
  Galaxies. p.~231 (\mn@eprint {arXiv} {astro-ph/0306021})

\bibitem[\protect\citeauthoryear{{Chua}, {Pillepich}, {Vogelsberger}  \&
  {Hernquist}}{{Chua} et~al.}{2019}]{Chua_2019}
{Chua} K. T.~E.,  {Pillepich} A.,  {Vogelsberger} M.,   {Hernquist} L.,  2019,
  \mn@doi [\mnras] {10.1093/mnras/sty3531}, \href
  {https://ui.adsabs.harvard.edu/abs/2019MNRAS.484..476C} {484, 476}

\bibitem[\protect\citeauthoryear{{Cole}}{{Cole}}{1991}]{Cole_1991}
{Cole} S.,  1991, \mn@doi [\apj] {10.1086/169600}, \href
  {https://ui.adsabs.harvard.edu/abs/1991ApJ...367...45C} {367, 45}

\bibitem[\protect\citeauthoryear{Cole, Lacey, Baugh  \& Frenk}{Cole
  et~al.}{2000}]{Cole_2000}
Cole S.,  Lacey C.~G.,  Baugh C.~M.,   Frenk C.~S.,  2000, \mn@doi [Monthly
  Notices of the Royal Astronomical Society]
  {10.1046/j.1365-8711.2000.03879.x}, 319, 168

\bibitem[\protect\citeauthoryear{Cooper et~al.,}{Cooper
  et~al.}{2010}]{Cooper_2010}
Cooper A.~P.,  et~al., 2010, \mn@doi [Monthly Notices of the Royal Astronomical
  Society] {10.1111/j.1365-2966.2010.16740.x}, 406, 744

\bibitem[\protect\citeauthoryear{Cooper, Cole, Frenk, Le~Bret  \&
  Pontzen}{Cooper et~al.}{2017}]{Cooper_2017}
Cooper A.~P.,  Cole S.,  Frenk C.~S.,  Le~Bret T.,   Pontzen A.,  2017, \mn@doi
  [Monthly Notices of the Royal Astronomical Society] {10.1093/mnras/stx955},
  469, 1691

\bibitem[\protect\citeauthoryear{{Courteau}}{{Courteau}}{1996}]{Courteau_1996}
{Courteau} S.,  1996, \mn@doi [\apjs] {10.1086/192281}, \href
  {https://ui.adsabs.harvard.edu/abs/1996ApJS..103..363C} {103, 363}

\bibitem[\protect\citeauthoryear{Croton et~al.,}{Croton
  et~al.}{2006}]{Croton_2006}
Croton D.~J.,  et~al., 2006, \mn@doi [Monthly Notices of the Royal Astronomical
  Society] {10.1111/j.1365-2966.2005.09675.x}, 365, 11

\bibitem[\protect\citeauthoryear{Croton et~al.,}{Croton
  et~al.}{2016}]{Croton_2016}
Croton D.~J.,  et~al., 2016, \mn@doi [The Astrophysical Journal Supplement
  Series] {10.3847/0067-0049/222/2/22}, 222, 22

\bibitem[\protect\citeauthoryear{{Cunningham} et~al.,}{{Cunningham}
  et~al.}{2021}]{Cunningham_2021}
{Cunningham} E.~C.,  et~al., 2021, arXiv e-prints, \href
  {https://ui.adsabs.harvard.edu/abs/2021arXiv211002957C} {p. arXiv:2110.02957}

\bibitem[\protect\citeauthoryear{{D'Onghia}, {Springel}, {Hernquist}  \&
  {Keres}}{{D'Onghia} et~al.}{2010}]{Donghia_2010}
{D'Onghia} E.,  {Springel} V.,  {Hernquist} L.,   {Keres} D.,  2010, \mn@doi
  [\apj] {10.1088/0004-637X/709/2/1138}, \href
  {https://ui.adsabs.harvard.edu/abs/2010ApJ...709.1138D} {709, 1138}

\bibitem[\protect\citeauthoryear{{D'Souza} \& {Bell}}{{D'Souza} \&
  {Bell}}{2018}]{DSouza_2018}
{D'Souza} R.,  {Bell} E.~F.,  2018, \mn@doi [\mnras] {10.1093/mnras/stx3081},
  \href {https://ui.adsabs.harvard.edu/abs/2018MNRAS.474.5300D} {474, 5300}

\bibitem[\protect\citeauthoryear{{Dai}, {Robertson}  \& {Madau}}{{Dai}
  et~al.}{2018}]{Dai_2018}
{Dai} B.,  {Robertson} B.~E.,   {Madau} P.,  2018, \mn@doi [\apj]
  {10.3847/1538-4357/aabb06}, \href
  {https://ui.adsabs.harvard.edu/abs/2018ApJ...858...73D} {858, 73}

\bibitem[\protect\citeauthoryear{{Davis}, {Efstathiou}, {Frenk}  \&
  {White}}{{Davis} et~al.}{1985}]{Davis_1985}
{Davis} M.,  {Efstathiou} G.,  {Frenk} C.~S.,   {White} S.~D.~M.,  1985,
  \mn@doi [\apj] {10.1086/163168}, \href
  {https://ui.adsabs.harvard.edu/abs/1985ApJ...292..371D} {292, 371}

\bibitem[\protect\citeauthoryear{{De Lucia}, {Kauffmann}  \& {White}}{{De
  Lucia} et~al.}{2004}]{DeLucia2004}
{De Lucia} G.,  {Kauffmann} G.,   {White} S. D.~M.,  2004, \mn@doi [\mnras]
  {10.1111/j.1365-2966.2004.07584.x}, \href
  {https://ui.adsabs.harvard.edu/abs/2004MNRAS.349.1101D} {349, 1101}

\bibitem[\protect\citeauthoryear{{Diemand}, {Kuhlen}, {Madau}, {Zemp}, {Moore},
  {Potter}  \& {Stadel}}{{Diemand} et~al.}{2008}]{Diemand_2008}
{Diemand} J.,  {Kuhlen} M.,  {Madau} P.,  {Zemp} M.,  {Moore} B.,  {Potter} D.,
    {Stadel} J.,  2008, \mn@doi [\nat] {10.1038/nature07153}, \href
  {https://ui.adsabs.harvard.edu/abs/2008Natur.454..735D} {454, 735}

\bibitem[\protect\citeauthoryear{{Drlica-Wagner} et~al.,}{{Drlica-Wagner}
  et~al.}{2020}]{Drlica-Wagner_2020}
{Drlica-Wagner} A.,  et~al., 2020, \mn@doi [\apj] {10.3847/1538-4357/ab7eb9},
  \href {https://ui.adsabs.harvard.edu/abs/2020ApJ...893...47D} {893, 47}

\bibitem[\protect\citeauthoryear{{Dubinski} \& {Carlberg}}{{Dubinski} \&
  {Carlberg}}{1991}]{Dubinski_1991}
{Dubinski} J.,  {Carlberg} R.~G.,  1991, \mn@doi [\apj] {10.1086/170451}, \href
  {https://ui.adsabs.harvard.edu/abs/1991ApJ...378..496D} {378, 496}

\bibitem[\protect\citeauthoryear{{Eggen}, {Lynden-Bell}  \& {Sandage}}{{Eggen}
  et~al.}{1962}]{Eggen_1962}
{Eggen} O.~J.,  {Lynden-Bell} D.,   {Sandage} A.~R.,  1962, \mn@doi [\apj]
  {10.1086/147433}, \href
  {https://ui.adsabs.harvard.edu/abs/1962ApJ...136..748E} {136, 748}

\bibitem[\protect\citeauthoryear{{Einasto}}{{Einasto}}{1965}]{Einasto_1965}
{Einasto} J.,  1965, Trudy Astrofizicheskogo Instituta Alma-Ata, \href
  {https://ui.adsabs.harvard.edu/abs/1965TrAlm...5...87E} {5, 87}

\bibitem[\protect\citeauthoryear{{Einasto} \& {Haud}}{{Einasto} \&
  {Haud}}{1989}]{Einasto_1989}
{Einasto} J.,  {Haud} U.,  1989, \aap, \href
  {https://ui.adsabs.harvard.edu/abs/1989A&A...223...89E} {223, 89}

\bibitem[\protect\citeauthoryear{{Engler} et~al.,}{{Engler}
  et~al.}{2021}]{Engler_2021}
{Engler} C.,  et~al., 2021, \mn@doi [\mnras] {10.1093/mnras/stab2437}, \href
  {https://ui.adsabs.harvard.edu/abs/2021MNRAS.507.4211E} {507, 4211}

\bibitem[\protect\citeauthoryear{{Erkal} et~al.,}{{Erkal}
  et~al.}{2018}]{Erkal_2018}
{Erkal} D.,  et~al., 2018, \mn@doi [\mnras] {10.1093/mnras/sty2518}, \href
  {https://ui.adsabs.harvard.edu/abs/2018MNRAS.481.3148E} {481, 3148}

\bibitem[\protect\citeauthoryear{{Ewald}}{{Ewald}}{1921}]{Ewald_1921}
{Ewald} P.~P.,  1921, \mn@doi [Annalen der Physik] {10.1002/andp.19213690304},
  \href {https://ui.adsabs.harvard.edu/abs/1921AnP...369..253E} {369, 253}

\bibitem[\protect\citeauthoryear{{Faltenbacher} \& {White}}{{Faltenbacher} \&
  {White}}{2010}]{Faltenbacher_2010}
{Faltenbacher} A.,  {White} S. D.~M.,  2010, \mn@doi [\apj]
  {10.1088/0004-637X/708/1/469}, \href
  {https://ui.adsabs.harvard.edu/abs/2010ApJ...708..469F} {708, 469}

\bibitem[\protect\citeauthoryear{{Font}, {Johnston}, {Bullock}  \&
  {Robertson}}{{Font} et~al.}{2006}]{Font_2006}
{Font} A.~S.,  {Johnston} K.~V.,  {Bullock} J.~S.,   {Robertson} B.~E.,  2006,
  \mn@doi [\apj] {10.1086/505131}, \href
  {https://ui.adsabs.harvard.edu/abs/2006ApJ...646..886F} {646, 886}

\bibitem[\protect\citeauthoryear{{Font}, {McCarthy}  \& {Belokurov}}{{Font}
  et~al.}{2021}]{Font_2021}
{Font} A.~S.,  {McCarthy} I.~G.,   {Belokurov} V.,  2021, \mn@doi [\mnras]
  {10.1093/mnras/stab1332}, \href
  {https://ui.adsabs.harvard.edu/abs/2021MNRAS.505..783F} {505, 783}

\bibitem[\protect\citeauthoryear{{Fritz}, {Di Cintio}, {Battaglia}, {Brook}  \&
  {Taibi}}{{Fritz} et~al.}{2020}]{Fritz_2020}
{Fritz} T.~K.,  {Di Cintio} A.,  {Battaglia} G.,  {Brook} C.,   {Taibi} S.,
  2020, \mn@doi [\mnras] {10.1093/mnras/staa1040}, \href
  {https://ui.adsabs.harvard.edu/abs/2020MNRAS.494.5178F} {494, 5178}

\bibitem[\protect\citeauthoryear{{Gao}, {White}, {Jenkins}, {Stoehr}  \&
  {Springel}}{{Gao} et~al.}{2004}]{Gao_2004}
{Gao} L.,  {White} S.~D.~M.,  {Jenkins} A.,  {Stoehr} F.,   {Springel} V.,
  2004, \mn@doi [\mnras] {10.1111/j.1365-2966.2004.08360.x}, \href
  {https://ui.adsabs.harvard.edu/abs/2004MNRAS.355..819G} {355, 819}

\bibitem[\protect\citeauthoryear{{Gao}, {Springel}  \& {White}}{{Gao}
  et~al.}{2005}]{Gao_2005}
{Gao} L.,  {Springel} V.,   {White} S. D.~M.,  2005, \mn@doi [\mnras]
  {10.1111/j.1745-3933.2005.00084.x}, \href
  {https://ui.adsabs.harvard.edu/abs/2005MNRAS.363L..66G} {363, L66}

\bibitem[\protect\citeauthoryear{{Garavito-Camargo}, {Besla}, {Laporte},
  {Johnston}, {G{\'o}mez}  \& {Watkins}}{{Garavito-Camargo}
  et~al.}{2019}]{Garavito_2019}
{Garavito-Camargo} N.,  {Besla} G.,  {Laporte} C. F.~P.,  {Johnston} K.~V.,
  {G{\'o}mez} F.~A.,   {Watkins} L.~L.,  2019, \mn@doi [\apj]
  {10.3847/1538-4357/ab32eb}, \href
  {https://ui.adsabs.harvard.edu/abs/2019ApJ...884...51G} {884, 51}

\bibitem[\protect\citeauthoryear{{Garavito Camargo}, {Besla}, {Johnston},
  {Weinberg}, {Laporte}, {Price-Whelan}  \& {Gomez}}{{Garavito Camargo}
  et~al.}{2020}]{Garavito_Camargo_2020}
{Garavito Camargo} N.,  {Besla} G.,  {Johnston} K.,  {Weinberg} M.,  {Laporte}
  C.,  {Price-Whelan} A.,   {Gomez} F.~A.,  2020, in American Astronomical
  Society Meeting Abstracts \#236. p. 123.04

\bibitem[\protect\citeauthoryear{Garrison-Kimmel, Boylan-Kolchin, Bullock  \&
  Lee}{Garrison-Kimmel et~al.}{2014}]{Garrison_Kimmel_2014}
Garrison-Kimmel S.,  Boylan-Kolchin M.,  Bullock J.~S.,   Lee K.,  2014,
  \mn@doi [Monthly Notices of the Royal Astronomical Society]
  {10.1093/mnras/stt2377}, 438, 2578

\bibitem[\protect\citeauthoryear{Garrison-Kimmel et~al.,}{Garrison-Kimmel
  et~al.}{2017}]{Garrison-Kimmel_2017}
Garrison-Kimmel S.,  et~al., 2017, \mn@doi [Monthly Notices of the Royal
  Astronomical Society] {10.1093/mnras/stx1710}, 471, 1709

\bibitem[\protect\citeauthoryear{{Genel}, {Fall}, {Hernquist}, {Vogelsberger},
  {Snyder}, {Rodriguez-Gomez}, {Sijacki}  \& {Springel}}{{Genel}
  et~al.}{2015}]{Genel_2015}
{Genel} S.,  {Fall} S.~M.,  {Hernquist} L.,  {Vogelsberger} M.,  {Snyder}
  G.~F.,  {Rodriguez-Gomez} V.,  {Sijacki} D.,   {Springel} V.,  2015, \mn@doi
  [\apjl] {10.1088/2041-8205/804/2/L40}, \href
  {https://ui.adsabs.harvard.edu/abs/2015ApJ...804L..40G} {804, L40}

\bibitem[\protect\citeauthoryear{Genel et~al.,}{Genel
  et~al.}{2019}]{Genel_2019}
Genel S.,  et~al., 2019, \mn@doi [The Astrophysical Journal]
  {10.3847/1538-4357/aaf4bb}, 871, 21

\bibitem[\protect\citeauthoryear{{Grand} et~al.,}{{Grand}
  et~al.}{2017}]{Grand_2017}
{Grand} R. J.~J.,  et~al., 2017, \mn@doi [\mnras] {10.1093/mnras/stx071}, \href
  {https://ui.adsabs.harvard.edu/abs/2017MNRAS.467..179G} {467, 179}

\bibitem[\protect\citeauthoryear{{Grand} et~al.,}{{Grand}
  et~al.}{2021}]{Grand_2021}
{Grand} R. J.~J.,  et~al., 2021, \mn@doi [\mnras] {10.1093/mnras/stab2492},
  \href {https://ui.adsabs.harvard.edu/abs/2021MNRAS.507.4953G} {507, 4953}

\bibitem[\protect\citeauthoryear{Graus, Bullock, Boylan-Kolchin  \&
  Nierenberg}{Graus et~al.}{2018}]{Graus_2018}
Graus A.~S.,  Bullock J.~S.,  Boylan-Kolchin M.,   Nierenberg A.~M.,  2018,
  \mn@doi [Monthly Notices of the Royal Astronomical Society]
  {10.1093/mnras/sty1924}, 480, 1322

\bibitem[\protect\citeauthoryear{Graus, Bullock, Kelley, Boylan-Kolchin,
  Garrison-Kimmel  \& Qi}{Graus et~al.}{2019}]{Graus_2019}
Graus A.~S.,  Bullock J.~S.,  Kelley T.,  Boylan-Kolchin M.,  Garrison-Kimmel
  S.,   Qi Y.,  2019, \mn@doi [Monthly Notices of the Royal Astronomical
  Society] {10.1093/mnras/stz1992}, 488, 4585

\bibitem[\protect\citeauthoryear{Hahn \& Abel}{Hahn \& Abel}{2011}]{Hahn_2011}
Hahn O.,  Abel T.,  2011, \mn@doi [Monthly Notices of the Royal Astronomical
  Society] {10.1111/j.1365-2966.2011.18820.x}, 415, 2101

\bibitem[\protect\citeauthoryear{Hargis, Willman  \& Peter}{Hargis
  et~al.}{2014}]{Hargis_2014}
Hargis J.~R.,  Willman B.,   Peter A. H.~G.,  2014, \mn@doi [The Astrophysical
  Journal] {10.1088/2041-8205/795/1/l13}, 795, L13

\bibitem[\protect\citeauthoryear{Harris et~al.,}{Harris
  et~al.}{2020}]{Harris_2020}
Harris C.~R.,  et~al., 2020, \mn@doi [Nature] {10.1038/s41586-020-2649-2}, 585,
  357

\bibitem[\protect\citeauthoryear{{Helmi}}{{Helmi}}{2020}]{Helmi_2020}
{Helmi} A.,  2020, \mn@doi [\araa] {10.1146/annurev-astro-032620-021917}, \href
  {https://ui.adsabs.harvard.edu/abs/2020ARA&A..58..205H} {58, 205}

\bibitem[\protect\citeauthoryear{{Helmi} \& {White}}{{Helmi} \&
  {White}}{1999}]{Helmi_1999}
{Helmi} A.,  {White} S. D.~M.,  1999, \mn@doi [\mnras]
  {10.1046/j.1365-8711.1999.02616.x}, \href
  {https://ui.adsabs.harvard.edu/abs/1999MNRAS.307..495H} {307, 495}

\bibitem[\protect\citeauthoryear{{Helmi}, {White}  \& {Springel}}{{Helmi}
  et~al.}{2002}]{Helmi_2002}
{Helmi} A.,  {White} S.~D.,   {Springel} V.,  2002, \mn@doi [\prd]
  {10.1103/PhysRevD.66.063502}, \href
  {https://ui.adsabs.harvard.edu/abs/2002PhRvD..66f3502H} {66, 063502}

\bibitem[\protect\citeauthoryear{{Hernquist}}{{Hernquist}}{1990}]{Hernquist_1990}
{Hernquist} L.,  1990, \mn@doi [\apj] {10.1086/168845}, \href
  {https://ui.adsabs.harvard.edu/abs/1990ApJ...356..359H} {356, 359}

\bibitem[\protect\citeauthoryear{{Hernquist}, {Bouchet}  \& {Suto}}{{Hernquist}
  et~al.}{1991}]{Hernquist_1991}
{Hernquist} L.,  {Bouchet} F.~R.,   {Suto} Y.,  1991, \mn@doi [\apjs]
  {10.1086/191530}, \href
  {https://ui.adsabs.harvard.edu/abs/1991ApJS...75..231H} {75, 231}

\bibitem[\protect\citeauthoryear{Hopkins}{Hopkins}{2015}]{Hopkins_2015}
Hopkins P.~F.,  2015, \mn@doi [Monthly Notices of the Royal Astronomical
  Society] {10.1093/mnras/stv195}, 450, 53

\bibitem[\protect\citeauthoryear{{Hopkins} et~al.,}{{Hopkins}
  et~al.}{2018}]{Hopkins_2018}
{Hopkins} P.~F.,  et~al., 2018, \mn@doi [\mnras] {10.1093/mnras/sty1690}, \href
  {https://ui.adsabs.harvard.edu/abs/2018MNRAS.480..800H} {480, 800}

\bibitem[\protect\citeauthoryear{{Jahn}, {Sales}, {Wetzel}, {Boylan-Kolchin},
  {Chan}, {El-Badry}, {Lazar}  \& {Bullock}}{{Jahn} et~al.}{2019}]{Jahn_2019}
{Jahn} E.~D.,  {Sales} L.~V.,  {Wetzel} A.,  {Boylan-Kolchin} M.,  {Chan}
  T.~K.,  {El-Badry} K.,  {Lazar} A.,   {Bullock} J.~S.,  2019, \mn@doi
  [\mnras] {10.1093/mnras/stz2457}, \href
  {https://ui.adsabs.harvard.edu/abs/2019MNRAS.489.5348J} {489, 5348}

\bibitem[\protect\citeauthoryear{Johnston, Bullock, Sharma, Font, Robertson  \&
  Leitner}{Johnston et~al.}{2008}]{Johnston_2008}
Johnston K.~V.,  Bullock J.~S.,  Sharma S.,  Font A.,  Robertson B.~E.,
  Leitner S.~N.,  2008, \mn@doi [The Astrophysical Journal] {10.1086/592228},
  689, 936

\bibitem[\protect\citeauthoryear{Jones, Oliphant, Peterson  et~al.}{Jones
  et~al.}{2001}]{Jones_2001}
Jones E.,  Oliphant T.,  Peterson P.,   et~al., 2001, {SciPy}: Open source
  scientific tools for {Python}, \url {http://www.scipy.org/}

\bibitem[\protect\citeauthoryear{{Kandrup} \& {Smith}}{{Kandrup} \&
  {Smith}}{1991}]{Kandrup_1991}
{Kandrup} H.~E.,  {Smith} Haywood J.,  1991, \mn@doi [\apj] {10.1086/170114},
  \href {https://ui.adsabs.harvard.edu/abs/1991ApJ...374..255K} {374, 255}

\bibitem[\protect\citeauthoryear{{Katz}}{{Katz}}{1991}]{Katz_1991}
{Katz} N.,  1991, \mn@doi [\apj] {10.1086/169696}, \href
  {https://ui.adsabs.harvard.edu/abs/1991ApJ...368..325K} {368, 325}

\bibitem[\protect\citeauthoryear{{Katz}, {Quinn}, {Bertschinger}  \&
  {Gelb}}{{Katz} et~al.}{1994}]{Katz_1994}
{Katz} N.,  {Quinn} T.,  {Bertschinger} E.,   {Gelb} J.~M.,  1994, \mn@doi
  [\mnras] {10.1093/mnras/270.1.L71}, \href
  {https://ui.adsabs.harvard.edu/abs/1994MNRAS.270L..71K} {270, L71}

\bibitem[\protect\citeauthoryear{Keller, Wadsley, Wang  \& Kruijssen}{Keller
  et~al.}{2018}]{Keller_2018}
Keller B.~W.,  Wadsley J.~W.,  Wang L.,   Kruijssen J. M.~D.,  2018, \mn@doi
  [Monthly Notices of the Royal Astronomical Society] {10.1093/mnras/sty2859},
  482, 2244

\bibitem[\protect\citeauthoryear{Kelley, Bullock, Garrison-Kimmel,
  Boylan-Kolchin, Pawlowski  \& Graus}{Kelley et~al.}{2019}]{Kelley_2019}
Kelley T.,  Bullock J.~S.,  Garrison-Kimmel S.,  Boylan-Kolchin M.,  Pawlowski
  M.~S.,   Graus A.~S.,  2019, \mn@doi [Monthly Notices of the Royal
  Astronomical Society] {10.1093/mnras/stz1553}, 487, 4409

\bibitem[\protect\citeauthoryear{{Kennicutt}}{{Kennicutt}}{1998}]{Kennicutt_1998}
{Kennicutt} Robert~C. J.,  1998, \mn@doi [\araa]
  {10.1146/annurev.astro.36.1.189}, \href
  {https://ui.adsabs.harvard.edu/abs/1998ARA&A..36..189K} {36, 189}

\bibitem[\protect\citeauthoryear{Kim, Peter  \& Hargis}{Kim
  et~al.}{2018}]{Kim_2018}
Kim S.~Y.,  Peter A. H.~G.,   Hargis J.~R.,  2018, \mn@doi [Phys. Rev. Lett.]
  {10.1103/PhysRevLett.121.211302}, 121, 211302

\bibitem[\protect\citeauthoryear{{Klypin}, {Kravtsov}, {Valenzuela}  \&
  {Prada}}{{Klypin} et~al.}{1999}]{Klypin_1999}
{Klypin} A.,  {Kravtsov} A.~V.,  {Valenzuela} O.,   {Prada} F.,  1999, \mn@doi
  [\apj] {10.1086/307643}, \href
  {https://ui.adsabs.harvard.edu/abs/1999ApJ...522...82K} {522, 82}

\bibitem[\protect\citeauthoryear{{Klypin}, {Yepes}, {Gottl{\"o}ber}, {Prada}
  \& {He{\ss}}}{{Klypin} et~al.}{2016}]{Klypin_2016}
{Klypin} A.,  {Yepes} G.,  {Gottl{\"o}ber} S.,  {Prada} F.,   {He{\ss}} S.,
  2016, \mn@doi [\mnras] {10.1093/mnras/stw248}, \href
  {https://ui.adsabs.harvard.edu/abs/2016MNRAS.457.4340K} {457, 4340}

\bibitem[\protect\citeauthoryear{Knebe, Wagner, Knollmann, Diekershoff  \&
  Krause}{Knebe et~al.}{2009}]{Knebe_2009}
Knebe A.,  Wagner C.,  Knollmann S.,  Diekershoff T.,   Krause F.,  2009,
  \mn@doi [The Astrophysical Journal] {10.1088/0004-637x/698/1/266}, 698, 266

\bibitem[\protect\citeauthoryear{Knebe et~al.,}{Knebe
  et~al.}{2011}]{Knebe_2011}
Knebe A.,  et~al., 2011, \mn@doi [Monthly Notices of the Royal Astronomical
  Society] {10.1111/j.1365-2966.2011.18858.x}, 415, 2293

\bibitem[\protect\citeauthoryear{{Kuhlen}, {Diemand}  \& {Madau}}{{Kuhlen}
  et~al.}{2007}]{Kuhlen_2007}
{Kuhlen} M.,  {Diemand} J.,   {Madau} P.,  2007, \mn@doi [\apj]
  {10.1086/522878}, \href
  {https://ui.adsabs.harvard.edu/abs/2007ApJ...671.1135K} {671, 1135}

\bibitem[\protect\citeauthoryear{Kuhlen, Madau  \& Silk}{Kuhlen
  et~al.}{2009}]{Kuhlen_2009}
Kuhlen M.,  Madau P.,   Silk J.,  2009, \mn@doi [Science]
  {10.1126/science.1174881}, 325, 970

\bibitem[\protect\citeauthoryear{{Kuhlen}, {Vogelsberger}  \&
  {Angulo}}{{Kuhlen} et~al.}{2012}]{Kuhlen_2012}
{Kuhlen} M.,  {Vogelsberger} M.,   {Angulo} R.,  2012, \mn@doi [Physics of the
  Dark Universe] {10.1016/j.dark.2012.10.002}, \href
  {https://ui.adsabs.harvard.edu/abs/2012PDU.....1...50K} {1, 50}

\bibitem[\protect\citeauthoryear{Le~Bret, Pontzen, Cooper, Frenk, Zolotov,
  Brooks, Governato  \& Parry}{Le~Bret et~al.}{2017}]{LeBret_2017}
Le~Bret T.,  Pontzen A.,  Cooper A.~P.,  Frenk C.,  Zolotov A.,  Brooks A.~M.,
  Governato F.,   Parry O.~H.,  2017, \mn@doi [Monthly Notices of the Royal
  Astronomical Society] {10.1093/mnras/stx552}, 468, 3212

\bibitem[\protect\citeauthoryear{Leroy, Garrison, Eisenstein, Joyce  \&
  Maleubre}{Leroy et~al.}{2020}]{Leroy_2020}
Leroy M.,  Garrison L.,  Eisenstein D.,  Joyce M.,   Maleubre S.,  2020,
  \mn@doi [Monthly Notices of the Royal Astronomical Society]
  {10.1093/mnras/staa3435}, 501, 5064

\bibitem[\protect\citeauthoryear{Li, Vogelsberger, Marinacci, Sales  \&
  Torrey}{Li et~al.}{2020}]{Li_2020}
Li H.,  Vogelsberger M.,  Marinacci F.,  Sales L.~V.,   Torrey P.,  2020,
  \mn@doi [Monthly Notices of the Royal Astronomical Society]
  {10.1093/mnras/staa3122}

\bibitem[\protect\citeauthoryear{Libeskind, Cole, Frenk, Okamoto  \&
  Jenkins}{Libeskind et~al.}{2006}]{Libeskind_2006}
Libeskind N.~I.,  Cole S.,  Frenk C.~S.,  Okamoto T.,   Jenkins A.,  2006,
  \mn@doi [Monthly Notices of the Royal Astronomical Society]
  {10.1111/j.1365-2966.2006.11205.x}, 374, 16

\bibitem[\protect\citeauthoryear{{Libeskind}, {Hoffman}, {Knebe}, {Steinmetz},
  {Gottl{\"o}ber}, {Metuki}  \& {Yepes}}{{Libeskind}
  et~al.}{2012}]{Libeskind_2012}
{Libeskind} N.~I.,  {Hoffman} Y.,  {Knebe} A.,  {Steinmetz} M.,
  {Gottl{\"o}ber} S.,  {Metuki} O.,   {Yepes} G.,  2012, \mn@doi [\mnras]
  {10.1111/j.1745-3933.2012.01222.x}, \href
  {https://ui.adsabs.harvard.edu/abs/2012MNRAS.421L.137L} {421, L137}

\bibitem[\protect\citeauthoryear{Ludlow, Schaye  \& Bower}{Ludlow
  et~al.}{2019}]{Ludlow_2019}
Ludlow A.~D.,  Schaye J.,   Bower R.,  2019, \mn@doi [Monthly Notices of the
  Royal Astronomical Society] {10.1093/mnras/stz1821}, 488, 3663

\bibitem[\protect\citeauthoryear{{Lynden-Bell}}{{Lynden-Bell}}{1967}]{Lynden-Bell_1967}
{Lynden-Bell} D.,  1967, \mn@doi [\mnras] {10.1093/mnras/136.1.101}, \href
  {https://ui.adsabs.harvard.edu/abs/1967MNRAS.136..101L} {136, 101}

\bibitem[\protect\citeauthoryear{{MacArthur}, {Courteau}  \&
  {Holtzman}}{{MacArthur} et~al.}{2003}]{MacArthur_2003}
{MacArthur} L.~A.,  {Courteau} S.,   {Holtzman} J.~A.,  2003, \mn@doi [\apj]
  {10.1086/344506}, \href
  {https://ui.adsabs.harvard.edu/abs/2003ApJ...582..689M} {582, 689}

\bibitem[\protect\citeauthoryear{{Mackereth} et~al.,}{{Mackereth}
  et~al.}{2019}]{Mackereth_2019}
{Mackereth} J.~T.,  et~al., 2019, \mn@doi [\mnras] {10.1093/mnras/sty2955},
  \href {https://ui.adsabs.harvard.edu/abs/2019MNRAS.482.3426M} {482, 3426}

\bibitem[\protect\citeauthoryear{{McConnachie}}{{McConnachie}}{2012}]{McConnachie_2012}
{McConnachie} A.~W.,  2012, \mn@doi [\aj] {10.1088/0004-6256/144/1/4}, \href
  {https://ui.adsabs.harvard.edu/abs/2012AJ....144....4M} {144, 4}

\bibitem[\protect\citeauthoryear{{McCord}}{{McCord}}{2017}]{McCord_2017}
{McCord} K.,  2017, PhD thesis, The University of Alabama

\bibitem[\protect\citeauthoryear{{Miyamoto} \& {Nagai}}{{Miyamoto} \&
  {Nagai}}{1975}]{Miyamoto_1975}
{Miyamoto} M.,  {Nagai} R.,  1975, \pasj, \href
  {https://ui.adsabs.harvard.edu/abs/1975PASJ...27..533M} {27, 533}

\bibitem[\protect\citeauthoryear{{Monachesi}, {G{\'o}mez}, {Grand},
  {Kauffmann}, {Marinacci}, {Pakmor}, {Springel}  \& {Frenk}}{{Monachesi}
  et~al.}{2016}]{Monachesi_2016_2}
{Monachesi} A.,  {G{\'o}mez} F.~A.,  {Grand} R. J.~J.,  {Kauffmann} G.,
  {Marinacci} F.,  {Pakmor} R.,  {Springel} V.,   {Frenk} C.~S.,  2016, \mn@doi
  [\mnras] {10.1093/mnrasl/slw052}, \href
  {https://ui.adsabs.harvard.edu/abs/2016MNRAS.459L..46M} {459, L46}

\bibitem[\protect\citeauthoryear{Monachesi et~al.,}{Monachesi
  et~al.}{2019}]{Monachesi_2019}
Monachesi A.,  et~al., 2019, \mn@doi [Monthly Notices of the Royal Astronomical
  Society] {10.1093/mnras/stz538}, 485, 2589

\bibitem[\protect\citeauthoryear{{Monaghan} \& {Lattanzio}}{{Monaghan} \&
  {Lattanzio}}{1985}]{Monaghan_1985}
{Monaghan} J.~J.,  {Lattanzio} J.~C.,  1985, \aap, \href
  {https://ui.adsabs.harvard.edu/abs/1985A&A...149..135M} {149, 135}

\bibitem[\protect\citeauthoryear{{Moore}, {Ghigna}, {Governato}, {Lake},
  {Quinn}, {Stadel}  \& {Tozzi}}{{Moore} et~al.}{1999}]{Moore_1999}
{Moore} B.,  {Ghigna} S.,  {Governato} F.,  {Lake} G.,  {Quinn} T.,  {Stadel}
  J.,   {Tozzi} P.,  1999, \mn@doi [\apjl] {10.1086/312287}, \href
  {https://ui.adsabs.harvard.edu/abs/1999ApJ...524L..19M} {524, L19}

\bibitem[\protect\citeauthoryear{More, Kravtsov, Dalal  \& Gottlöber}{More
  et~al.}{2011}]{More_2011}
More S.,  Kravtsov A.~V.,  Dalal N.,   Gottlöber S.,  2011, \mn@doi [The
  Astrophysical Journal Supplement Series] {10.1088/0067-0049/195/1/4}, 195, 4

\bibitem[\protect\citeauthoryear{{Munshi}, {Brooks}, {Christensen},
  {Applebaum}, {Holley-Bockelmann}, {Quinn}  \& {Wadsley}}{{Munshi}
  et~al.}{2019}]{Munshi_2019}
{Munshi} F.,  {Brooks} A.~M.,  {Christensen} C.,  {Applebaum} E.,
  {Holley-Bockelmann} K.,  {Quinn} T.~R.,   {Wadsley} J.,  2019, \mn@doi [\apj]
  {10.3847/1538-4357/ab0085}, \href
  {https://ui.adsabs.harvard.edu/abs/2019ApJ...874...40M} {874, 40}

\bibitem[\protect\citeauthoryear{{Nadler} et~al.,}{{Nadler}
  et~al.}{2021}]{Nadler_2021}
{Nadler} E.~O.,  et~al., 2021, \mn@doi [\prl] {10.1103/PhysRevLett.126.091101},
  \href {https://ui.adsabs.harvard.edu/abs/2021PhRvL.126i1101N} {126, 091101}

\bibitem[\protect\citeauthoryear{{Naidu} et~al.,}{{Naidu}
  et~al.}{2021}]{Naidu_2021}
{Naidu} R.~P.,  et~al., 2021, arXiv e-prints, \href
  {https://ui.adsabs.harvard.edu/abs/2021arXiv210303251N} {p. arXiv:2103.03251}

\bibitem[\protect\citeauthoryear{{Navarro} \& {White}}{{Navarro} \&
  {White}}{1994}]{Navarro_1994}
{Navarro} J.~F.,  {White} S. D.~M.,  1994, \mn@doi [\mnras]
  {10.1093/mnras/267.2.401}, \href
  {https://ui.adsabs.harvard.edu/abs/1994MNRAS.267..401N} {267, 401}

\bibitem[\protect\citeauthoryear{{Navarro}, {Frenk}  \& {White}}{{Navarro}
  et~al.}{1996}]{Navarro_1996}
{Navarro} J.~F.,  {Frenk} C.~S.,   {White} S. D.~M.,  1996, \mn@doi [\apj]
  {10.1086/177173}, \href
  {https://ui.adsabs.harvard.edu/abs/1996ApJ...462..563N} {462, 563}

\bibitem[\protect\citeauthoryear{{Navarro}, {Frenk}  \& {White}}{{Navarro}
  et~al.}{1997}]{Navarro_1997}
{Navarro} J.~F.,  {Frenk} C.~S.,   {White} S. D.~M.,  1997, \mn@doi [\apj]
  {10.1086/304888}, \href
  {https://ui.adsabs.harvard.edu/abs/1997ApJ...490..493N} {490, 493}

\bibitem[\protect\citeauthoryear{{Nelson} et~al.,}{{Nelson}
  et~al.}{2015}]{Nelson_2015}
{Nelson} D.,  et~al., 2015, \mn@doi [Astronomy and Computing]
  {10.1016/j.ascom.2015.09.003}, \href
  {https://ui.adsabs.harvard.edu/abs/2015A&C....13...12N} {13, 12}

\bibitem[\protect\citeauthoryear{{Newton}, {Cautun}, {Jenkins}, {Frenk}  \&
  {Helly}}{{Newton} et~al.}{2018}]{Newton_2018}
{Newton} O.,  {Cautun} M.,  {Jenkins} A.,  {Frenk} C.~S.,   {Helly} J.~C.,
  2018, \mn@doi [\mnras] {10.1093/mnras/sty1085}, \href
  {https://ui.adsabs.harvard.edu/abs/2018MNRAS.479.2853N} {479, 2853}

\bibitem[\protect\citeauthoryear{O\~{n}orbe, Garrison-Kimmel, Maller, Bullock,
  Rocha  \& Hahn}{O\~{n}orbe et~al.}{2013}]{Onorbe_2020}
O\~{n}orbe J.,  Garrison-Kimmel S.,  Maller A.~H.,  Bullock J.~S.,  Rocha M.,
  Hahn O.,  2013, \mn@doi [Monthly Notices of the Royal Astronomical Society]
  {10.1093/mnras/stt2020}, 437, 1894

\bibitem[\protect\citeauthoryear{{Obreja}, {Buck}  \& {Macci{\`o}}}{{Obreja}
  et~al.}{2022}]{Obreja_2022}
{Obreja} A.,  {Buck} T.,   {Macci{\`o}} A.~V.,  2022, \mn@doi [\aap]
  {10.1051/0004-6361/202140983}, \href
  {https://ui.adsabs.harvard.edu/abs/2022A&A...657A..15O} {657, A15}

\bibitem[\protect\citeauthoryear{{Oser}, {Ostriker}, {Naab}, {Johansson}  \&
  {Burkert}}{{Oser} et~al.}{2010}]{Oser_2010}
{Oser} L.,  {Ostriker} J.~P.,  {Naab} T.,  {Johansson} P.~H.,   {Burkert} A.,
  2010, \mn@doi [\apj] {10.1088/0004-637X/725/2/2312}, \href
  {https://ui.adsabs.harvard.edu/abs/2010ApJ...725.2312O} {725, 2312}

\bibitem[\protect\citeauthoryear{{Peebles}}{{Peebles}}{1974}]{Peebles_1974}
{Peebles} P.~J.~E.,  1974, \mn@doi [\apjl] {10.1086/181462}, \href
  {https://ui.adsabs.harvard.edu/abs/1974ApJ...189L..51P} {189, L51}

\bibitem[\protect\citeauthoryear{{Peebles}}{{Peebles}}{1982}]{Peebles_1982}
{Peebles} P.~J.~E.,  1982, \mn@doi [\apjl] {10.1086/183911}, \href
  {https://ui.adsabs.harvard.edu/abs/1982ApJ...263L...1P} {263, L1}

\bibitem[\protect\citeauthoryear{{Planck Collaboration} et~al.,}{{Planck
  Collaboration} et~al.}{2016}]{Planck_2016}
{Planck Collaboration} et~al., 2016, \mn@doi [\aap]
  {10.1051/0004-6361/201525830}, \href
  {https://ui.adsabs.harvard.edu/abs/2016A&A...594A..13P} {594, A13}

\bibitem[\protect\citeauthoryear{{Pohlen} \& {Trujillo}}{{Pohlen} \&
  {Trujillo}}{2006}]{Pohlen_2006}
{Pohlen} M.,  {Trujillo} I.,  2006, \mn@doi [\aap]
  {10.1051/0004-6361:20064883}, \href
  {https://ui.adsabs.harvard.edu/abs/2006A&A...454..759P} {454, 759}

\bibitem[\protect\citeauthoryear{{Power}, {Navarro}, {Jenkins}, {Frenk},
  {White}, {Springel}, {Stadel}  \& {Quinn}}{{Power} et~al.}{2003}]{Power_2003}
{Power} C.,  {Navarro} J.~F.,  {Jenkins} A.,  {Frenk} C.~S.,  {White} S.~D.~M.,
   {Springel} V.,  {Stadel} J.,   {Quinn} T.,  2003, \mn@doi [\mnras]
  {10.1046/j.1365-8711.2003.05925.x}, \href
  {https://ui.adsabs.harvard.edu/abs/2003MNRAS.338...14P} {338, 14}

\bibitem[\protect\citeauthoryear{{Prada}, {Forero-Romero}, {Grand}, {Pakmor}
  \& {Springel}}{{Prada} et~al.}{2019}]{Prada_2019}
{Prada} J.,  {Forero-Romero} J.~E.,  {Grand} R. J.~J.,  {Pakmor} R.,
  {Springel} V.,  2019, \mn@doi [\mnras] {10.1093/mnras/stz2873}, \href
  {https://ui.adsabs.harvard.edu/abs/2019MNRAS.490.4877P} {490, 4877}

\bibitem[\protect\citeauthoryear{{Price}}{{Price}}{2007}]{Price_2007}
{Price} D.~J.,  2007, \mn@doi [\pasa] {10.1071/AS07022}, \href
  {https://ui.adsabs.harvard.edu/abs/2007PASA...24..159P} {24, 159}

\bibitem[\protect\citeauthoryear{{Putman}, {Zheng}, {Price-Whelan}, {Grcevich},
  {Johnson}, {Tollerud}  \& {Peek}}{{Putman} et~al.}{2021}]{Putman_2021}
{Putman} M.~E.,  {Zheng} Y.,  {Price-Whelan} A.~M.,  {Grcevich} J.,  {Johnson}
  A.~C.,  {Tollerud} E.,   {Peek} J. E.~G.,  2021, \mn@doi [\apj]
  {10.3847/1538-4357/abe391}, \href
  {https://ui.adsabs.harvard.edu/abs/2021ApJ...913...53P} {913, 53}

\bibitem[\protect\citeauthoryear{Rashkov, Madau, Kuhlen  \& Diemand}{Rashkov
  et~al.}{2012}]{Rashkov_2012}
Rashkov V.,  Madau P.,  Kuhlen M.,   Diemand J.,  2012, \mn@doi [The
  Astrophysical Journal] {10.1088/0004-637x/745/2/142}, 745, 142

\bibitem[\protect\citeauthoryear{{Rey} \& {Starkenburg}}{{Rey} \&
  {Starkenburg}}{2022}]{Rey_2022}
{Rey} M.~P.,  {Starkenburg} T.~K.,  2022, \mn@doi [\mnras]
  {10.1093/mnras/stab3709}, \href
  {https://ui.adsabs.harvard.edu/abs/2022MNRAS.510.4208R} {510, 4208}

\bibitem[\protect\citeauthoryear{{Ricotti}, {Polisensky}  \&
  {Cleland}}{{Ricotti} et~al.}{2022}]{Ricotti_2022}
{Ricotti} M.,  {Polisensky} E.,   {Cleland} E.,  2022, \mn@doi [\mnras]
  {10.1093/mnras/stac1485}, \href
  {https://ui.adsabs.harvard.edu/abs/2022MNRAS.515..302R} {515, 302}

\bibitem[\protect\citeauthoryear{Robertson, Bullock, Font, Johnston  \&
  Hernquist}{Robertson et~al.}{2005}]{Robertson_2005}
Robertson B.,  Bullock J.~S.,  Font A.~S.,  Johnston K.~V.,   Hernquist L.,
  2005, \mn@doi [The Astrophysical Journal] {10.1086/452619}, 632, 872

\bibitem[\protect\citeauthoryear{{Ruggiero}}{{Ruggiero}}{2017}]{Ruggiero_2017}
{Ruggiero} R.,  2017, {galstep: Initial conditions for spiral galaxy
  simulations} (\mn@eprint {ascl} {1711.007})

\bibitem[\protect\citeauthoryear{{Samuel} et~al.,}{{Samuel}
  et~al.}{2020}]{Samuel_2020}
{Samuel} J.,  et~al., 2020, \mn@doi [\mnras] {10.1093/mnras/stz3054}, \href
  {https://ui.adsabs.harvard.edu/abs/2020MNRAS.491.1471S} {491, 1471}

\bibitem[\protect\citeauthoryear{Sanderson et~al.,}{Sanderson
  et~al.}{2018}]{Sanderson_2018}
Sanderson R.~E.,  et~al., 2018, \mn@doi [The Astrophysical Journal]
  {10.3847/1538-4357/aaeb33}, 869, 12

\bibitem[\protect\citeauthoryear{{Schaye} et~al.,}{{Schaye}
  et~al.}{2015}]{Schaye_2015}
{Schaye} J.,  et~al., 2015, \mn@doi [\mnras] {10.1093/mnras/stu2058}, \href
  {https://ui.adsabs.harvard.edu/abs/2015MNRAS.446..521S} {446, 521}

\bibitem[\protect\citeauthoryear{{Schechter}}{{Schechter}}{1976}]{Schechter_1976}
{Schechter} P.,  1976, \mn@doi [\apj] {10.1086/154079}, \href
  {https://ui.adsabs.harvard.edu/abs/1976ApJ...203..297S} {203, 297}

\bibitem[\protect\citeauthoryear{{Schneider}, {Frenk}  \& {Cole}}{{Schneider}
  et~al.}{2012}]{Schneider_2012}
{Schneider} M.~D.,  {Frenk} C.~S.,   {Cole} S.,  2012, \mn@doi [\jcap]
  {10.1088/1475-7516/2012/05/030}, \href
  {https://ui.adsabs.harvard.edu/abs/2012JCAP...05..030S} {2012, 030}

\bibitem[\protect\citeauthoryear{{Sharma}, {Steinmetz}  \&
  {Bland-Hawthorn}}{{Sharma} et~al.}{2012}]{Sharma_2012}
{Sharma} S.,  {Steinmetz} M.,   {Bland-Hawthorn} J.,  2012, \mn@doi [\apj]
  {10.1088/0004-637X/750/2/107}, \href
  {https://ui.adsabs.harvard.edu/abs/2012ApJ...750..107S} {750, 107}

\bibitem[\protect\citeauthoryear{{Shu}}{{Shu}}{1978}]{Shu_1978}
{Shu} F.~H.,  1978, \mn@doi [\apj] {10.1086/156470}, \href
  {https://ui.adsabs.harvard.edu/abs/1978ApJ...225...83S} {225, 83}

\bibitem[\protect\citeauthoryear{Silk \& Mamon}{Silk \&
  Mamon}{2012}]{Silk_2012}
Silk J.,  Mamon G.~A.,  2012, \mn@doi [Research in Astronomy and Astrophysics]
  {10.1088/1674-4527/12/8/004}, 12, 917

\bibitem[\protect\citeauthoryear{{Simon}}{{Simon}}{2019}]{Simon_2019}
{Simon} J.~D.,  2019, \mn@doi [\araa] {10.1146/annurev-astro-091918-104453},
  \href {https://ui.adsabs.harvard.edu/abs/2019ARA&A..57..375S} {57, 375}

\bibitem[\protect\citeauthoryear{Smith, Flynn, Candlish, Fellhauer  \&
  Gibson}{Smith et~al.}{2015}]{Smith_2015}
Smith R.,  Flynn C.,  Candlish G.~N.,  Fellhauer M.,   Gibson B.~K.,  2015,
  \mn@doi [Monthly Notices of the Royal Astronomical Society]
  {10.1093/mnras/stv228}, 448, 2934

\bibitem[\protect\citeauthoryear{{Springel}}{{Springel}}{2005}]{Springel_2005}
{Springel} V.,  2005, \mn@doi [\mnras] {10.1111/j.1365-2966.2005.09655.x},
  \href {https://ui.adsabs.harvard.edu/abs/2005MNRAS.364.1105S} {364, 1105}

\bibitem[\protect\citeauthoryear{Springel \& Hernquist}{Springel \&
  Hernquist}{2003}]{Springel_2003}
Springel V.,  Hernquist L.,  2003, \mn@doi [Monthly Notices of the Royal
  Astronomical Society] {10.1046/j.1365-8711.2003.06206.x}, 339, 289

\bibitem[\protect\citeauthoryear{{Springel}, {White}, {Tormen}  \&
  {Kauffmann}}{{Springel} et~al.}{2001}]{Springel_2001}
{Springel} V.,  {White} S. D.~M.,  {Tormen} G.,   {Kauffmann} G.,  2001,
  \mn@doi [\mnras] {10.1046/j.1365-8711.2001.04912.x}, \href
  {https://ui.adsabs.harvard.edu/abs/2001MNRAS.328..726S} {328, 726}

\bibitem[\protect\citeauthoryear{{Springel}, {Di Matteo}  \&
  {Hernquist}}{{Springel} et~al.}{2005}]{Springel_2005b}
{Springel} V.,  {Di Matteo} T.,   {Hernquist} L.,  2005, \mn@doi [\mnras]
  {10.1111/j.1365-2966.2005.09238.x}, \href
  {https://ui.adsabs.harvard.edu/abs/2005MNRAS.361..776S} {361, 776}

\bibitem[\protect\citeauthoryear{Springel et~al.,}{Springel
  et~al.}{2008}]{Springel_2008}
Springel V.,  et~al., 2008, \mn@doi [Monthly Notices of the Royal Astronomical
  Society] {10.1111/j.1365-2966.2008.14066.x}, 391, 1685

\bibitem[\protect\citeauthoryear{{Springel}, {Pakmor}, {Zier}  \&
  {Reinecke}}{{Springel} et~al.}{2020}]{Springel_2020}
{Springel} V.,  {Pakmor} R.,  {Zier} O.,   {Reinecke} M.,  2020, arXiv
  e-prints, \href {https://ui.adsabs.harvard.edu/abs/2020arXiv201003567S} {p.
  arXiv:2010.03567}

\bibitem[\protect\citeauthoryear{Srisawat et~al.,}{Srisawat
  et~al.}{2013}]{Srisawat_2013}
Srisawat C.,  et~al., 2013, \mn@doi [Monthly Notices of the Royal Astronomical
  Society] {10.1093/mnras/stt1545}, 436, 150

\bibitem[\protect\citeauthoryear{{Teklu}, {Remus}, {Dolag}, {Beck}, {Burkert},
  {Schmidt}, {Schulze}  \& {Steinborn}}{{Teklu} et~al.}{2015}]{Teklu_2015}
{Teklu} A.~F.,  {Remus} R.-S.,  {Dolag} K.,  {Beck} A.~M.,  {Burkert} A.,
  {Schmidt} A.~S.,  {Schulze} F.,   {Steinborn} L.~K.,  2015, \mn@doi [\apj]
  {10.1088/0004-637X/812/1/29}, \href
  {https://ui.adsabs.harvard.edu/abs/2015ApJ...812...29T} {812, 29}

\bibitem[\protect\citeauthoryear{{Tollerud}, {Bullock}, {Strigari}  \&
  {Willman}}{{Tollerud} et~al.}{2008}]{Tollerud_2008}
{Tollerud} E.~J.,  {Bullock} J.~S.,  {Strigari} L.~E.,   {Willman} B.,  2008,
  \mn@doi [\apj] {10.1086/592102}, \href
  {https://ui.adsabs.harvard.edu/abs/2008ApJ...688..277T} {688, 277}

\bibitem[\protect\citeauthoryear{{Tremaine}, {Henon}  \&
  {Lynden-Bell}}{{Tremaine} et~al.}{1986}]{Tremaine_1986}
{Tremaine} S.,  {Henon} M.,   {Lynden-Bell} D.,  1986, \mn@doi [\mnras]
  {10.1093/mnras/219.2.285}, \href
  {https://ui.adsabs.harvard.edu/abs/1986MNRAS.219..285T} {219, 285}

\bibitem[\protect\citeauthoryear{Triani, Sinha, Croton, Pacifici  \&
  Dwek}{Triani et~al.}{2020}]{Triani_2020}
Triani D.~P.,  Sinha M.,  Croton D.~J.,  Pacifici C.,   Dwek E.,  2020, \mn@doi
  [Monthly Notices of the Royal Astronomical Society] {10.1093/mnras/staa446},
  493, 2490

\bibitem[\protect\citeauthoryear{{Trowland}, {Lewis}  \&
  {Bland-Hawthorn}}{{Trowland} et~al.}{2013}]{Trowland_2013}
{Trowland} H.~E.,  {Lewis} G.~F.,   {Bland-Hawthorn} J.,  2013, \mn@doi [\apj]
  {10.1088/0004-637X/762/2/72}, \href
  {https://ui.adsabs.harvard.edu/abs/2013ApJ...762...72T} {762, 72}

\bibitem[\protect\citeauthoryear{{Turk}, {Smith}, {Oishi}, {Skory}, {Skillman},
  {Abel}  \& {Norman}}{{Turk} et~al.}{2011}]{Turk_2011}
{Turk} M.~J.,  {Smith} B.~D.,  {Oishi} J.~S.,  {Skory} S.,  {Skillman} S.~W.,
  {Abel} T.,   {Norman} M.~L.,  2011, \mn@doi [The Astrophysical Journal
  Supplement Series] {10.1088/0067-0049/192/1/9}, \href
  {https://ui.adsabs.harvard.edu/abs/2011ApJS..192....9T} {192, 9}

\bibitem[\protect\citeauthoryear{{Vasiliev}, {Belokurov}  \&
  {Erkal}}{{Vasiliev} et~al.}{2021}]{Vasiliev_2021}
{Vasiliev} E.,  {Belokurov} V.,   {Erkal} D.,  2021, \mn@doi [\mnras]
  {10.1093/mnras/staa3673}, \href
  {https://ui.adsabs.harvard.edu/abs/2021MNRAS.501.2279V} {501, 2279}

\bibitem[\protect\citeauthoryear{{Vera-Ciro} \& {Helmi}}{{Vera-Ciro} \&
  {Helmi}}{2013}]{Vera_Ciro_2013}
{Vera-Ciro} C.,  {Helmi} A.,  2013, \mn@doi [\apjl]
  {10.1088/2041-8205/773/1/L4}, \href
  {https://ui.adsabs.harvard.edu/abs/2013ApJ...773L...4V} {773, L4}

\bibitem[\protect\citeauthoryear{{Vera-Ciro}, {Sales}, {Helmi}, {Frenk},
  {Navarro}, {Springel}, {Vogelsberger}  \& {White}}{{Vera-Ciro}
  et~al.}{2011}]{Vera_Ciro_2011}
{Vera-Ciro} C.~A.,  {Sales} L.~V.,  {Helmi} A.,  {Frenk} C.~S.,  {Navarro}
  J.~F.,  {Springel} V.,  {Vogelsberger} M.,   {White} S. D.~M.,  2011, \mn@doi
  [\mnras] {10.1111/j.1365-2966.2011.19134.x}, \href
  {https://ui.adsabs.harvard.edu/abs/2011MNRAS.416.1377V} {416, 1377}

\bibitem[\protect\citeauthoryear{Vogelsberger et~al.,}{Vogelsberger
  et~al.}{2014a}]{Vogelsberger_2014}
Vogelsberger M.,  et~al., 2014a, \mn@doi [Monthly Notices of the Royal
  Astronomical Society] {10.1093/mnras/stu1536}, 444, 1518

\bibitem[\protect\citeauthoryear{{Vogelsberger} et~al.,}{{Vogelsberger}
  et~al.}{2014b}]{Vogelsberger_2014_2}
{Vogelsberger} M.,  et~al., 2014b, \mn@doi [\nat] {10.1038/nature13316}, \href
  {https://ui.adsabs.harvard.edu/abs/2014Natur.509..177V} {509, 177}

\bibitem[\protect\citeauthoryear{{Vogelsberger}, F.  \&
  Puchwein}{{Vogelsberger} et~al.}{2020}]{Vogelsberger_2020}
{Vogelsberger} M.,  F. M.,   Puchwein T. P. T.~E.,  2020, \mn@doi [Nature
  Reviews Physics] {10.1038/s42254-019-0127-2}, 2, 42

\bibitem[\protect\citeauthoryear{{Wadepuhl} \& {Springel}}{{Wadepuhl} \&
  {Springel}}{2011}]{Wadepuhl_2011}
{Wadepuhl} M.,  {Springel} V.,  2011, \mn@doi [\mnras]
  {10.1111/j.1365-2966.2010.17576.x}, \href
  {https://ui.adsabs.harvard.edu/abs/2011MNRAS.410.1975W} {410, 1975}

\bibitem[\protect\citeauthoryear{{Walsh}, {Willman}  \& {Jerjen}}{{Walsh}
  et~al.}{2009}]{Walsh_2009}
{Walsh} S.~M.,  {Willman} B.,   {Jerjen} H.,  2009, \mn@doi [\aj]
  {10.1088/0004-6256/137/1/450}, \href
  {https://ui.adsabs.harvard.edu/abs/2009AJ....137..450W} {137, 450}

\bibitem[\protect\citeauthoryear{{Wang}, {Dutton}, {Stinson}, {Macci{\`o}},
  {Penzo}, {Kang}, {Keller}  \& {Wadsley}}{{Wang} et~al.}{2015}]{Wang_2015}
{Wang} L.,  {Dutton} A.~A.,  {Stinson} G.~S.,  {Macci{\`o}} A.~V.,  {Penzo} C.,
   {Kang} X.,  {Keller} B.~W.,   {Wadsley} J.,  2015, \mn@doi [\mnras]
  {10.1093/mnras/stv1937}, \href
  {https://ui.adsabs.harvard.edu/abs/2015MNRAS.454...83W} {454, 83}

\bibitem[\protect\citeauthoryear{{Warren}, {Quinn}, {Salmon}  \&
  {Zurek}}{{Warren} et~al.}{1992}]{Warren_1992}
{Warren} M.~S.,  {Quinn} P.~J.,  {Salmon} J.~K.,   {Zurek} W.~H.,  1992,
  \mn@doi [\apj] {10.1086/171937}, \href
  {https://ui.adsabs.harvard.edu/abs/1992ApJ...399..405W} {399, 405}

\bibitem[\protect\citeauthoryear{Wetzel, Hopkins, hoon Kim,
  Faucher-Gigu{\`{e}}re, Kere{\v{s}}  \& Quataert}{Wetzel
  et~al.}{2016}]{Wetzel_2016}
Wetzel A.~R.,  Hopkins P.~F.,  hoon Kim J.,  Faucher-Gigu{\`{e}}re C.-A.,
  Kere{\v{s}} D.,   Quataert E.,  2016, \mn@doi [The Astrophysical Journal]
  {10.3847/2041-8205/827/2/l23}, 827, L23

\bibitem[\protect\citeauthoryear{{Wetzel} et~al.,}{{Wetzel}
  et~al.}{2022}]{Wetzel_2022}
{Wetzel} A.,  et~al., 2022, arXiv e-prints, \href
  {https://ui.adsabs.harvard.edu/abs/2022arXiv220206969W} {p. arXiv:2202.06969}

\bibitem[\protect\citeauthoryear{{White}}{{White}}{1984}]{White_1984}
{White} S.~D.~M.,  1984, \mn@doi [\apj] {10.1086/162573}, \href
  {https://ui.adsabs.harvard.edu/abs/1984ApJ...286...38W} {286, 38}

\bibitem[\protect\citeauthoryear{{White} \& {Frenk}}{{White} \&
  {Frenk}}{1991}]{White_1991}
{White} S. D.~M.,  {Frenk} C.~S.,  1991, \mn@doi [\apj] {10.1086/170483}, \href
  {https://ui.adsabs.harvard.edu/abs/1991ApJ...379...52W} {379, 52}

\bibitem[\protect\citeauthoryear{{White} \& {Rees}}{{White} \&
  {Rees}}{1978}]{White_1978}
{White} S.~D.~M.,  {Rees} M.~J.,  1978, \mn@doi [\mnras]
  {10.1093/mnras/183.3.341}, \href
  {https://ui.adsabs.harvard.edu/abs/1978MNRAS.183..341W} {183, 341}

\bibitem[\protect\citeauthoryear{{Yu} et~al.,}{{Yu} et~al.}{2020}]{Yu_2020}
{Yu} S.,  et~al., 2020, \mn@doi [\mnras] {10.1093/mnras/staa522}, \href
  {https://ui.adsabs.harvard.edu/abs/2020MNRAS.494.1539Y} {494, 1539}

\bibitem[\protect\citeauthoryear{{Zaritsky} \& {White}}{{Zaritsky} \&
  {White}}{1994}]{Zaritsky_1994}
{Zaritsky} D.,  {White} S.~D.~M.,  1994, in European Southern Observatory
  Conference and Workshop Proceedings. p.~355

\bibitem[\protect\citeauthoryear{{Zaritsky}, {Conroy}, {Zhang}, {Naidu},
  {Bonaca}, {Caldwell}, {Cargile}  \& {Johnson}}{{Zaritsky}
  et~al.}{2020}]{Zaritsky_2020}
{Zaritsky} D.,  {Conroy} C.,  {Zhang} H.,  {Naidu} R.~P.,  {Bonaca} A.,
  {Caldwell} N.,  {Cargile} P.~A.,   {Johnson} B.~D.,  2020, \mn@doi [\apj]
  {10.3847/1538-4357/ab5b93}, \href
  {https://ui.adsabs.harvard.edu/abs/2020ApJ...888..114Z} {888, 114}

\bibitem[\protect\citeauthoryear{{Zel'Dovich}}{{Zel'Dovich}}{1970}]{zeldovich_1970}
{Zel'Dovich} Y.~B.,  1970, \aap, \href
  {https://ui.adsabs.harvard.edu/abs/1970A&A.....5...84Z} {500, 13}

\bibitem[\protect\citeauthoryear{Zemp, Gnedin, Gnedin  \& Kravtsov}{Zemp
  et~al.}{2011}]{Zemp_2011}
Zemp M.,  Gnedin O.~Y.,  Gnedin N.~Y.,   Kravtsov A.~V.,  2011, \mn@doi [The
  Astrophysical Journal Supplement Series] {10.1088/0067-0049/197/2/30}, 197,
  30

\bibitem[\protect\citeauthoryear{Zhu, Marinacci, Maji, Li, Springel  \&
  Hernquist}{Zhu et~al.}{2016}]{Zhu_2016}
Zhu Q.,  Marinacci F.,  Maji M.,  Li Y.,  Springel V.,   Hernquist L.,  2016,
  \mn@doi [Monthly Notices of the Royal Astronomical Society]
  {10.1093/mnras/stw374}, 458, 1559

\bibitem[\protect\citeauthoryear{{Zjupa} \& {Springel}}{{Zjupa} \&
  {Springel}}{2017}]{Zjupa_2017}
{Zjupa} J.,  {Springel} V.,  2017, \mn@doi [\mnras] {10.1093/mnras/stw2945},
  \href {https://ui.adsabs.harvard.edu/abs/2017MNRAS.466.1625Z} {466, 1625}

\bibitem[\protect\citeauthoryear{{de Jong}}{{de Jong}}{1996}]{deJong_1996}
{de Jong} R.~S.,  1996, \aap, \href
  {https://ui.adsabs.harvard.edu/abs/1996A&A...313...45D} {313, 45}

\makeatother
\end{thebibliography}





\appendix

\section{Orbit test}
\label{sec:OrbitTest}

\begin{figure}
 	\includegraphics[width=\columnwidth]{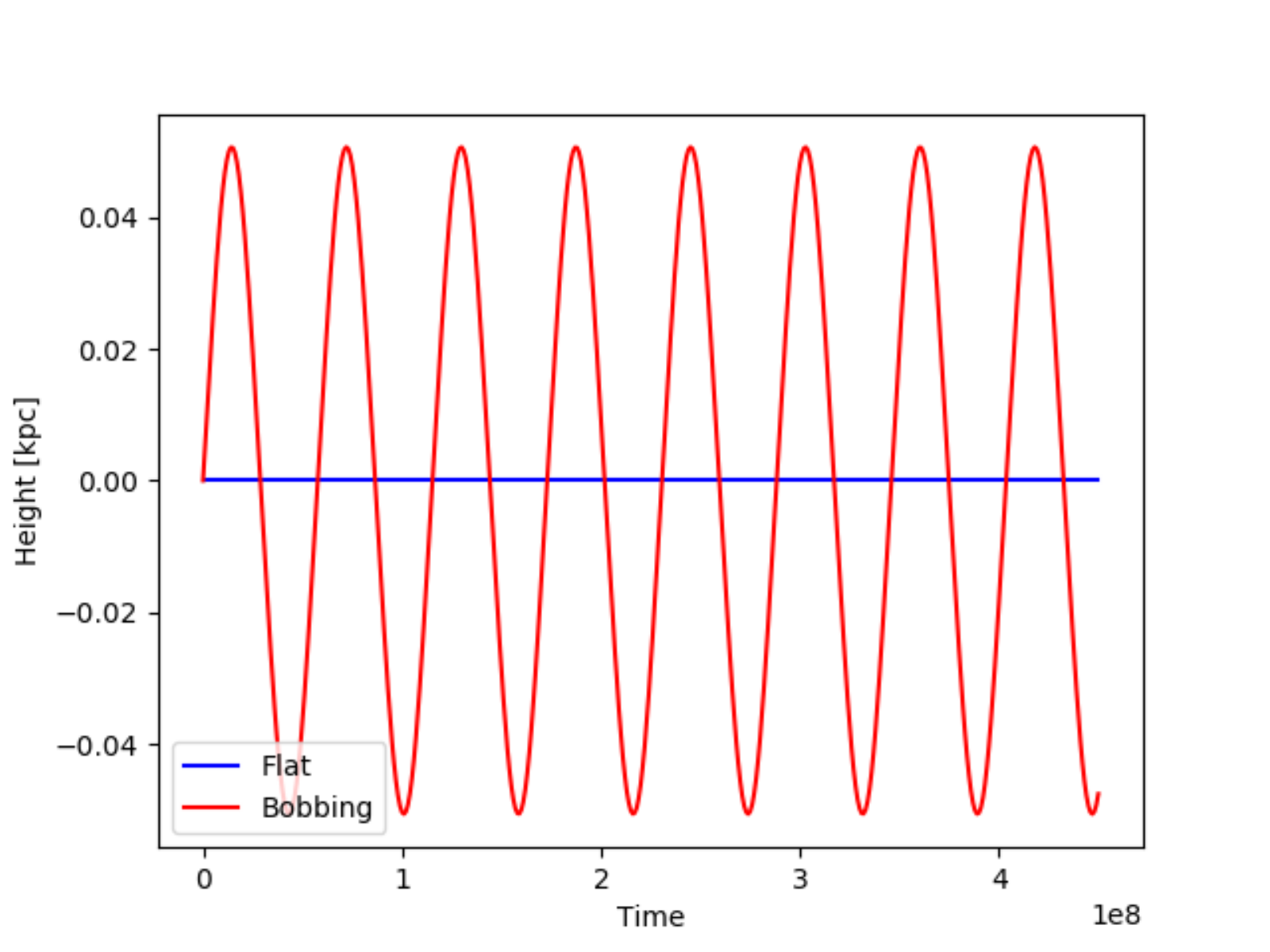}
     \caption{Evolution of a particle under the influence of the analytic potential. This figure shows two orbits of the two particles orbiting around the center at exactly the circular velocity (blue) and with a velocity component perpendicular to the disk (red). We can see the flat rotation and a stable oscillating orbit as expected.}
     \label{fig:OrbitTest}
 \end{figure}

One of the most important tests of the galactic potential embedded in the code is the orbit test. We test orbits both with CoSANG and with an independent orbit integrator.
The initial condition is produced manually and consists of a handful of particles with precise velocities and positions.
We configured CoSANG for non-cosmological simulation. We put test particles with different initial velocities and positions and we let them evolve under this potential. The orbit of these test particles are well-understood (\citealp{Binney_2008}). These tests confirm that the new analytic potential is consistent with known solutions.
We performed multiple orbit tests; Figure \ref{fig:OrbitTest} shows two orbits for a particle orbiting around the center at the expected circular velocity without any perpendicular component (blue) and with a velocity component perpendicular to the disk (red). We can see the flat rotation and a stable oscillating orbits. We recover the same results with the orbit integrator and with full CoSANG, both consistent with analytic orbits.

\section{Redshift evolution}
\label{sec:redshiftevolution}
\begin{figure*}
	\includegraphics[width=\textwidth]{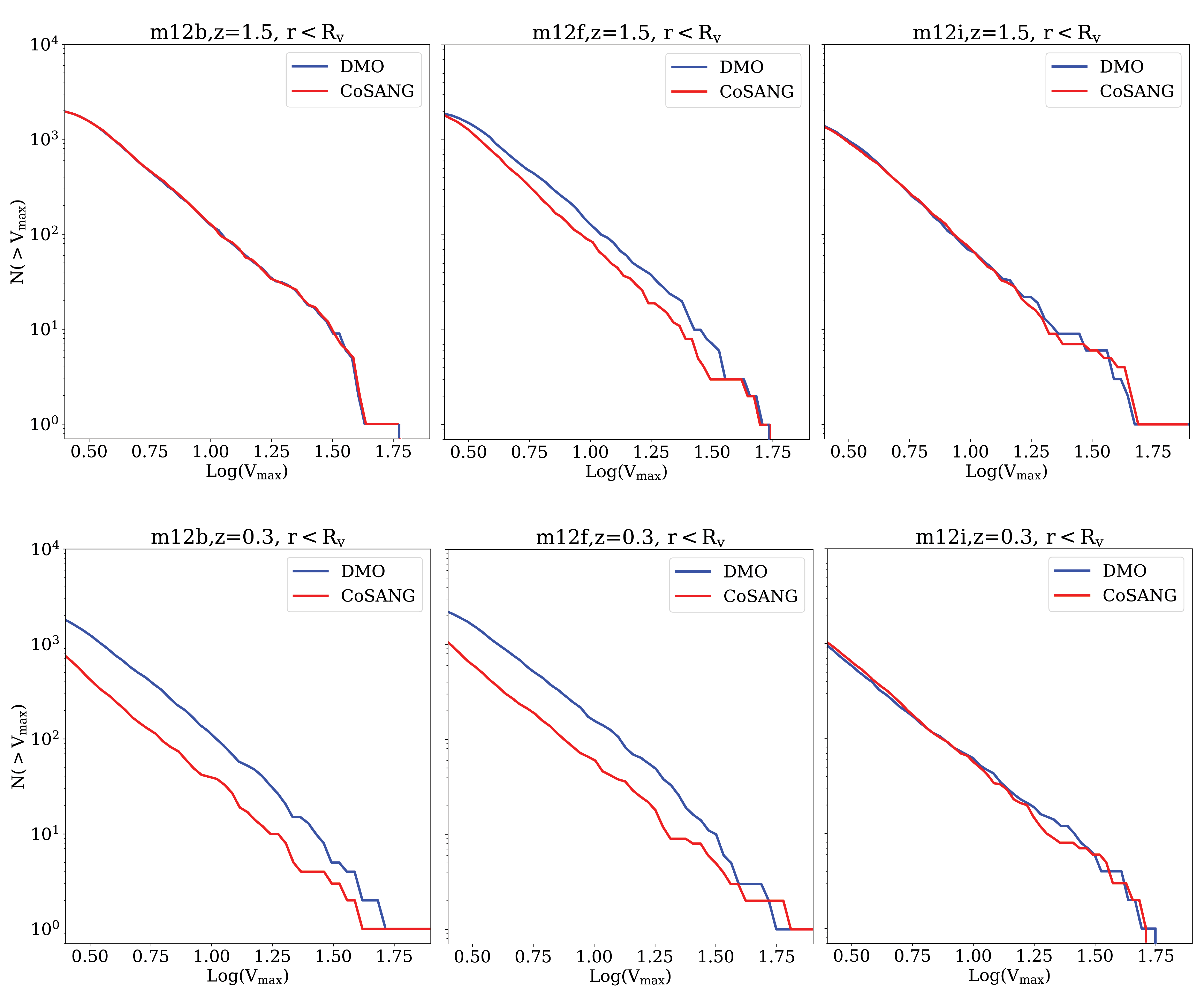}
    \caption{The redshift evolution of the cumulative $V_{\mathrm{max}}$ distribution. Redshift 1.5 (top) and redshift 0.3 (bottom) for m12b (left), m12f (middle) and m12i (right). The difference in the distribution is negligible at redshift 1.5 and becomes more significant by redshift 0.3.}
    \label{fig:vmax_z}
\end{figure*}

\begin{figure*}
	\includegraphics[width=\textwidth]{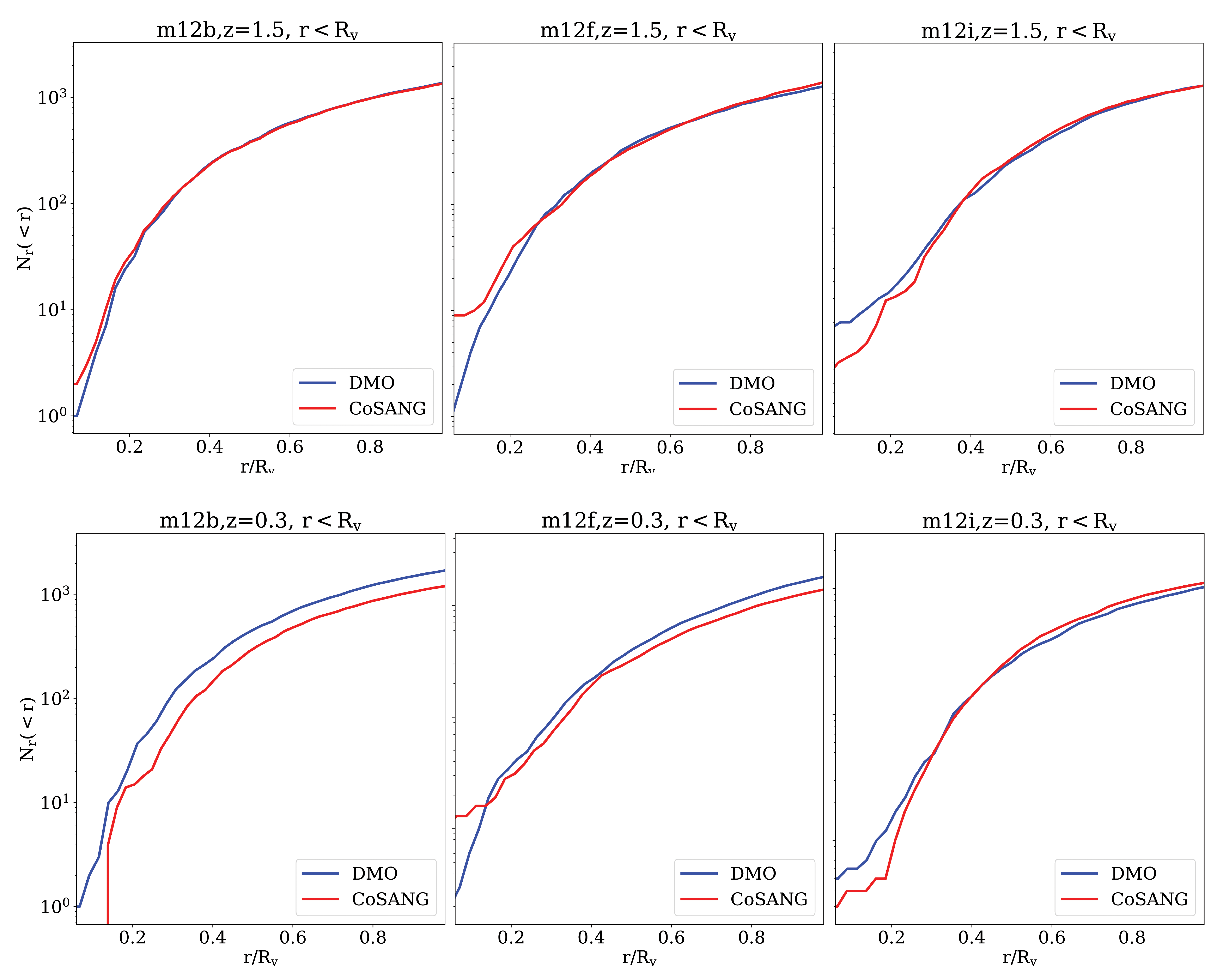}
    \caption{Redshift evolution of the radial distribution of subhaloes within the virial radius of m12b (left), m12f (middle) and m12i (right) in DMO (blue) and CoSANG (red) at redshift 1.5 (top) and redshift 0.3 (bottom). CoSANG and DMO are almost identical at redshift 1.5. In m12b the deviation between the models increases as the galaxy grows. In m12f and m12i this deviation is less significant.}
    \label{fig:RadialAbundanceRedshift}
\end{figure*}

The cumulative $V_\mathrm{max}$ (\ref{fig:vmax_z}) and the radial abundance of subhaloes (\ref{fig:RadialAbundanceRedshift}) at two different redshifts for both DMO (blue) and CoSANG (red) and all three models m12b (right), m12f (middle) and m12i (left). There is no significant difference between DMO and CoSANG at redshift 1.5 and higher. The difference  increases after redshift 1.5.

\bsp	
\label{lastpage}
\end{document}